\documentclass{jpp}

\usepackage{graphicx}
\usepackage{natbib}

\usepackage{graphics}
\usepackage{epsfig}
\usepackage{bm}
\usepackage{amsmath,amssymb}
\usepackage{stmaryrd}

\def\ov#1{\overline{#1}}

\def\wt#1{\widetilde{#1}}
\def\vb#1{\mbox{\boldmath$#1$}}
\def\pd#1#2{\frac{\partial #1}{\partial #2}}

\def\wh#1{\widehat{#1}}
\def\bdot{\,\vb{\cdot}\,}
\def\btimes{\,\vb{\times}\,}

\def\bhat{\wh{{\sf b}}}
\def\cal#1{\mathcal{#1}}

\def\exd{{\sf d}}
\def\bhat{\wh{{\sf b}}}

\newcommand{\bc}{\begin{center}}
\newcommand{\ec}{\end{center}}
\newcommand{\bt}{\begin{tabbing}}
\newcommand{\et}{\end{tabbing}}
\newcommand{\be}{\begin{equation}}
\newcommand{\ee}{\end{equation}}
\newcommand{\ba}{\begin{eqnarray}}
\newcommand{\ea}{\end{eqnarray}}

\title[Equivalent Higher-order Guiding-center Hamiltonian Theories]{Equivalent Higher-order Guiding-center Hamiltonian Theories}

\author[Brizard and Tronko]%
{A.~J.~Brizard$^1$%
  \thanks{Email address for correspondence: abrizard@smcvt.edu}\ns
\and N.~Tronko$^2$}

\affiliation{$^1$Department of Physics, Saint Michael's College, Colchester, VT 05439, USA\\[\affilskip]
$^2$Max-Planck-Institut f\"{u}r Plasmaphysik, 85748 Garching, Germany}

\date{?; revised ?; accepted ?. - To be entered by editorial office}
\begin{document}

\maketitle

\begin{abstract}
Equivalent guiding-center Hamiltonian theories are constructed based on higher-order Lie-transform perturbation methods. Higher-order guiding-center theories are distinguished on the basis of whether correction terms associated with magnetic-field nonuniformity appear either in the guiding-center symplectic (Poisson-bracket) structure, in the guiding-center Hamiltonian, or both. These theories are called equivalent because they describe the same guiding-center magnetic-moment invariant. The present work presents the detailed derivations of results that were summarized in a recent paper by Tronko and Brizard [Phys.~Plasmas {\bf 22}, 112507 (2015)].
\end{abstract}

\section{Introduction}

The concept of equivalent Hamiltonian theories plays a useful role in gyrokinetic theory \citep{Hahm_Lee_Brizard_1988, Brizard_Hahm_2007}, where magnetic-field perturbations either appear in the gyrocenter Hamiltonian (in the Hamiltonian representation) or the gyrocenter Poisson bracket (in the symplectic representation). These gyrokinetic theories are said to be equivalent since they both use the same definition for the gyrocenter magnetic moment \citep{Brizard_Hahm_2007}. In the Hamiltonian representation, the parallel gyrocenter momentum $\Pi_{\|}$ that appears in the gyrocenter symplectic structure is interpreted as a canonical momentum and, thus, the gyrocenter Hamiltonian is expressed in terms of the gauge-independent combination $\Pi_{\|} - (e/c)\,A_{1\|{\rm gc}}$, where $A_{1\|{\rm gc}}$ denotes the parallel component of the perturbed magnetic vector potential (expressed in terms of guiding-center coordinates). In the symplectic representation, on the other hand, the magnetic perturbation term $A_{1\|{\rm gc}}$ appears explicitly in the gyrocenter symplectic structure, which also includes the parallel gyrocenter (kinetic) momentum $p_{\|} = mv_{\|}$. In the symplectic representation of gyrokinetic theory, the parallel component of the inductive electric field now appears explicitly in the gyrocenter equation of motion for $p_{\|}$, while it is absent in the gyrocenter equation of motion for $\Pi_{\|}$ in the Hamiltonian representation. Further discussion of the equivalent representations of gyrokinetic theory can be found in the review paper of \cite{Brizard_Hahm_2007}.

The purpose of the present paper is to construct equivalent higher-order guiding-center Hamiltonian theories in which higher-order corrections associated with magnetic-field nonuniformity appear either in the guiding-center Poisson bracket (symplectic structure) or the guiding-center Hamiltonian. The main results presented here were summarized in a recent paper by \cite{Tronko_Brizard_2015} and a recent review of guiding-center Hamiltonian theory was presented by \cite{Cary_Brizard_2009}. The material contained in this manuscript is presented in tutorial form, with detailed calculations appearing in the literature for the first time. 

The remainder of the paper is organized as follows. In Sec.~\ref{sec:gcHam}, we present a summary of the general formulation of guiding-center Hamiltonian theory, in which corrections associated with magnetic-field nonuniformity appear at all orders in the guiding-center Hamiltonian and/or the guiding-center Poisson bracket. In Sec.~\ref{sec:Lie}, the formulation of Lie-transform perturbation theory for the Lagrange one-form is presented up to fourth order in the ordering parameter $\epsilon$, which are explicitly solved in Secs.~\ref{sec:first}-\ref{sec:fourth}. The ordering parameter $\epsilon$ is used in the renormalization of the electric charge $e \rightarrow e/\epsilon$ that appears in the {\it macroscopic} view of guiding-center dynamics, in which the magnetic-nonuniformity length scale is finite while the gyroradius is small. In Secs.~\ref{sec:Jacobian}-\ref{sec:push_Lag}, we present the Jacobian and Lagrangian constraints that establish the consistency of the guiding-center phase-space transformation. In Sec.~\ref{sec:gc_pol_canonical}, we derive the guiding-center polarization directly from the guiding-center transformation, which further constrains the transformation, and discuss the conservation of the guiding-center toroidal canonical momentum. In Sec.~\ref{sec:sum}, we summarize our work. Lastly, the Appendices \ref{sec:dyadic}-\ref{sec:Pgcphi_gyro} provide a wealth of results that support the material presented in the text.

\section{\label{sec:gcHam}Guiding-center Hamiltonian Theory}

Guiding-center Hamiltonian dynamics is expressed in terms of a guiding-center Hamiltonian function that depends on the guiding-center position ${\bf X}$, the guiding-center parallel momentum $p_{\|}$, and the guiding-center gyroaction $J \equiv \mu\,B/\Omega$ (defined in terms of the guiding-center magnetic moment $\mu$ and the gyrofrequency $\Omega = eB/mc$ for a particle of mass $m$ and charge $e$); it is, however, independent of the gyroangle $\theta$ at all orders. Since the guiding-center phase-space coordinates are non-canonical coordinates, a noncanonical guiding-center Poisson bracket is also needed. For the sake of simplicity, the guiding-center Hamiltonian theory is presented here for a time-independent nonuniform magnetic field in the absence of an electric field.

In this Section, we summarize the results of the Hamiltonian formulation of guiding-center dynamics in a nonuniform magnetic field \citep{Tronko_Brizard_2015}. Here, the guiding-center Hamiltonian is defined as
\begin{equation}
H_{\rm gc} \;\equiv\; \frac{p_{\|}^{2}}{2m} \;+\; \Psi,
\label{eq:Hamiltonian_gc}
\end{equation}
where the effective guiding-center potential energy
\begin{equation}
\Psi \;=\;  \sum_{n = 0}^{\infty}\epsilon^{n}\,\Psi_{n} \;\equiv\; J\,\Omega \;+\; \epsilon\,\Psi_{1} \;+\; \epsilon^{2}\,\Psi_{2} \;+\; \cdots
\label{eq:Psi_def}
\end{equation}
is defined in terms of higher-order corrections $\Psi_{n}$ ($n \geq 1$) that vanish in a uniform magnetic field. The guiding-center symplectic structure, on the other hand, is expressed in terms of the guiding-center Poincar\'{e}-Cartan one-form
\begin{eqnarray}
\Gamma_{\rm gc} & \equiv & \left( \frac{e}{\epsilon c}\,{\bf A} \;+\; \vb{\Pi} \right)\bdot\exd{\bf X} \;+\; \epsilon\,J\left(\exd\theta \;-\; {\bf R}\bdot\exd{\bf X}\right), 
\label{eq:Gamma_gc}
\end{eqnarray}
where the symplectic guiding-center momentum
\begin{equation}
\vb{\Pi} \;=\; \sum_{n = 0}^{\infty}\epsilon^{n}\,\vb{\Pi}_{n} \;\equiv\; p_{\|}\,\bhat \;+\; \epsilon\,\vb{\Pi}_{1} \;+\; \epsilon^{2}\,\vb{\Pi}_{2} + \cdots 
\label{eq:Pi_def}
\end{equation}
is expressed in terms of the gyroangle-independent vector terms $\vb{\Pi}_{n}$ ($n \geq 1$), which contain corrections due to magnetic-field nonuniformity, and the presence of the gyrogauge vector ${\bf R} \equiv
\nabla\wh{\sf 1}\bdot\wh{\sf 2}$ (where $\bhat \equiv \wh{\sf 1}\btimes\wh{\sf 2}$) guarantees that the guiding-center one-form \eqref{eq:Gamma_gc} is gyrogauge-invariant (see App.~\ref{sec:dyadic}). 

\subsection{Guiding-center Euler-Lagrange equations}

The guiding-center Hamiltonian \eqref{eq:Hamiltonian_gc} and the guiding-center Poincar\'{e}-Cartan one-form \eqref{eq:Gamma_gc} can be used to construct the guiding-center phase-space Lagrangian 
\begin{eqnarray}
\Lambda_{\rm gc} \;\equiv\; \Gamma_{\rm gc} \;-\; H_{\rm gc}\;\exd t & \equiv & \left[ \left(\frac{e}{\epsilon c}\,{\bf A} + p_{\|}\,\bhat +  \sum_{n=1}^{\infty} \epsilon^{n}\,\vb{\Pi}_{n} \right)\bdot\exd{\bf X} \;+\; \epsilon\;J\;\left(\exd\theta \;-\frac{}{} {\bf R}\bdot\exd{\bf X}\right) \right] \nonumber \\
 &  &-\; \left(\frac{p_{\|}^{2}}{2m} \;+\; J\,\Omega \;+\; \sum_{n=1}^{\infty} \epsilon^{n}\,\Psi_{n}\right)\;\exd t.
 \label{eq:Lambda_gc}
\end{eqnarray}
By using the guiding-center variational principle
\begin{equation}
0 \;=\; \int \delta\Lambda_{\rm gc} \;=\; \int \delta Z^{\alpha} \left[ (\vb{\omega}_{\rm gc})_{\alpha\beta}\;dZ^{\beta} \;-\; \pd{H_{\rm gc}}{Z^{\alpha}}\;
dt \right],
\end{equation}
we obtain the guiding-center Euler-Lagrange equations
\begin{equation}
(\vb{\omega}_{\rm gc})_{\alpha\beta}\;\frac{d_{\rm gc}Z^{\beta}}{dt} \;=\; \pd{H_{\rm gc}}{Z^{\alpha}}.
\label{eq:EL_gc}
\end{equation}
Here, the matrix elements $(\vb{\omega}_{\rm gc})_{\alpha\beta}$ are the components of the guiding-center Lagrange two-form
\begin{equation} 
\vb{\omega}_{\rm gc} \equiv \exd\Gamma_{\rm gc} =  \frac{eB^{*k}}{2\,\epsilon c}\,\varepsilon_{ijk}\;\exd X^{i}\wedge\exd X^{j} +
\exd p_{\|}\wedge{\sf b}^{*}\bdot\exd{\bf X} + \epsilon\;{\bf R}^{*}\bdot\exd{\bf X}\wedge\exd J + \epsilon\;\exd J\wedge\exd\theta, 
\label{eq:omega_gc}
\end{equation}
where $\varepsilon_{ijk}$ denotes the Levi-Civita tensor and we use the definitions
\begin{eqnarray}
{\bf B}^{*} & \equiv & \nabla\btimes\left[ {\bf A} \;+\; \frac{c}{e}\,\left( \epsilon\,\vb{\Pi} \;-\frac{}{}
\epsilon^{2}\;J\;{\bf R} \right) \right] \;=\; {\bf B} \;+\; \frac{\epsilon c}{e}\,p_{\|}\;\nabla\btimes\bhat \;+\; \cdots, \label{eq:Bstar_def} \\
{\sf b}^{*} & \equiv & \pd{\vb{\Pi}}{p_{\|}} \;=\; \bhat \;+\; \epsilon\;\pd{\vb{\Pi}_{1}}{p_{\|}} \;+\; \epsilon^{2}\;\pd{\vb{\Pi}_{2}}{p_{\|}} \;+\; \cdots, 
\label{eq:bstar_def} \\
{\bf R}^{*} & \equiv & {\bf R} \;-\; \epsilon^{-1}\;\pd{\vb{\Pi}}{J} \;=\; {\bf R} \;-\; \pd{\vb{\Pi}_{1}}{J} \;-\; \epsilon\,\pd{\vb{\Pi}_{2}}{J} \;+\; \cdots. \label{eq:Rstar_def}
\end{eqnarray}
We note that the fields ${\bf B}^{*}$ and ${\sf b}^{*}$ satisfy the identities
\begin{equation}
\left. \begin{array}{rcl}
\nabla\bdot{\bf B}^{*} & \equiv & 0 \\
 &  & \\
\partial{\bf B}^{*}/\partial p_{\|} & \equiv & \epsilon\,(c/e)\,\nabla\btimes{\sf b}^{*}
\end{array} \right\},
\label{eq:Bbhat_star_id}
\end{equation}
which will play an important role in the properties of the guiding-center Poisson bracket.

Using the components of the guiding-center Lagrange two-form \eqref{eq:omega_gc}, the guiding-center Euler-Lagrange equations \eqref{eq:EL_gc} become
\begin{eqnarray}
\frac{d_{\rm gc}{\bf X}}{dt} \btimes\frac{e{\bf B}^{*}}{\epsilon\,c} \;-\; {\sf b}^{*}\;\frac{d_{\rm gc}p_{\|}}{dt} \;+\; \epsilon\,{\bf R}^{*}\;\frac{d_{\rm gc}J}{dt} & = & \nabla H_{\rm gc}, \label{eq:ELgc_X} \\
{\sf b}^{*}\bdot\frac{d_{\rm gc}{\bf X}}{dt} & = & \pd{H_{\rm gc}}{p_{\|}}, \label{eq:ELgc_p} \\
\epsilon\;\frac{d_{\rm gc}\theta}{dt} \;-\; \epsilon\,{\bf R}^{*}\bdot\frac{d_{\rm gc}{\bf X}}{dt} & = & \pd{H_{\rm gc}}{J}, \label{eq:ELgc_J} \\
-\;\epsilon\;\frac{d_{\rm gc}J}{dt}  & = & \pd{H_{\rm gc}}{\theta}. \label{eq:ELgc_theta}
\end{eqnarray}
Since the guiding-center Hamiltonian \eqref{eq:Hamiltonian_gc} is explicitly independent of the gyroangle $\theta$, the guiding-center gyroaction $J$ is, therefore, an exact guiding-center invariant in Eqs.~\eqref{eq:ELgc_X} and \eqref{eq:ELgc_theta}: 
\begin{equation}
\frac{d_{\rm gc}J}{dt} \;=\; -\;\epsilon^{-1}\,\pd{H_{\rm gc}}{\theta} \;\equiv\; 0;
\label{eq:gc_J_dot}
\end{equation}
it is still, however, an adiabatic invariant of the exact particle dynamics. The guiding-center equation for the gyroangle $\theta$ is given by 
Eq.~\eqref{eq:ELgc_J} as
\begin{equation}
\frac{d_{\rm gc}\theta}{dt} \;=\; \epsilon^{-1}\pd{H_{\rm gc}}{J} \;+\; {\bf R}^{*}\bdot\frac{d_{\rm gc}{\bf X}}{dt},
\label{eq:gc_theta_dot}
\end{equation}
where the first term includes the lowest-order gyrofrequency as well as higher-order corrections due to magnetic-field nonuniformity, which are also included in the second term (involving the gyrogauge vector ${\bf R}$). 

The remaining guiding-center equations of motion for the guiding-center position ${\bf X}$ and the guiding-center parallel kinetic momentum $p_{\|}$ are obtained from Eqs.~\eqref{eq:ELgc_X}-\eqref{eq:ELgc_p} as follows. First, we take the cross-product of Eq.~\eqref{eq:ELgc_X} with ${\sf b}^{*}$ and use Eq.~\eqref{eq:ELgc_p} to obtain the guiding-center equation of motion for ${\bf X}$:
\begin{equation}
\frac{d_{\rm gc}{\bf X}}{dt} \;=\; \pd{H_{\rm gc}}{p_{\|}}\;\frac{{\bf B}^{*}}{B_{\|}^{**}} \;+\; \epsilon\;\frac{c\,{\sf b}^{*}}{eB_{\|}^{**}}\btimes\nabla H_{\rm gc},
\label{eq:gc_X_dot}
\end{equation}
where
\begin{equation}
B_{\|}^{**} \;\equiv\; {\sf b}^{*}\bdot{\bf B}^{*} = \left( \bhat + \epsilon\pd{\vb{\Pi}_{1}}{p_{\|}} + \cdots \right)\bdot{\bf B}^{*} \;=\; B_{\|}^{*} \;+\; \epsilon\;\pd{\vb{\Pi}_{1}}{p_{\|}}\bdot{\bf B}^{*} + \cdots.
\label{eq:B||star_def}
\end{equation}
Next, the guiding-center equation of motion for $p_{\|}$ is obtained by taking the dot-product of Eq.~\eqref{eq:ELgc_X} with ${\bf B}^{*}$, which yields
\begin{equation}
\frac{d_{\rm gc}p_{\|}}{dt} \;=\; -\;\frac{{\bf B}^{*}}{B_{\|}^{**}}\bdot\nabla H_{\rm gc}.
\label{eq:gc_p_dot}
\end{equation}
We note that Eqs.~\eqref{eq:gc_X_dot} and \eqref{eq:gc_p_dot} satisfy the guiding-center Liouville theorem
\begin{equation}
\nabla\bdot\left( B_{\|}^{**}\;\frac{d_{\rm gc}{\bf X}}{dt}\right) \;+\; \pd{}{p_{\|}}\left(B_{\|}^{**}\;\frac{d_{\rm gc}p_{\|}}{dt}\right) \;=\; 0,
\label{eq:gc_Liouville}
\end{equation}
which follows from the identities \eqref{eq:Bbhat_star_id}. 

\subsection{Guiding-center Hamilton equations}

We now wish to show that the guiding-center equations of motion \eqref{eq:gc_J_dot}-\eqref{eq:gc_X_dot} and \eqref{eq:gc_p_dot} can be expressed in Hamiltonian form in terms of a guiding-center Poisson bracket $\{\;,\;\}_{\rm gc}$ constructed from the guiding-center Lagrange two-form \eqref{eq:omega_gc}.

First, we note that the Lagrange component-matrix \eqref{eq:omega_gc} is invertible if the guiding-center Jacobian does not vanish:
\begin{equation}
{\cal J}_{\rm gc} \;\equiv\; \sqrt{{\rm det}(\vb{\omega}_{\rm gc})} \;=\; \epsilon\;{\sf b}^{*}\bdot\left(\frac{e}{\epsilon\,c}\;{\bf B}^{*}\right) \;\equiv\; 
\frac{e}{c}\;B_{\|}^{**} \;\neq\; 0,
\label{eq:Jac_gc}
\end{equation}
which is consistent with the guiding-center ordering itself. We note here that, since $B^{**}_{\|} = B\,(1 + \epsilon\,\varrho_{\|}\,\bhat\bdot\nabla\btimes\bhat + \cdots)$ up to the first
order in $\epsilon$, we might be concerned with the possibility of $B^{**}_{\|}$ vanishing if the parallel guiding-center velocity $v_{\|}$ is large enough, i.e., when $|v_{\|}| =
|v_{\|{\rm cr}}| \equiv L_{\tau} \Omega$, where $L_{\tau} \equiv |\bhat\bdot\nabla\btimes\bhat|^{-1}$. If we introduce the ordering $L_{\tau} = \rho_{\rm th}\,\epsilon_{B}^{-1}$ 
expressed in terms of the thermal gyroradius $\rho_{\rm th} = v_{\rm th}/\Omega$, we then obtain the ordering $|v_{\|{\rm cr}}|/v_{\rm th} = \epsilon_{B}^{-1} \gg 1$. Hence, 
only (extreme) superthermal parallel guiding-center motion would cause the guiding-center Jacobian to become singular. Under those circumstances, however, the standard 
guiding-center orderings would also break down (e.g., the curvature-drift motion would now be ordered at $\epsilon_{B}^{-1}$), and thus, the guiding-center theory would become 
invalid for these superthermal particles. In any case, we note that the removal of this singularity can also be accomplished by a process of regularization of guiding-center theory \citep{CRW_1985, CRPW_1986}.

Next, we invert the guiding-center Lagrange matrix defined in Eq.~\eqref{eq:omega_gc} to construct the guiding-center Poisson matrix with components $J_{\rm gc}^{\alpha\beta}$, such that $J_{\rm gc}^{\alpha\nu}\,(\omega_{\rm gc})_{\nu\beta} \equiv \delta^{\alpha}_{\;\beta}$. Lastly, we construct the guiding-center Poisson bracket $\{F,\; G\}_{\rm gc} \equiv (\partial F/\partial Z^{\alpha})\,J_{\rm gc}^{\alpha\beta}\,(\partial G/\partial Z^{\beta})$:
\begin{eqnarray}
\left\{ F,\frac{}{} G\right\}_{\rm gc} & = & \epsilon^{-1} \left( \pd{F}{\theta}\,\pd{G}{J} \;-\; \pd{F}{J}\,\pd{G}{\theta} \right) \;+\;
\frac{{\bf B}^{*}}{B_{\|}^{**}}\bdot\left(\nabla^{*}F\;\pd{G}{p_{\|}} \;-\; \pd{F}{p_{\|}}\;\nabla^{*}G \right) \nonumber \\
 &  &-\; \frac{\epsilon\,c{\sf b}^{*}}{e\,B_{\|}^{**}}\bdot\nabla^{*}F\btimes\nabla^{*}G,
\label{eq:PB_gc_star}
\end{eqnarray}
where the modified gradient operator $\nabla^{*} \equiv \nabla + {\bf R}^{*}\partial/\partial\theta$ is gyrogauge-invariant. The derivation procedure of the guiding-center Poisson bracket \eqref{eq:PB_gc_star} guarantees that it satisfies the standard Poisson-bracket properties, while the guiding-center Jacobian \eqref{eq:Jac_gc} can be used to write Eq.~\eqref{eq:PB_gc_star} in phase-space divergence form
\begin{equation}
\left\{ F,\frac{}{} G\right\}_{\rm gc} \;=\; \frac{1}{{\cal J}_{\rm gc}}\;\pd{}{Z^{\alpha}}\left({\cal J}_{\rm gc}\;F\frac{}{} \left\{ Z^{\alpha},\;
G\right\}_{\rm gc}\right).
\label{eq:PBgc_div}
\end{equation} 
The guiding-center equations of motion \eqref{eq:gc_J_dot}-\eqref{eq:gc_X_dot} and \eqref{eq:gc_p_dot} can thus be expressed in Hamiltonian form as 
\begin{equation}
\frac{d_{\rm gc}Z^{\alpha}}{dt} \;\equiv\; \{ Z^{\alpha},\; H_{\rm gc}\}_{\rm gc}.
\label{eq:Zgc_Ham}
\end{equation}

Lastly, it will later be useful to expand Eqs.~\eqref{eq:gc_X_dot} and \eqref{eq:gc_p_dot} in powers of $\epsilon$ as
\begin{equation}
\frac{d_{\rm gc}Z^{\alpha}}{dt} \;\equiv\; \sum_{n = 0}^{\infty}\;\epsilon^{n}\;\frac{d^{*}_{n}Z^{\alpha}}{dt},
\end{equation}
so that, without expanding $B_{\|}^{**}$, we find the guiding-center velocity at the first three orders:
\begin{eqnarray}
\frac{d^{*}_{0}{\bf X}}{dt} & = & \frac{p_{\|}}{m}\;\frac{\bf B}{B_{\|}^{**}},
\label{eq:dotXgc_0} \\
\frac{d^{*}_{1}{\bf X}}{dt} & = & \pd{\Psi_{1}}{p_{\|}}\frac{\bf B}{B_{\|}^{**}} \;+\; \frac{c}{e\,B_{\|}^{**}} \left( \frac{p_{\|}^{2}}{m}\;
\nabla\btimes\bhat + \bhat\btimes\nabla(J\Omega) \right),
\label{eq:dotXgc_1} \\
\frac{d^{*}_{2}{\bf X}}{dt} & = & \pd{\Psi_{2}}{p_{\|}}\frac{\bf B}{B_{\|}^{**}} \;+\; \frac{c}{e\,B_{\|}^{**}} \left[ \frac{p_{\|}}{m}\;\nabla\btimes(
\vb{\Pi}_{1} - J\,{\bf R}) \;+\; \pd{\vb{\Pi}_{1}}{p_{\|}}\btimes\nabla(J\,\Omega) \right] \nonumber \\
 &  &+\;   \frac{c}{e\,B_{\|}^{**}} \left( p_{\|}\,\pd{\Psi_{1}}{p_{\|}}\;\nabla\btimes\bhat \;+\; \bhat\btimes\nabla\Psi_{1} \right),
\label{eq:dotXgc_2}
\end{eqnarray}
where Eq.~\eqref{eq:dotXgc_0} represents the parallel motion along magnetic-field lines, Eq.~\eqref{eq:dotXgc_1} represents the first-order parallel Ba\~{n}os drift velocity \citep{Banos_1967}, defined as $\partial\Psi_{1}/\partial p_{\|}$, as well as the perpendicular magnetic drift velocities, and Eq.~\eqref{eq:dotXgc_2} represents second-order corrections. The guiding-center force equation, on the other hand, is given at the first three orders as 
\begin{eqnarray}
\frac{d^{*}_{0}p_{\|}}{dt} & = & -\;\frac{{\bf B}}{B_{\|}^{**}}\bdot\nabla(J\,\Omega) \;=\; J\,\Omega\;(\nabla\bdot\bhat)\;\frac{B}{B_{\|}^{**}},
\label{eq:dotpgc_0} \\
\frac{d^{*}_{1}p_{\|}}{dt} & = & -\;\frac{cp_{\|}}{e\,B_{\|}^{**}}\,\nabla\btimes\bhat\bdot\nabla(J\,\Omega) \;-\; \frac{{\bf B}}{B_{\|}^{**}}\bdot
\nabla\Psi_{1},
\label{eq:dotpgc_1} \\
\frac{d^{*}_{2}p_{\|}}{dt} & = & -\;\frac{c}{e\,B_{\|}^{**}}\left[p_{\|}\;\nabla\btimes\bhat\bdot\nabla\Psi_{1} + \nabla\btimes(\vb{\Pi}_{1} - J\,
{\bf R}) \bdot\nabla(J\,\Omega) \right] - \frac{{\bf B}}{B_{\|}^{**}}\bdot\nabla\Psi_{2},
\label{eq:dotpgc_2} 
\end{eqnarray}
which includes the standard magnetic mirror force \eqref{eq:dotpgc_0} as well as higher-order corrections \eqref{eq:dotpgc_1}-\eqref{eq:dotpgc_2}. These guiding-center equations satisfy the guiding-center Liouville theorem \eqref{eq:gc_Liouville} separately:
\[ \nabla\bdot\left( B_{\|}^{**}\;\frac{d^{*}_{n}{\bf X}}{dt}\right) \;+\; \pd{}{p_{\|}}\left(B_{\|}^{**}\;\frac{d^{*}_{n}p_{\|}}{dt}\right) \;=\; 0 \]at each order $\epsilon^{n}$ (for $n = 0,1,2,...$). We will also need the expression
\begin{equation}
\frac{d_{\rm gc}\theta}{dt} = \epsilon^{-1}\,\Omega \;+\; \pd{\Psi_{1}}{J} \;+\; \frac{d_{\rm gc}{\bf X}}{dt}\bdot\left( {\bf R} - \pd{\vb{\Pi}_{1}}{J} \right) \;+\; \cdots, 
\label{eq:dotthetagc_epsilon}
\end{equation}
where the first term $(\epsilon^{-1}\Omega)$ is dominant while the remaining terms (except for the gyrogauge term) vanish in a uniform magnetic field. 

\subsection{Equivalent guiding-center representations}

The guiding-center Hamiltonian \eqref{eq:Hamiltonian_gc} and the guiding-center phase-space Lagrangian \eqref{eq:Gamma_gc} are defined in terms of the scalar field $\Psi$ and the vector field $\vb{\Pi}$. In a purely {\it Hamiltonian} representation: 
\begin{equation}
\left. \begin{array}{rcl}
\vb{\Pi} & \equiv & p_{\|}\,\bhat \\
 &  & \\
 \Psi & \equiv & J\,\Omega + \epsilon\,\Psi_{1} + \epsilon^{2}\,\Psi_{2} + \cdots
 \end{array} \right\},
 \label{eq:Ham_rep}
 \end{equation}
the vector field $\vb{\Pi}$ is independent of the gyroaction $J$, while the scalar field $ \Psi$ contains all the correction terms associated with the nonuniformity of the magnetic field. Hence, in the Hamiltonian representation, the vector field ${\sf b}^{*}$ in Eq.~\eqref{eq:gc_X_dot} is ${\sf b}^{*} \equiv \partial\vb{\Pi}/\partial p_{\|} = \bhat$ while ${\bf R}^{*} \equiv {\bf R} - \partial\vb{\Pi}/\partial J = {\bf R}$ in 
Eq.~\eqref{eq:gc_theta_dot}. This representation simplifies the guiding-center Poisson bracket \eqref{eq:PB_gc_star} at the expense of the guiding-center Hamiltonian \eqref{eq:Hamiltonian_gc}.

In a purely {\it symplectic} representation:
\begin{equation}
\left. \begin{array}{rcl}
\vb{\Pi} & \equiv & p_{\|}\,\bhat \;+\; \epsilon\,\vb{\Pi}_{1} + \cdots \\
 &  & \\
 \Psi & \equiv & J\,\Omega
 \end{array} \right\},
 \label{eq:Symp_rep}
 \end{equation}
the scalar field $\Psi$ is independent of the parallel momentum $p_{\|}$, while the vector field $\vb{\Pi} $ contains all the correction terms associated with the nonuniformity of the magnetic field. Hence, the vector fields 
${\sf b}^{*}$ and ${\bf R}^{*}$ are defined in terms of the expressions \eqref{eq:bstar_def}-\eqref{eq:Rstar_def}, respectively. This representation simplifies the guiding-center Hamiltonian \eqref{eq:Hamiltonian_gc} at the expense of  the guiding-center Poisson bracket \eqref{eq:PB_gc_star}.

In the perturbation analysis presented below, it will be shown that a purely symplectic representation is impossible, i.e., $\Psi_{n} = 0$ cannot be chosen at all orders $n \geq 2$ in $\epsilon$
[for example, see Eq.~\eqref{eq:Hamiltonian_constraint_2} for $n = 2$]. Instead, it is sometimes convenient to use a mixed representation in which the guiding-center phase-space Lagrangian \eqref{eq:Lambda_gc} is expressed with the symplectic part and the Hamiltonian part contain magnetic-nonuniformity correction terms. In standard guiding-center and gyrokinetic theories, for example, we find $\Psi_{1} \equiv 0$ and 
$\vb{\Pi}_{1} \neq 0$ at first order, which will also be adopted in the present work, while $\Psi_{2} \neq 0$ must be chosen at second order.

\section{\label{sec:Lie}Guiding-center phase-space Lagrangian: Lie-transform Derivation}

The derivation of the guiding-center Hamiltonian \eqref{eq:Hamiltonian_gc} and the guiding-center phase-space Lagrangian \eqref{eq:Gamma_gc} by Lie-transform phase-space Lagrangian perturbation method is based on a phase-space transformation from the (local) particle phase-space coordinates $z_{0}^{\alpha} = ({\bf x}, p_{\|0}; J_{0},\theta_{0})$ to the guiding-center phase-space coordinates $Z^{\alpha} = ({\bf X}, p_{\|}; J, \theta)$ generated by the vector fields $({\sf G}_{1}, {\sf G}_{2}, \cdots)$:
\begin{equation}
Z^{\alpha} \;=\; z_{0}^{\alpha} \;+\; \epsilon\,G_{1}^{\alpha} \;+\; \epsilon^{2}\,\left( G_{2}^{\alpha} + \frac{1}{2}\,{\sf G}_{1}\cdot\exd 
G_{1}^{\alpha}\right) \;+\; \cdots,
\label{eq:z_bar_z} 
\end{equation}
and its inverse
\begin{equation}
z_{0}^{\alpha} \;=\; Z^{\alpha} \;-\; \epsilon\,G_{1}^{\alpha} \;-\; \epsilon^{2}\,\left( G_{2}^{\alpha} - \frac{1}{2}\,{\sf G}_{1}\cdot\exd 
G_{1}^{\alpha}\right) \;+\; \cdots.
\label{eq:zz_bar} 
\end{equation} 
Here, we adopt the macroscopic view in which the electric charge $e$ is renormalized $e \rightarrow e/\epsilon$, with the ordering $\epsilon \ll 1$ being consistent with the strong magnetic-field limit. The local particle phase-space coordinates $z_{0}^{\alpha}$ are defined as the particle position ${\bf x}$ and the particle momentum 
\begin{equation}
{\bf p}_{0} \;\equiv\; p_{\|0}\,\bhat({\bf x}) \;+\; {\bf p}_{\bot0}(J_{0},\theta_{0};{\bf x})
\label{eq:p_local}
\end{equation}
is decomposed into parallel and perpendicular components $p_{\|0} \equiv {\bf p}_{0}\bdot\bhat$ and ${\bf p}_{\bot0} \equiv \bhat\btimes({\bf p}_{0}
\btimes\bhat)$, respectively, with $J_{0} \equiv |{\bf p}_{\bot0}|^{2}/(2m\Omega)$ and $\partial{\bf p}_{\bot0}/\partial\theta_{0} \equiv {\bf p}_{\bot0}\btimes\bhat$.

The guiding-center Jacobian \eqref{eq:Jac_gc} associated with the phase-space transformation \eqref{eq:z_bar_z} is defined as
\begin{eqnarray}
{\cal J}_{\rm gc} & \equiv & {\cal J}_{0} \;-\; \pd{}{z^{\alpha}}\left[ {\cal J}_{0}\frac{}{} \left(\epsilon\,G_{1}^{\alpha} \;+\frac{}{} \epsilon^{2}\,
G_{2}^{\alpha} + \cdots\right) \;-\; \frac{\epsilon^{2}}{2}\;G_{1}^{\alpha}\;\pd{}{z^{\beta}}\left({\cal J}_{0}\frac{}{} G_{1}^{\beta} + \cdots\right) 
\;+\; \cdots \right] \nonumber \\
 & = & {\cal J}_{0} \;+\; \epsilon\,{\cal J}_{1} \;+\; \epsilon^{2}\;{\cal J}_{2} \;+\; \cdots
\label{eq:Jacobian_Lie}
\end{eqnarray}
where ${\cal J}_{0} \equiv e\,B({\bf x})/c$ is the Jacobian for the transformation from the particle phase-space coordinates $z^{\alpha} = ({\bf x},{\bf p})$ to the local particle phase-space coordinates $z_{0}^{\alpha}$ (i.e., $d^{3}x\,d^{3}p = {\cal J}_{0}\,d^{3}x\,dp_{\|0} dJ_{0} d\theta_{0}$). 

Next, the effective guiding-center potential energy \eqref{eq:Psi_def} is defined in terms of the guiding-center transformation as
\begin{eqnarray}
\Psi_{1} - \frac{p_{\|}}{m}\,\Pi_{1\|} & \equiv & -\,\Omega \left( G_{1}^{J} \;+\; J\;G_{1}^{\bf x}\bdot\nabla\ln B \right) \;-\; \frac{p_{\|}}{m}\;
\left( G_{1}^{p_{\|}} \;+\; \Pi_{1\|} \right),
\label{eq:Psi_1} \\
\Psi_{2} - \frac{p_{\|}}{m}\,\Pi_{2\|} & \equiv & -\,\Omega \left( G_{2}^{J} + J\;G_{2}^{\bf x}\bdot\nabla\ln B \right) - \frac{p_{\|}}{m}\;
\left( G_{2}^{p_{\|}} + \Pi_{2\|} \right) - \frac{1}{2}\,{\sf G}_{1}\cdot\exd\Psi_{1},
\label{eq:Psi_2}
\end{eqnarray}
where the last term in Eq.~\eqref{eq:Psi_2} depends on the choice for $\Psi_{1}$ made in Eq.~\eqref{eq:Psi_1}. We note here that the components $G_{n}^{J}$ ($n \geq 1$) will be chosen independently from the representation $(\Psi_{n}, \Pi_{n\|})$ at all orders in order to guarantee equivalent guiding-center Hamiltonian theories.

Beginning with the local particle phase-space Lagrangian 
\begin{equation}
\gamma \;\equiv\; \left( \frac{e}{\epsilon\,c}\;{\bf A} \;+\; {\bf p}_{0}\right)\bdot\exd{\bf x} \;=\; \epsilon^{-1}\,\gamma_{0} \;+\; \gamma_{1},
\label{eq:Gamma_particle}
\end{equation}
we derive the new (guiding-center) phase-space Lagrangian $\Gamma_{\rm gc}$:
\begin{eqnarray}
\Gamma_{\rm gc} & = & {\sf T}_{\rm gc}^{-1}\gamma \;+\; \exd S \;\equiv\; \epsilon^{-1} \left( \Gamma_{0} \;+\; \epsilon\,\Gamma_{1} \;+\; \epsilon^{2}\,\Gamma_{2} \;+\; \cdots \right),
\label{eq:ovgamma_Lie}
\end{eqnarray}
where each perturbation term $\Gamma_{n} \equiv \Gamma_{n\alpha}\,\exd Z^{\alpha} + \exd S_{n}$ is expressed in terms of the symplectic components 
$\Gamma_{n\alpha}$ and the $n$th-order component of the phase-space gauge function $S \equiv S_{1} +\epsilon\,S_{2} + \cdots$. In 
Eq.~(\ref{eq:ovgamma_Lie}), the {\it push-forward} operator ${\sf T}_{\rm gc}^{-1} \equiv \cdots \exp(-\,\epsilon^{2}\pounds_{2})\,\exp(-\,\epsilon
\pounds_{1})$ is defined in terms of the product of Lie-transforms $\exp(-\epsilon^{n}\,\pounds_{n})$, where the $n$th-order Lie derivative $\pounds_{n}$ is generated by the $n$th-order vector field ${\sf G}_{n}$. According to Cartan's formula \citep{RGL_1982}, the Lie derivative $\pounds_{G}$ of a one-form $\gamma$ yields the one-form 
\begin{eqnarray}
\pounds_{G}\gamma & \equiv & \iota_{G}\cdot\exd\gamma \;+\; \exd(\iota_{G}\cdot\gamma) \;=\; G^{\alpha}\,\omega_{\alpha\beta}\;\exd z^{\beta} \;+\; \exd\left( G^{\alpha}\,\gamma_{\alpha}\right).
\label{eq:pounds_n}
\end{eqnarray}
Note that, according to this formula, the exterior derivative $\exd$ and the Lie derivative $\pounds_{G}$ commute, i.e., $\pounds_{G}(\exd
\gamma) = \exd(\pounds_{G}\gamma)$. Furthermore, an arbitrary exact exterior derivative $\exd S$ can be added to the push-forward 
${\sf T}_{\rm gc}^{-1}\gamma$ in Eq.~(\ref{eq:ovgamma_Lie}) without affecting the guiding-center two-form 
\begin{eqnarray}
\vb{\omega}_{\rm gc} & \equiv & \exd\Gamma_{\rm gc} \;=\; \exd\left({\sf T}_{\rm gc}^{-1}\gamma\right) \;+\; \exd^{2}S \;=\; {\sf T}_{\rm gc}^{-1}\left(\exd\gamma\right) \;=\; {\sf T}_{\rm gc}^{-1}\vb{\omega},
\label{eq:ovomega_def}
\end{eqnarray}
since $\exd^{2}$ for any k-form vanishes and the push-forward ${\sf T}_{\rm gc}^{-1}$ commutes with $\exd$ (because all functions of Lie derivatives do).

When the push-forward ${\sf T}_{\rm gc}^{-1}$ and the phase-space gauge function $S$ are expanded in powers of $\epsilon$ in 
Eq.~\eqref{eq:ovgamma_Lie}, we obtain the zeroth-order equation
\begin{equation}
\Gamma_{0} \;=\; \gamma_{0} \;\equiv\; \frac{e}{c}\;{\bf A}({\bf X})\bdot\exd{\bf X}, 
\label{eq:ovgamma_0}
\end{equation}
the first-order equation
\begin{equation}
\Gamma_{1} \;=\; \gamma_{1} \;-\; \pounds_{1}\;\gamma_{0} \;+\; \exd S_{1} \;\equiv\; \gamma_{1} \;-\; \iota_{1}\cdot\vb{\omega}_{0} \;+\; 
\exd \sigma_{1}, 
\label{eq:ovgamma_1}
\end{equation}
the second-order equation
\begin{eqnarray}
\Gamma_{2} & = & -\; \pounds_{2}\;\gamma_{0} \;-\; \pounds_{1}\;\gamma_{1} \;+\; \frac{1}{2}\;\pounds_{1}^{2}\;\gamma_{0} \;+\; 
\exd S_{2} \nonumber \\
 & \equiv & -\;\iota_{2}\cdot\omega_{0} \;-\; \frac{1}{2}\;\iota_{1}\cdot\left(\vb{\omega}_{1} \;+\frac{}{} \vb{\omega}_{{\rm gc}1} \right) 
\;+\; \exd \sigma_{2}, 
\label{eq:ovgamma_2}
\end{eqnarray}
the third-order equation
\begin{eqnarray}
\Gamma_{3} & = & -\; \pounds_{3}\;\gamma_{0} \;-\; \pounds_{2}\;\gamma_{1} \;+\; \frac{1}{2}\;\pounds_{1}^{2}\;\gamma_{1} \;+\; \pounds_{2}\pounds_{1}\;\gamma_{0} \;-\; \frac{1}{6}\;\pounds_{1}^{3}\;\gamma_{0} \;+\; \exd S_{3} 
\nonumber \\
 & \equiv & -\iota_{3}\cdot\vb{\omega}_{0} \;-\; \iota_{2}\cdot\vb{\omega}_{{\rm gc}1} \;+\; \frac{\iota_{1}}{3}\cdot\exd\left( \iota_{1}\cdot
\vb{\omega}_{1} + \frac{\iota_{1}}{2}\cdot\vb{\omega}_{{\rm gc}1} \right) \;+\; \exd \sigma_{3},
\label{eq:ovgamma_3}
\end{eqnarray}
and the fourth-order equation
\begin{eqnarray}
\Gamma_{4} & = & -\;\pounds_{4}\;\gamma_{0} \;+\; \pounds_{3}\left(\pounds_{1}\gamma_{0} \;-\; \gamma_{1}\right) \;+\; \pounds_{2}\left( 
\pounds_{1}\gamma_{1} \;-\; \frac{1}{2}\,\pounds_{1}^{2}\gamma_{0} \;+\; \frac{1}{2}\,\pounds_{2}\gamma_{0}\right) \nonumber \\
 &  &-\; \frac{1}{6}\,\pounds_{1}^{3}\left( \gamma_{1} \;-\; \frac{1}{4}\,\pounds_{1}\gamma_{0}\right) \;+\; \exd S_{4} 
\label{eq:ovgamma_4} \\
 & \equiv & -\,\iota_{4}\cdot\vb{\omega}_{0} \,-\, \iota_{3}\cdot\vb{\omega}_{{\rm gc}1} \;-\; \frac{\iota_{2}}{2}\cdot\left[ \vb{\omega}_{{\rm gc}2} 
\,-\, \frac{1}{2}\; \exd\left( \iota_{1}\cdot\vb{\omega}_{1} \,+\frac{}{} \iota_{1}\cdot\vb{\omega}_{{\rm gc}1} \right) \right] \nonumber \\
 &  &-\; \frac{\iota_{1}}{8}\cdot\exd\left[ \iota_{1}\cdot\exd\left( \iota_{1}\cdot\vb{\omega}_{1} \,+\, \frac{\iota_{1}}{3}\cdot\vb{\omega}_{{\rm gc}1}\right) \right] \,+\, \exd \sigma_{4}, 
\nonumber
\end{eqnarray}
where $\iota_{n}\cdot\vb{\omega}_{k} = G_{n}^{\alpha}\;\omega_{k\alpha\beta}\;\exd z^{\beta}$ and, since $\pounds_{n}\gamma_{k} = \iota_{n}\cdot
\vb{\omega}_{k} + \exd(G_{n}^{\alpha}\gamma_{k\alpha})$,  we have redefined the phase-space gauge functions $S_{n} \rightarrow  \sigma_{n}$ by absorbing all exact exterior derivatives: $\exd(\cdots) + \exd S_{n} \equiv \exd \sigma_{n}$ (i.e., $ \sigma_{1} \equiv S_{1} - G_{1}^{\alpha}\gamma_{0\alpha}$). The phase-space gauge functions $\sigma_{n}$ in Eqs.~\eqref{eq:ovgamma_1}-\eqref{eq:ovgamma_4} are generally considered to be gyroangle-dependent functions (i.e., $\langle\sigma_{n}\rangle = 0$) but it is not a strict requirement. Note also that we use results obtained at lower orders to simplify expressions at each higher order (i.e., at second order, we use $\pounds_{1}\gamma_{1} - \frac{1}{2}\,\pounds_{1}^{2}\gamma_{0} = \frac{1}{2}\,\pounds_{1}\gamma_{1} + \frac{1}{2}\,\pounds_{1}\Gamma_{1}$).

In Eqs.~(\ref{eq:ovgamma_1})-(\ref{eq:ovgamma_3}), we need to evaluate the contractions $\iota_{n}\cdot\vb{\omega}_{0}$ generated by the vector fields 
$({\sf G}_{1}, {\sf G}_{2}, \cdots)$ on the zeroth-order two-form:
\begin{eqnarray} 
\vb{\omega}_{0} & = & \exd\gamma_{0} \;=\; \frac{e}{c}\,\pd{A_{j}}{x^{i}}\;\exd X^{i}\;\wedge\;\exd X^{j} \;\equiv\; \frac{1}{2}\;\omega_{0\,ij}\;
\exd X^{i}\;\wedge\;\exd X^{j},
\label{eq:omega_0}
\end{eqnarray}
where $\omega_{0\,ij} \equiv \varepsilon_{ijk}\,(e/c)\,B^{k}$ is defined in terms of the magnetic field ${\bf B} \equiv \nabla\btimes{\bf A}$. Using the contraction formula \eqref{eq:pounds_n}, we obtain the $n^{th}$-order expression
\begin{equation}
\iota_{n}\cdot\vb{\omega}_{0} \;=\; \frac{e}{c}\;{\bf B}\btimes G_{n}^{{\bf x}}\bdot\exd{\bf X},
\label{eq:iota_n0}
\end{equation}
where $G_{n}^{{\bf x}}$ denote the spatial components of the $n$th-order generating vector field ${\sf G}_{n}$.

Similarly, in Eqs.~(\ref{eq:ovgamma_2})-(\ref{eq:ovgamma_3}), we need to evaluate the contractions $\iota_{n}\cdot\vb{\omega}_{1}$ generated by the vector fields $({\sf G}_{1}, {\sf G}_{2}, \cdots)$ on the first-order two-form $\vb{\omega}_{1} = \exd\gamma_{1}$. When evaluated explicitly, we obtain the
$(n + 1)^{\rm th}$-order expression
\begin{eqnarray}
\iota_{n}\cdot\vb{\omega}_{1} & \equiv & D_{n}(p_{\|0}\,\bhat + {\bf p}_{\bot0})\bdot\exd{\bf X} \;-\; G_{n}^{\bf x}\bdot\left(\bhat\;\exd p_{\|0} + 
\pd{{\bf p}_{\bot0}}{J_{0} }\;\exd J_{0}  + \pd{{\bf p}_{\bot0}}{\theta_{0} }\;\exd\theta_{0}  \right),
\label{eq:iota_n1}
\end{eqnarray}
where the spatial components are expressed in terms of the operator $D_{n}({\bf C})$ defined as
\begin{eqnarray}
D_{n}({\bf C}) & \equiv & \left( G_{n}^{p_{\|}}\,\pd{{\bf C}}{p_{\|0}} + G_{n}^{J}\,\pd{{\bf C}}{J_{0}} + G_{n}^{\theta}\,\pd{{\bf C}}{\theta_{0}}\right) \;-\; G_{n}^{\bf x}\btimes\nabla\btimes
{\bf C},
\label{eq:Pn_def}
\end{eqnarray}
where ${\bf C}$ is an arbitrary vector function on guiding-center phase space. In what follows, unless it is necessary, we will omit writing the subscript $0$ on 
local particle phase-space coordinates, i.e., $p_{\|0}$ is written as $p_{\|}$.

In the next sections (Secs.~\ref{sec:first}-\ref{sec:fourth}), we will progressively solve for the components $G_{n}^{\alpha}$ $(n \geq 1$) up to second order in $\epsilon_{B}$: at order $\epsilon_{B}^{0}$, we will obtain 
$G_{1}^{\bf x}$; at order $\epsilon_{B}$, we will obtain $(G_{1}^{p_{\|}}, G_{1}^{J}, G_{1}^{\theta})$ and $G_{2}^{\bf x}$; and at order $\epsilon_{B}^{2}$, we will obtain $(G_{2}^{p_{\|}}, G_{2}^{J}, G_{2}^{\theta})$ and $G_{3}^{\bf x}$.

\section{\label{sec:first}First-order Perturbation Analysis}

We begin our perturbation analysis by considering the first-order guiding-center symplectic one-form (\ref{eq:ovgamma_1}), which is now explicitly written as
\begin{eqnarray}
\Gamma_{1} & = & \left( p_{\|}\;\bhat \;+\; {\bf p}_{\bot}\right)\bdot\exd{\bf X} \;-\; \frac{e}{c}\;{\bf B}\btimes G_{1}^{{\bf x}}\bdot\exd{\bf X} \;+\; \exd \sigma_{1} \nonumber \\
 & = & p_{\|}\;\bhat\bdot\exd{\bf X} \;+\; \left( {\bf p}_{\bot} \;-\; \frac{e}{c}\;{\bf B}\btimes G_{1}^{{\bf x}} \right)\bdot\exd{\bf X} \;+\; \exd \sigma_{1} \label{eq:ovgamma1_Lie_prime} \\
 & \equiv & p_{\|}\,\bhat\bdot\exd{\bf X},
\label{eq:ovgamma1_Lie}
\end{eqnarray}
where we have separated the terms that are independent and dependent on the gyroangle $\theta$ in Eq.~\eqref{eq:ovgamma1_Lie_prime}. It is immediately clear that the first-order phase-space gauge function $\sigma_{1}$ is not needed to remove the gyroangle dependence on the right side of Eq.~(\ref{eq:ovgamma1_Lie_prime}), and thus we set 
$\sigma_{1} \equiv 0$. 

The spatial components $G_{1}^{{\bf x}}$ of the first-order generating vector field ${\sf G}_{1}$ is determined by the condition 
\[ {\bf p}_{\bot} \;-\; (e/c)\,{\bf B}\btimes G_{1}^{{\bf x}} \;\equiv\; 0, \]
which removes the gyroangle dependence in the first-order phase-space Lagrangian (\ref{eq:ovgamma1_Lie_prime}). This condition can easily be solved as
\begin{equation}
G_{1}^{{\bf x}} \;=\; \left( \bhat\bdot G_{1}^{{\bf x}}\right)\;\bhat \;-\; \frac{c\bhat}{eB}\btimes{\bf p}_{\bot} \;\equiv\; G_{1\|}^{\bf x}\;\bhat
\;-\; \vb{\rho}_{0},
\label{eq:G1_x}
\end{equation}
where $G_{1\|}^{\bf x} \equiv \bhat\bdot G_{1}^{{\bf x}}$ denotes the parallel component of $G_{1}^{{\bf x}}$, which is undetermined at this order. Here, the 
gyroangle-dependent vector $\vb{\rho}_{0}$ represents the lowest-order displacement between the particle position ${\bf x}$ and the guiding-center position 
${\bf X} = {\bf x} - \vb{\rho}_{0}$.  

With $\sigma_{1} \equiv 0$ in Eq.~\eqref{eq:ovgamma1_Lie_prime} and $G_{1}^{{\bf x}}$ defined by Eq.~\eqref{eq:G1_x}, the resulting first-order guiding-center phase-space Lagrangian is given Eq.~(\ref{eq:ovgamma1_Lie}), where all spatially-dependent fields are now evaluated at the guiding-center position ${\bf X}$. Hence, we obtain the $n$th-order contraction 
\begin{equation}
\iota_{n}\cdot\omega_{{\rm gc}1} \;\equiv\; D_{n}(p_{\|}\,\bhat)\bdot\exd{\bf X} \;-\; G_{n\|}^{\bf x}\;\exd p_{\|},
\label{eq:iotan_1gc}
\end{equation}
where $G_{n\|}^{\bf x} \equiv \bhat\bdot G_{n}^{{\bf x}}$ denotes the parallel component of $G_{n}^{{\bf x}}$, and the spatial components in 
Eq.~\eqref{eq:iotan_1gc} are
\begin{eqnarray}
D_{n}(p_{\|}\,\bhat) & = & \left( G_{n}^{p_{\|}} \;-\frac{}{} p_{\|}\,\vb{\kappa}\bdot G_{n}^{\bf x} \right)\;\bhat \;+\; p_{\|}\;\left( \tau\,\bhat\btimes G_{n}^{{\bf x}} \;+\; G_{n\|}^{\bf x}\;\vb{\kappa} \right),
\label{eq:ovP_n_def}
\end{eqnarray}
where the curl of $\bhat$:
\begin{equation}
\nabla\btimes\bhat \;\equiv\; \tau\,\bhat \;+\; \bhat\btimes\vb{\kappa} 
\label{eq:curl_bhat}
\end{equation}
is written in terms of the magnetic twist $\tau \equiv \bhat\bdot\nabla\btimes\bhat$ (which is proportional to the plasma current density flowing along magnetic-field lines) and the magnetic curvature $\vb{\kappa} \equiv \bhat\bdot\nabla\bhat$ (which is perpendicular to $\bhat$: $\bhat\bdot\vb{\kappa} \equiv 0$).

\section{\label{sec:second}Second-order Perturbation Analysis}

We now proceed with the second-order guiding-center symplectic one-form~(\ref{eq:ovgamma_2}), which is explicitly expressed as
\begin{eqnarray} 
\Gamma_{2} & = & -\;\left[ \frac{e}{c}{\bf B}\btimes G_{2}^{{\bf x}} \;+\; D_{1}({\bf P}_{2}) \right] \bdot\exd{\bf X} \;+\; \frac{1}{2}\,G_{1}^{{\bf x}}\bdot\left( \pd{{\bf p}_{\bot}}{J}\,\exd J + \pd{{\bf p}_{\bot}}{\theta}\;\exd\theta \right) \nonumber \\
 & = & -\;\left[ \frac{e}{c}{\bf B}\btimes G_{2}^{{\bf x}} \;+\; D_{1}({\bf P}_{2}) \right] \bdot\exd{\bf X} 
\;+\; J\;\exd\theta \label{eq:ovgamma2_Lie_prime} \\
 & \equiv & \vb{\Pi}_{1}\bdot\exd{\bf X} \;+\; J\;\left(\exd\theta \;-\frac{}{} {\bf R}\bdot\exd{\bf X}\right),
\label{eq:ovgamma2_Lie}
\end{eqnarray}
where we use the notation
\begin{equation}
{\bf P}_{2} \;\equiv\; p_{\|}\,\bhat + \frac{1}{2}\,{\bf p}_{\bot},
\label{eq:P_2}
\end{equation}
and we used $\sigma_{2} \equiv 0$ with $G_{1}^{{\bf x}}\bdot\partial{\bf p}_{\bot}/\partial J = 0$ and $G_{1}^{{\bf x}}\bdot\partial{\bf p}_{\bot}/\partial\theta = 2\,J$. Since 
$G_{1\|}^{\bf x} \equiv -\,\partial\sigma_{2}/\partial p_{\|} \equiv 0$, the spatial component of ${\sf G}_{1}$ is now exactly
\begin{equation}
G_{1}^{{\bf x}} \;=\; -\;\vb{\rho}_{0},
\label{eq:G1x_gc}
\end{equation}
i.e., to lowest order, the displacement from the particle position ${\bf x}$ to the guiding-center position ${\bf X}$ is perpendicular to ${\bf B}$. 

Using Eqs.~\eqref{eq:Pn_def} and \eqref{eq:ovP_n_def} for $n = 1$, with Eq.~\eqref{eq:P_2}, we find
\begin{eqnarray}
D_{1}({\bf P}_{2}) & = & \left( G_{1}^{p_{\|}} \;+\frac{}{} p_{\|}\;\vb{\rho}_{0}\bdot\vb{\kappa} \right)\;\bhat \;+\; p_{\|}\,\tau\; \pd{\vb{\rho}_{0}}{\theta} \;+\; J \left[ 
{\bf R} \;-\; \left( \frac{\tau}{2} \;+\; \alpha_{1}\right) \bhat \right] \nonumber \\
 &  &+\; \frac{1}{2}\;\left( G_{1}^{ J} \;-\frac{}{}  J\;\vb{\rho}_{0}\bdot\nabla\ln B \right)\;\pd{{\bf p}_{\bot}}{ J} \;+\; \frac{1}{2}\;\left(
G_{1}^{\theta} \;+\frac{}{} \vb{\rho}_{0}\bdot{\bf R} \right)\;\pd{{\bf p}_{\bot}}{\theta},
\label{eq:KovK_2}
\end{eqnarray}
where $\alpha_{1} \equiv {\sf a}_{1}:\nabla\bhat$ is defined in App.~\ref{sec:dyadic} (here, $\langle\alpha_{1}\rangle = 0$). We note that ${\bf R}$ appears in Eq.~\eqref{eq:ovgamma2_Lie} in order to satisfy the property of gyrogauge invariance. With this choice, we obtain the vector equation
\begin{equation}
J\;{\bf R} \;-\; \vb{\Pi}_{1} \;\equiv\; \frac{e}{c}\;{\bf B}\btimes G_{2}^{{\bf x}} \;+\; D_{1}({\bf P}_{2}).
\label{eq:space_2}
\end{equation}
From the parallel components of Eq.~\eqref{eq:space_2}, we obtain the first-order component
\begin{equation}
G_{1}^{p_{\|}} \;=\; -\;p_{\|}\;\vb{\rho}_{0}\bdot\vb{\kappa} \;+\; J \left( \frac{\tau}{2} \;+\; \alpha_{1} \right) \;-\; \Pi_{1\|},
\label{eq:G1_vpar}
\end{equation}
where $\Pi_{1\|} \equiv \bhat\bdot\vb{\Pi}_{1}$. By using the definition \eqref{eq:Psi_1}, on the other hand, we obtain the first-order component
\begin{eqnarray}
G_{1}^{J} & \equiv & J\;\vb{\rho}_{0}\bdot\nabla\ln B \;-\; \varrho_{\|}\; G_{1}^{p_{\|}} \;-\; \Psi_{1}/\Omega \nonumber \\
 & = & \vb{\rho}_{0}\bdot\left( J\;\nabla\ln B \;+\; \frac{p_{\|}^{2}}{m\Omega}\;\vb{\kappa} \right) \;-\; J\,\varrho_{\|}\; \left( \frac{\tau}{2} \;+\; \alpha_{1} \right) \;+\; \left( \varrho_{\|}\,\Pi_{1\|} - \frac{\Psi_{1}}{\Omega}\right),
\label{eq:G1_mu} 
\end{eqnarray}
where we introduced the notation $\varrho_{\|} \equiv p_{\|}/(m\Omega)$. The gyroangle-averaged part of Eq.~\eqref{eq:G1_mu} yields
the first-order guiding-center Hamiltonian constraint
\begin{equation}
\frac{p_{\|}}{m}\,\Pi_{1\|} \;-\; \Psi_{1} \;=\; \Omega\left( \langle G_{1}^{J}\rangle \;-\; \frac{1}{2}\,J\;\varrho_{\|}\,\tau \right),
\label{eq:Banos_mu}
\end{equation} 
while the gyroangle-dependent part yields $\wt{G}_{1}^{J} \equiv G_{1}^{J} - \langle G_{1}^{J}\rangle$. In the next Section, we will discuss how $\Pi_{1\|}$ and 
$\Psi_{1}$ may be chosen once the gyroangle-averaged part $\langle G_{1}^{J}\rangle$ is determined.

Lastly, from the perpendicular components of Eq.~(\ref{eq:space_2}), we find
\begin{eqnarray}
G_{2}^{{\bf x}} & = & G_{2\|}^{\bf x}\;\bhat \;+\; \vb{\rho}_{0} \left( \varrho_{\|}\;\tau \right) \;+\; \frac{1}{2} \left( G_{1}^{J} -\frac{}{} J\;
\vb{\rho}_{0}\bdot\nabla\ln B \right) \pd{\vb{\rho}_{0}}{J} \nonumber \\
 &  &+\; \frac{1}{2} \left( G_{1}^{\theta} + \vb{\rho}_{0}\bdot{\bf R} \right) \pd{\vb{\rho}_{0}}{\theta} \;-\; \vb{\Pi}_{1}\btimes\frac{\bhat}{m\Omega},
\label{eq:G2_x}
\end{eqnarray}
where $G_{2\|}^{\bf x} \equiv \bhat\bdot G_{2}^{{\bf x}}$ denotes the parallel component of $G_{2}^{\bf x}$. In Eqs.~\eqref{eq:Banos_mu}-\eqref{eq:G2_x}, we need to obtain an expression for $\langle G_{1}^{J}\rangle$ as well as $G_{2\|}^{\bf x}$ and $G_{1}^{\theta}$ in order to complete the guiding-center transformation at first order in $\epsilon_{B}$.

\section{\label{sec:third}Third-order Perturbation Analysis}

The third-order guiding-center symplectic one-form (\ref{eq:ovgamma_3}) is explicitly given in terms of the spatial components
\begin{equation}
\Gamma_{3{\bf x}} = D_{1}^{2}({\bf P}_{3}) \;-\; \frac{e}{c}\,{\bf B}\btimes G_{3}^{{\bf x}} \;-\; D_{2}\left(p_{\|}\,\bhat\right) \;+\; 
\nabla\sigma_{3} \;\equiv\; \vb{\Pi}_{2},
\label{eq:Gamma_3_x}
\end{equation}
where 
\begin{equation}
{\bf P}_{3} \;\equiv\; \frac{1}{2}\,p_{\|}\bhat + \frac{1}{3}\,{\bf p}_{\bot},
\label{eq:P_3} 
\end{equation}
and the momentum components are now chosen to vanish exactly
\begin{eqnarray} 
\Gamma_{3{\bf p}} & \equiv & \left[ G_{2\|}^{\bf x} \;+\; \pd{D_{1}({\bf P}_{3})}{p_{\|}}\bdot\vb{\rho}_{0} \;+\; \pd{\sigma_{3}}{p_{\|}} \right]\;\exd 
p_{\|} + \left[ \frac{2}{3}\,G_{1}^{\theta} + \pd{D_{1}({\bf P}_{3})}{J}\bdot\vb{\rho}_{0} + \pd{\sigma_{3}}{J} \right]\;\exd J \nonumber \\
 &  &+\; \left[ -\,\frac{2}{3}\,G_{1}^{J} + \pd{D_{1}({\bf P}_{3})}{\theta}\bdot\vb{\rho}_{0} + \pd{\sigma_{3}}{\theta} \right]\;\exd \theta \;\equiv\; 0.
\label{eq:Gamma_3_p}
\end{eqnarray}
In Eqs.~\eqref{eq:Gamma_3_x}-\eqref{eq:Gamma_3_p}, we find
\begin{eqnarray}
D_{1}({\bf P}_{3}) & = & \frac{1}{2}\;G_{1}^{p_{\|}}\,\bhat \;+\; \frac{1}{3} \left(G_{1}^{J}\;\pd{{\bf p}_{\bot}}{J} + G_{1}^{\theta}\;
\pd{{\bf p}_{\bot}}{\theta}\right) \;+\; \vb{\rho}_{0}\btimes\nabla\btimes{\bf P}_{3}, \label{eq:D1_3rd} \\
D_{1}^{2}({\bf P}_{3}) & \equiv &  \frac{1}{2}\,D_{1}^{2}(p_{\|}\bhat) + \frac{1}{3}\,D_{1}^{2}({\bf p}_{\bot}),
\label{eq:D1_2_3rd}
\end{eqnarray}
and
\begin{eqnarray}
D_{2}(p_{\|}\,\bhat) & = & \left( G_{2}^{p_{\|}} \;-\frac{}{} p_{\|}\,\vb{\kappa}\bdot G_{2}^{\bf x} \right)\;\bhat \;+\; p_{\|}\;\left( \tau\,\bhat\btimes G_{2}^{{\bf x}} \;+\; G_{2\|}^{\bf x}\;\vb{\kappa} \right).
\label{eq:ovP_2_def}
\end{eqnarray}

\subsection{Momentum components}

If we use the fact that $\partial\vb{\rho}_{0}/\partial p_{\|} = 0$, then the $p_{\|}$-component of Eq.~\eqref{eq:Gamma_3_p} suggests that we define the new gauge function
\begin{equation}
\ov{\sigma}_{3} \;\equiv\; \sigma_{3} \;+\; D_{1}({\bf P}_{3})\bdot\vb{\rho}_{0} \;=\; \sigma_{3}
\;-\; \frac{2}{3}\;J\,G_{1}^{\theta},
\label{eq:sigma3_ov}
\end{equation}
where the last expression follows from Eq.~\eqref{eq:D1_3rd}. Using the new gauge function \eqref{eq:sigma3_ov}, the momentum components \eqref{eq:Gamma_3_p}, therefore, become
\begin{eqnarray}
\Gamma_{3{\bf p}} & = & \left( G_{2\|}^{\bf x} \;+\; \pd{\ov{\sigma}_{3}}{p_{\|}} \right)\;\exd p_{\|} \;+\; \left( G_{1}^{\theta} \;+\; 
\pd{\ov{\sigma}_{3}}{J} \right)\;\exd J \nonumber \\
 &  &+\; \left[ \pd{\ov{\ov{\sigma}}_{3}}{\theta} \;-\; \left(G_{1}^{J} \;+\frac{}{} J\;\varrho_{\|}\tau\right)
\right]\;\exd \theta,
\label{eq:ovgamma_3gc_final}
\end{eqnarray}
where, using Eq.~\eqref{eq:D1_3rd}, we introduced the identities
\begin{eqnarray*}
D_{1}({\bf P}_{3})\bdot\pd{\vb{\rho}_{0}}{J} & \equiv & -\;\frac{1}{3}\,G_{1}^{\theta}, \\
D_{1}({\bf P}_{3})\bdot\pd{\vb{\rho}_{0}}{\theta} & \equiv & \frac{1}{3}\,G_{1}^{J} \;+\; \frac{2J\bhat}{m\Omega}\bdot\nabla\btimes{\bf P}_{3},
\end{eqnarray*}
where $2\bhat\bdot\nabla\btimes{\bf P}_{3} = p_{\|}\,\tau + (2/3) \bhat\bdot\nabla\btimes{\bf p}_{\bot}$, so that we can also introduce yet another gauge function
\begin{equation}
\ov{\ov{\sigma}}_{3} \;\equiv\; \ov{\sigma}_{3} \;-\; \frac{1}{3} \left( 2\,J\;\vb{\rho}_{0}\bdot{\bf R} \;+\; J\;\pd{\vb{\rho}_{0}}{\theta}\bdot\nabla
\ln B\right)
\label{eq:sigma3_ov_ov}
\end{equation}
in the $\theta$-component of Eq.~\eqref{eq:Gamma_3_p}. By requiring that the momentum components \eqref{eq:ovgamma_3gc_final} vanish, we now obtain the definitions
\begin{eqnarray}
G_{1}^{J} & \equiv & -\;J\;\varrho_{\|}\tau \;+\; \pd{\ov{\ov{\sigma}}_{3}}{\theta}, \label{eq:G1_J_eq} \\
G_{2\|}^{\bf x} & \equiv & -\;\pd{\ov{\sigma}_{3}}{p_{\|}}, \label{eq:G2_xpar_eq} \\
G_{1}^{\theta} & \equiv & -\;\pd{\ov{\sigma}_{3}}{J}. \label{eq:G1_theta_eq}
\end{eqnarray}
From Eq.~\eqref{eq:G1_J_eq}, we immediately conclude that $\langle G_{1}^{J}\rangle$ must be defined as
\begin{equation}
\langle G_{1}^{J}\rangle \;\equiv\; -\;J\;\varrho_{\|}\tau,
\label{eq:Banos_J}
\end{equation} 
so that
\begin{equation}
G_{1}^{J} \;=\; \vb{\rho}_{0}\bdot\left( J\;\nabla\ln B + \frac{p_{\|}^{2}\,\vb{\kappa}}{m\Omega}\right) - J\,\varrho_{\|}\,(\tau + \alpha_{1}).
\label{eq:G1_J}
\end{equation}
By comparing Eq.~\eqref{eq:Banos_mu} with Eq.~\eqref{eq:Banos_J}, we therefore obtain 
\begin{equation}
\frac{p_{\|}}{m}\;\Pi_{1\|} \;-\; \Psi_{1} \;\equiv\; -\;J\,\Omega\;\left(\frac{1}{2}\;\varrho_{\|}\,\tau\right),
\label{eq:Ham_constraint_1}
\end{equation}
which yields an infinite number of choices for $(\Pi_{1\|}, \Psi_{1})$. One possible choice for $(\Pi_{1\|},\Psi_{1})$ is $\Pi_{1\|} = \frac{1}{2}\,J\,\tau$ and $\Psi_{1} = J\,\Omega\;(\varrho_{\|}\tau)$, which allows the 
first-order Ba\~{n}os parallel drift velocity $\partial\Psi_{1}/\partial p_{\|} = J\,\tau/m$ to be explicitly included in Eq.~\eqref{eq:dotXgc_1}. We note here that, since the right side of Eq.~\eqref{eq:Ham_constraint_1} is linear in $p_{\|}$, we may also choose $\Psi_{1} \equiv 0$ without making $\Pi_{1\|}$ singular in $p_{\|}$. 

In accordance with standard guiding-center Hamiltonian theory \citep{RGL_1983,  Cary_Brizard_2009}, we therefore choose the first-order guiding-center symplectic representation
\begin{equation}
\left. \begin{array}{rcl}
\Psi_{1} & \equiv & 0 \\
 &  & \\
\Pi_{1\|} & \equiv & -\;\frac{1}{2}\,J\,\tau
\end{array} \right\}
\label{eq:Pi1||_def}
\end{equation} 
in what follows, so that Eq.~\eqref{eq:G1_vpar} becomes
\begin{equation}
G_{1}^{p_{\|}} \;=\; -\;p_{\|}\;\vb{\rho}_{0}\bdot\vb{\kappa} \;+\; J \left( \tau \;+\; \alpha_{1} \right)
\label{eq:G1_p||} 
\end{equation}
in the symplectic representation.

Using Eq.~\eqref{eq:G1_J}, Eq.~\eqref{eq:G1_J_eq} yields a differential equation for $\ov{\ov{\sigma}}_{3}$:
\[ \pd{\ov{\ov{\sigma}}_{3}}{\theta} \;=\; \vb{\rho}_{0}\bdot\left(J\,\nabla\ln B + \frac{p_{\|}^{2}}{m\,\Omega}\,\vb{\kappa}\right) - J\;
\varrho_{\|}\,\alpha_{1}, \]
whose solution is
\begin{eqnarray}
\ov{\ov{\sigma}}_{3} & = & -\,\pd{\vb{\rho}_{0}}{\theta}\bdot\left(J\,\nabla\ln B + \frac{p_{\|}^{2}}{m\,\Omega}\,\vb{\kappa}\right) - 
J\;\varrho_{\|}\,\alpha_{2},
\label{eq:ovov_sigma3}
\end{eqnarray}
where we used $\alpha_{1} \equiv \partial\alpha_{2}/\partial\theta$ (see App.~\ref{sec:dyadic}) and we assumed that $\langle\ov{\ov{\sigma}}_{3}\rangle = 0$. Next, we use Eq.~\eqref{eq:sigma3_ov_ov} to obtain
\begin{eqnarray}
\ov{\sigma}_{3} & = & \frac{2}{3}\,J \left( \vb{\rho}_{0}\bdot{\bf R} \;-\; \pd{\vb{\rho}_{0}}{\theta}\bdot\nabla\ln B \right) \;-\; \varrho_{\|} 
\left(p_{\|}\;\pd{\vb{\rho}_{0}}{\theta}\bdot\vb{\kappa} \;+\; J\;\alpha_{2} \right).
\label{eq:ov_sigma3}
\end{eqnarray}
from which we obtain the remaining components \eqref{eq:G2_xpar_eq}-\eqref{eq:G1_theta_eq}:
\begin{eqnarray}
G_{2\|}^{\bf x} & \equiv & -\,\pd{\ov{\sigma}_{3}}{p_{\|}} \;=\; 2\;\varrho_{\|}\;\pd{\vb{\rho}_{0}}{\theta}\bdot\vb{\kappa} \;+\; \frac{J\;\alpha_{2}}{m\Omega}, \label{eq:G2par_def} \\
G_{1}^{\theta} & \equiv & -\,\pd{\ov{\sigma}_{3}}{J} \;=\; -\;\vb{\rho}_{0}\bdot{\bf R} \;+\; \varrho_{\|}\;\alpha_{2} \;+\; \pd{\vb{\rho}_{0}}{\theta}\bdot
\left(\nabla\ln B \;+\; \frac{p_{\|}^{2}\;\vb{\kappa}}{2\,m\;J\Omega} \right). \label{eq:G1_theta_def}
\end{eqnarray}
By combining Eqs.~\eqref{eq:ov_sigma3} and \eqref{eq:G1_theta_def} into Eq.~\eqref{eq:sigma3_ov}, we also obtain the expression for $\sigma_{3}$:
\begin{equation}
\sigma_{3} \;=\; \ov{\sigma}_{3} \;+\; \frac{2J}{3}\;G_{1}^{\theta} \;\equiv\; -\;\frac{1}{3}\,p_{\|}\;G_{2\|}^{\bf x},
\label{eq:sigma3_sol}
\end{equation}
where $G_{2\|}^{\bf x}$ is expressed in Eq.~\eqref{eq:G2par_def}. Lastly, the second-order spatial component $G_{2}^{\bf x}$ is now explicitly expressed as
\begin{eqnarray}
G_{2}^{\bf x} & = & \left( 2\,\varrho_{\|}\;\pd{\vb{\rho}_{0}}{\theta}\bdot\vb{\kappa} \;+\; \frac{J\,\alpha_{2}}{m\Omega}\right) \bhat \;+\;
\frac{1}{2} \left[ \frac{p_{\|}^{2}}{m\Omega}\;(\vb{\rho}_{0}\bdot\vb{\kappa}) \;+\; J\,\varrho_{\|}\;(3 \tau - \alpha_{1}) \right] \pd{\vb{\rho}_{0}}{J}
\nonumber \\
 &  &+\; \frac{1}{2} \left[ \varrho_{\|}\,\alpha_{2} \;+\; \pd{\vb{\rho}_{0}}{\theta}\bdot \left( \nabla\ln B \;+\; 
\frac{p_{\|}^{2}\,\vb{\kappa}}{2m\Omega\,J}\right) \right]\; \pd{\vb{\rho}_{0}}{\theta} \;-\; \vb{\Pi}_{1}\btimes\frac{\bhat}{m\Omega},
\label{eq:G2x_exp}
\end{eqnarray}
from which we obtain the gyroangle-averaged expression
\begin{equation}
\langle G_{2}^{\bf x}\rangle \;=\; -\; \vb{\Pi}_{1}\btimes\frac{\bhat}{m\Omega} \;+\; \frac{1}{2} \left( \frac{J}{m\Omega}\;\nabla_{\bot}\ln B \;+\;
\varrho_{\|}^{2}\;\vb{\kappa} \right).
\label{eq:G2_x_ave}
\end{equation}

\subsection{Spatial components}

The remaining components of the third-order one-form \eqref{eq:Gamma_3_x} are 
\begin{eqnarray}
\Gamma_{3{\bf x}} \;\equiv\; \vb{\Pi}_{2} & = & -\;\frac{e}{c}\,{\bf B}\btimes \left[ G_{3}^{\bf x} \;-\frac{}{} \varrho_{\|}\,
\left(\tau\;G_{2\bot}^{\bf x} \;+\frac{}{}G_{2\|}^{\bf x}\;\bhat\btimes\vb{\kappa} \right) \right] \nonumber \\
 &  &-\; \left( G_{2}^{p_{\|}} - p_{\|}\,G_{2}^{\bf x}\bdot\vb{\kappa} \right)\bhat + D_{1}^{2}({\bf P}_{3}) + \nabla \sigma_{3},
\label{eq:Gamma_3gc_x}
\end{eqnarray}
which is now used to determine the second-order $(\epsilon_{B}^{2})$ components $G_{2}^{p_{\|}}$ and $G_{3\bot}^{\bf x}$. 

The parallel spatial component of Eq.~\eqref{eq:Gamma_3gc_x} yields the expression for $G_{2}^{p_{\|}}$:
\begin{eqnarray}
G_{2}^{p_{\|}} & = & p_{\|}\;\vb{\kappa}\frac{}{}\bdot G_{2}^{\bf x} \;+\; \bhat\bdot\left[D_{1}^{2}({\bf P}_{3}) + \nabla\sigma_{3} - \vb{\Pi}_{2}\right],
\label{eq:G2_p||}
\end{eqnarray}
where $\sigma_{3}$ is defined in Eq.~\eqref{eq:sigma3_sol}, and
\begin{equation}
\langle G_{2}^{p_{\|}}\rangle \;=\; -\;\varrho_{\|}\;\bhat\btimes\vb{\kappa}\bdot\vb{\Pi}_{1} \;+\; \varrho_{\|}\,\vb{\kappa}\bdot\left(J\;\nabla\ln B 
\;+\; \frac{p_{\|}^{2}\vb{\kappa}}{m\Omega}\right) \;-\; \Pi_{2\|} \;+\; \bhat\bdot\left\langle D_{1}^{2}({\bf P}_{3})\right\rangle,
\label{eq:G2_p_ave}
\end{equation}
where $\Pi_{2\|} \equiv \bhat\bdot\vb{\Pi}_{2}$ and Eq.~\eqref{eq:D1_2_3rd} (see App.~\ref{sec:D_12}) gives the expression
\begin{equation} 
\bhat\bdot\left\langle D_{1}^{2}({\bf P}_{3})\right\rangle \;=\; -\;J\;\varrho_{\|}\;\left(\frac{1}{2}\,\tau^{2}\;-\; \langle\alpha_{1}^{2}\rangle\right),
\label{eq:D1_3}
\end{equation}
where $\langle\alpha_{1}^{2}\rangle$ is given in App.~\ref{sec:dyadic}. With $G_{2}^{\bf x}$ and $G_{2}^{p_{\|}}$ given by Eqs.~\eqref{eq:G2x_exp} and \eqref{eq:G2_p||}, the second-order component $G_{2}^{J}$ is now obtained from the definition 
\eqref{eq:Psi_2}:
\begin{eqnarray}
G_{2}^{J} & \equiv & -\;J\,G_{2}^{\bf x}\bdot\nabla\ln B \;-\; \varrho_{\|}\;G_{2}^{p_{\|}} \;-\; \Psi_{2}/\Omega \label{eq:G_2_mu_tilde} \\
 & = & -\,G_{2}^{\bf x}\bdot\left(J\;\nabla\ln B \;+\; \frac{p_{\|}^{2}\,\vb{\kappa}}{m\,\Omega} \right) \;-\; \varrho_{\|}\bhat\bdot
\left[D_{1}^{2}({\bf P}_{3}) + \nabla\sigma_{3}\right] \;-\; \frac{1}{\Omega} \left( \Psi_{2} \;-\; \frac{p_{\|}}{m}\;\Pi_{2\|} \right),
\nonumber
\end{eqnarray}
where we used the first-order symplectic representation \eqref{eq:Pi1||_def}: $\Psi_{1} \equiv 0$. The gyroangle-averaged contribution of Eq.~\eqref{eq:G_2_mu_tilde} yields
\begin{eqnarray}
\left\langle G_{2}^{J}\right\rangle & = & -\; \left\langle G_{2}^{\bf x}\right\rangle\bdot\left( J\;\nabla\ln B \;+\; 
\frac{p_{\|}^{2}}{m\Omega}\;\vb{\kappa} \right) \;-\;  \varrho_{\|}\,\bhat\bdot\left\langle D_{1}^{2}({\bf P}_{3})\right\rangle \;-\; \frac{1}{\Omega} 
\left( \Psi_{2} \;-\; \frac{p_{\|}}{m}\;\Pi_{2\|} \right) \nonumber \\
 & = & \varrho_{\|}\;\Pi_{2\|} \;-\; \frac{1}{\Omega} \left( \Psi_{2} \;+\; \frac{m}{2}\,|{\bf v}_{\rm gc}|^{2} \;-\; \vb{\Pi}_{1}\bdot{\bf v}_{\rm gc}
\right) \;+\; J\;\varrho_{\|}^{2} \left( \frac{1}{2}\,\tau^{2} \;-\; \langle\alpha_{1}^{2}\rangle\right),
\label{eq:Psi2_ave_1}
\end{eqnarray}
which becomes the second-order guiding-center Hamiltonian constraint:
\begin{equation}
\frac{p_{\|}}{m}\;\Pi_{2\|} \;-\; \Psi_{2} \;=\; \Omega\;\left[ \langle G_{2}^{J}\rangle \;-\; J\;\varrho_{\|}^{2} \left( \frac{1}{2}\,\tau^{2} 
\;-\; \langle\alpha_{1}^{2}\rangle \right) \right] \;+\; \frac{m}{2}\,|{\bf v}_{\rm gc}|^{2} \;-\; \vb{\Pi}_{1}\bdot{\bf v}_{\rm gc},
\label{eq:Ham_constraint_2_initial}
\end{equation}
where $\langle G_{2}^{J}\rangle$ will be calculated at fourth order in the Lie-transform perturbation analysis, and the lowest-order guiding-center drift velocity is defined as
\begin{equation}
{\bf v}_{\rm gc} \;\equiv\; \frac{\bhat}{m}\btimes\left( J\;\nabla\ln B + \frac{p_{\|}^{2}}{m\Omega}\;\vb{\kappa} \right).
\label{eq:vgc_def}
\end{equation}
In the next Section, we will derive another expression for the gyroangle-averaged component $\langle G_{2}^{J}\rangle$, which will once again be independent of the choice of representation.

Lastly, the perpendicular spatial components of Eq.~\eqref{eq:Gamma_3gc_x} yields 
\begin{eqnarray}
G_{3}^{\bf x} & = & G_{3\|}^{\bf x}\bhat + G_{2\|}^{\bf x}\left(\varrho_{\|}\frac{}{}\nabla\btimes\bhat\right) - G_{2}^{\bf x}
\left(\varrho_{\|}\frac{}{}\tau \right) - \frac{c\bhat}{eB}\btimes\left[D_{1}^{2}({\bf P}_{3}) +\frac{}{} \nabla\sigma_{3} - \vb{\Pi}_{2}\right],
\label{eq:G3_x_sol}
\end{eqnarray}
where the parallel component $G_{3\|}^{\bf x}$ is determined at the fourth order.

\section{\label{sec:fourth}Fourth-order Perturbation Analysis}

At second order in magnetic-field nonuniformity, the missing components $G_{3\|}^{\bf x}$, $G_{2}^{\theta}$, and $\langle G_{2}^{J}\rangle$ are calculated at fourth order. The fourth-order guiding-center symplectic one-form (\ref{eq:ovgamma_4}) is explicitly expressed in five parts. The first part is
\begin{equation}
-\,\imath_{4}\cdot\vb{\omega}_{0} \;=\; -\;\frac{e}{c}\;{\bf B}\btimes G_{4}^{\bf x}\bdot\exd{\bf X},
\label{eq:four_1}
\end{equation}
the second part is
\begin{equation}
-\,\imath_{3}\cdot\vb{\omega}_{{\rm gc}1} \;=\; -\;D_{3}\left(p_{\|}\,\bhat\right)\bdot\exd{\bf X} \;+\; G_{3\|}^{\bf x}\;\exd p_{\|},
\label{eq:four_2}
\end{equation}
the third part is
\begin{eqnarray}
-\,\frac{1}{2}\;\imath_{2}\cdot\vb{\omega}_{{\rm gc}2} & = & \frac{1}{2}\;D_{2}\left(J\;{\bf R} - \vb{\Pi}_{1}\right)\bdot\exd{\bf X} \;-\; 
\frac{1}{2}\;G_{2}^{J}\;\exd\theta \;+\; \frac{1}{2} \left(G_{2}^{\theta} \;-\frac{}{} G_{2}^{\bf x}\bdot{\bf R}^{*}\right)\;\exd J,
\label{eq:four_3}
\end{eqnarray}
where ${\bf R}^{*} \equiv {\bf R} - \partial\vb{\Pi}_{1}/\partial J$, the fourth part is
\begin{eqnarray}
\frac{1}{4}\,\imath_{2}\cdot\exd\left[\imath_{1}\cdot\left(\vb{\omega}_{1} \;+\frac{}{} \vb{\omega}_{{\rm gc}1}\right)\right] & = & \frac{1}{2}\;D_{2}\left[ D_{1}({\bf P}_{2})\right]\bdot\exd{\bf X} \;-\; \frac{1}{2} \left( G_{2}^{J}\;\exd\theta \;-\frac{}{}G_{2}^{\theta}\;\exd J \right) \nonumber \\
 &  &-\; \frac{1}{2}G_{2}^{\bf x}\bdot\pd{D_{1}({\bf P}_{2})}{u^{a}}\;\exd u^{a},
\label{eq:four_4}
\end{eqnarray}
and the fifth part is
\begin{eqnarray}
-\,\frac{1}{8}\;\imath_{1}\cdot\exd\left[\imath_{1}\cdot\exd\left(\imath_{1}\cdot\omega_{1} \;+\; \frac{1}{3}\,\imath_{1}\cdot\vb{\omega}_{{\rm gc}1}\right)\right] & = & -\;\frac{1}{2}\;D_{1}^{3}({\bf P}_{4})\bdot\exd{\bf X} \;-\; \frac{1}{2}\;\pd{D_{1}^{2}({\bf P}_{4})}{u^{a}}\bdot\vb{\rho}_{0}\;\exd u^{a} \nonumber \\
 &  &+\; \frac{1}{4}\,\left(\exd G_{1}^{\theta}\;G_{1}^{J} \;-\frac{}{} \exd G_{1}^{J}\;G_{1}^{\theta}\right)\nonumber \\
 &  &-\; \frac{1}{2}\left( {\sf G}_{1}\cdot\exd F_{1a}\;\exd u^{a} \;-\frac{}{} \exd F_{1a}\;G_{1}^{a} \right) \nonumber \\
 &  &+\; \frac{1}{4}\,{\sf G}_{1}\cdot\left(\exd G_{1}^{J}\;\exd\theta \;-\frac{}{} \exd G_{1}^{\theta}\;\exd J \right),
\label{eq:four_5}
\end{eqnarray}
where 
\begin{equation}
{\bf P}_{4} \;\equiv\; \frac{1}{3}\,p_{\|}\,\bhat \;+\; \frac{1}{4}\,{\bf p}_{\bot},
\label{eq:P_4} 
\end{equation}
and the momentum coordinates are labeled as $u^{a} = (p_{\|}, J, \theta)$ in Eqs.~\eqref{eq:four_4}-\eqref{eq:four_5}, with the momentum-space vector components $F_{1a}$ defined as $F_{1a} \equiv 
(\partial D_{1}({\bf P}_{4})/\partial u^{a})\bdot\vb{\rho}_{0}$ in Eq.~\eqref{eq:four_5}.

We now combine these parts to write the components of the fourth-order guiding-center symplectic one-form (\ref{eq:ovgamma_4}) as
\begin{eqnarray}
\Gamma_{4{\bf x}} & \equiv & \vb{\Pi}_{3} \;=\; \nabla\sigma_{4} \;-\;\frac{e}{c}\;{\bf B}\btimes G_{4}^{\bf x} \;-\; D_{3}\left(p_{\|}\,\bhat\right) \;+\; \frac{1}{2}\;D_{2}\left(J\;{\bf R} - \vb{\Pi}_{1}\right) \label{eq:ovGamma4_x} \\
 &  &+\; \frac{1}{2}\;D_{2}\left[D_{1}({\bf P}_{2})\right] \;-\; \frac{1}{2}\;\left[ D_{1}^{3}({\bf P}_{4}) \;-\; \nabla F_{1a}\;G_{1}^{a} \right] \;+\; \frac{1}{4} \left(G_{1}^{J}\;\nabla G_{1}^{\theta} \;-\frac{}{} G_{1}^{\theta}\;\nabla G_{1}^{J} \right), 
\nonumber \\
\Gamma_{4p_{\|}}  & \equiv &  0 \;=\; \pd{\sigma_{4}}{p_{\|}} \;+\; G_{3\|}^{\bf x} \;-\; \frac{1}{2}\;G_{2}^{\bf x}\bdot\pd{D_{1}({\bf P}_{2})}{p_{\|}} \;-\; \frac{1}{2}\;\vb{\rho}_{0}\bdot\pd{D_{1}^{2}({\bf P}_{4})}{p_{\|}} \label{eq:ovGamma4_p} \\
 &  &-\; \frac{1}{2} \left( {\sf G}_{1}\cdot\exd F_{1p_{\|}} \;-\; \pd{F_{1a}}{p_{\|}}\;G_{1}^{a} \right) \;+\; \frac{1}{4} 
\left(G_{1}^{J}\;\pd{G_{1}^{\theta}}{p_{\|}} \;-\frac{}{} G_{1}^{\theta}\;\pd{G_{1}^{J}}{p_{\|}} \right), 
\nonumber \\
\Gamma_{4J} & \equiv &  0 \;=\; \pd{\sigma_{4}}{J} \;+\; G_{2}^{\theta} \;-\; \frac{1}{2}\;G_{2}^{\bf x}\bdot\left[ {\bf R}^{*} \;+\; 
\pd{D_{1}({\bf P}_{2})}{J} \right] \;-\; \frac{1}{2}\;\vb{\rho}_{0}\bdot\pd{D_{1}^{2}({\bf P}_{4})}{J} \label{eq:ovGamma4_J} \\
 &  &-\; \frac{1}{2} \left( {\sf G}_{1}\cdot\exd F_{1J} \;-\; \pd{F_{1a}}{J}\;G_{1}^{a} \right) \;+\; \frac{1}{4} 
\left(G_{1}^{J}\;\pd{G_{1}^{\theta}}{J} \;-\frac{}{} G_{1}^{\theta}\;\pd{G_{1}^{J}}{J} \;-\; {\sf G}_{1}\cdot\exd G_{1}^{\theta}\right), 
\nonumber \\
\Gamma_{4\theta}  & \equiv &  0 \;=\; \pd{\sigma_{4}}{\theta} \;-\; G_{2}^{J} \;-\; \frac{1}{2}\;G_{2}^{\bf x}\bdot\pd{D_{1}({\bf P}_{2})}{\theta} \;-\; 
\frac{1}{2}\;\vb{\rho}_{0}\bdot\pd{D_{1}^{2}({\bf P}_{4})}{\theta} \label{eq:ovGamma4_theta} \\
 &  &-\; \frac{1}{2} \left( {\sf G}_{1}\cdot\exd F_{1\theta} \;-\; \pd{F_{1a}}{\theta}\;G_{1}^{a} \right) \;+\; \frac{1}{4} 
\left(G_{1}^{J}\;\pd{G_{1}^{\theta}}{\theta} \;-\frac{}{} G_{1}^{\theta}\;\pd{G_{1}^{J}}{\theta} \;+\; {\sf G}_{1}\cdot\exd G_{1}^{J}\right),
\nonumber
\end{eqnarray}
where the momentum components are once again assumed to vanish identically. Hence, the components $G_{4\bot}^{\bf x}$ and $G_{3}^{p_{\|}}$ are obtained from Eq.~\eqref{eq:ovGamma4_x}, the component $G_{3\|}^{\bf x}$ is obtained from Eq.~\eqref{eq:ovGamma4_p}, and the components $(G_{2}^{\theta}, G_{2}^{J})$ are obtained from Eqs.~\eqref{eq:ovGamma4_J}-\eqref{eq:ovGamma4_theta}, respectively. We note that all components, except for $\langle G_{2}^{J}\rangle$, require a solution for the scalar field $\sigma_{4}$, while $(G_{4\bot}^{\bf x}, G_{3}^{p_{\|}})$ also involve $\vb{\Pi}_{3}$.

From the condition $\Gamma_{4\theta} \equiv 0$ in Eq.~\eqref{eq:ovGamma4_theta}, we obtain a differential equation for $\partial\sigma_{4}/\partial\theta$ with 
$\wt{G}_{2}^{J} \equiv G_{2}^{J} - \langle G_{2}^{J}\rangle$ defined by Eq.~\eqref{eq:G_2_mu_tilde}, which yields the solution for $\sigma_{4}$. The missing component $\langle G_{2}^{J}\rangle$ in 
Eq.~\eqref{eq:ovGamma4_theta}, on the other hand, is defined as
\begin{eqnarray}
\langle G_{2}^{J}\rangle & \equiv & \frac{1}{2} \left\langle \pd{G_{2}^{\bf x}}{\theta}\bdot D_{1}({\bf P}_{2})\right\rangle \;+\; \frac{1}{4} \left\langle 
G_{1}^{J}\;\pd{G_{1}^{\theta}}{\theta} \;-\; G_{1}^{\theta}\;\pd{G_{1}^{J}}{\theta} \;+\;
{\sf G}_{1}\cdot\exd G_{1}^{J} \right\rangle \nonumber \\
 &  &+\; \frac{1}{2} \left\langle \pd{\vb{\rho}_{0}}{\theta}\bdot D_{1}^{2}({\bf P}_{4})\right\rangle
\;-\; \frac{1}{2} \left\langle {\sf G}_{1}\cdot\exd F_{1\theta} \;-\; G_{1}^{a}\;\pd{F_{1a}}{\theta}\right\rangle,
\label{eq:G2_J_ave}
\end{eqnarray}
where all components of the first-order generating vector field $(G_{1}^{\bf x},G_{1}^{p_{\|}},G_{1}^{J},G_{1}^{\theta})$ have been calculated at lower orders. After several calculations detailed in App.~\ref{sec:G2_J} [for example, see Eq.~\eqref{eq:G2_J_Ham_app}], we obtain
\begin{eqnarray}
\langle G_{2}^{J}\rangle & = & \frac{J^{2}}{2m\Omega} \left[ \frac{\tau^{2}}{2} + \bhat\bdot\nabla\btimes{\bf R} - \langle\alpha_{1}^{2}
\rangle - \frac{\bhat}{2}\bdot\nabla\btimes\left(\bhat\btimes\nabla\ln B\right) \right] \nonumber \\
 &  &-\; \frac{J}{2}\,\varrho_{\|}^{2} \left[ \vb{\kappa}\bdot\left(3\,\vb{\kappa} \;-\frac{}{} \nabla\ln B\right) \;+\frac{}{} \nabla\bdot\vb{\kappa} - \tau^{2} \right].
\label{eq:G2_J_Ham}
\end{eqnarray} 
Once again, we note that the component $G_{2}^{J} \equiv \wt{G}_{2}^{J} + \langle G_{2}^{J}\rangle$, respectively defined in terms of Eqs.~\eqref{eq:G_2_mu_tilde} and 
Eq.~\eqref{eq:G2_J_Ham}, is independent of the fields $(\Pi_{n\|}, \Psi_{n})$, for $n = 1, 2$, which establishes the equivalency of the Hamiltonian guiding-center theories 
summarized in Sec.~\ref{sec:gcHam}.

When compared with Eq.~\eqref{eq:Psi2_ave_1}, we now obtain the second-order guiding-center Hamiltonian constraint
\begin{eqnarray}
\frac{p_{\|}}{m}\;\Pi_{2\|} \;-\; \Psi_{2} & = &  \frac{m}{2}\,|{\bf v}_{\rm gc}|^{2} \;-\; \vb{\Pi}_{1}\bdot{\bf v}_{\rm gc} \;+\; \frac{J}{2}\,\Omega\;\varrho_{\|}^{2} \left[ 2\,\langle\alpha_{1}^{2}\rangle \;-\; \vb{\kappa}\bdot(3\,\vb{\kappa} - \nabla\ln B) \;-\frac{}{} \nabla\bdot\vb{\kappa} \right] \nonumber \\
 &  &+\; \frac{J^{2}}{2m} \left[ \frac{1}{2}\,\tau^{2} \;+\; \bhat\bdot\nabla\btimes{\bf R} \;-\; \langle\alpha_{1}^{2}
\rangle \;-\; \frac{1}{2}\;\bhat\bdot\nabla\btimes\left(\bhat\btimes\nabla\ln B\right) \right] \nonumber \\
 & \equiv & -\;J\,\Omega\left( \frac{J}{2\,m\Omega}\;\beta_{2\bot} \;+\; \frac{1}{2}\,\varrho_{\|}^{2}\;\beta_{2\|} \right) \;+\; \frac{p_{\|}^{2}}{2m}\;\left(
 \varrho_{\|}^{2}\frac{}{}|\vb{\kappa}|^{2}\right) \;-\; \vb{\Pi}_{1}\bdot{\bf v}_{\rm gc},
\label{eq:Hamiltonian_constraint_2}
\end{eqnarray}
where the second-order functions $\beta_{2\bot}$ and $\beta_{2\|}$ depend only on the guiding-center position
\begin{eqnarray}
\beta_{2\bot} & = & -\,\frac{1}{2}\,\tau^{2} - \bhat\bdot\nabla\btimes{\bf R} + \langle\alpha_{1}^{2}\rangle + \frac{1}{2}\;\bhat\bdot\nabla\btimes\left(\bhat\btimes\nabla\ln B\right) - \left|\bhat\btimes\nabla\ln B\right|^{2},
\label{eq:beta2_perp} \\
\beta_{2\|} & = & -\,2\;\langle\alpha_{1}^{2}\rangle \;-\; 3\;\vb{\kappa}\bdot\left(\nabla\ln B \;-\frac{}{} \vb{\kappa}\right) \;+\; 
\nabla\bdot\vb{\kappa}.
\label{eq:beta2_par}
\end{eqnarray}
The definitions of $\langle\alpha_{1}^{2}\rangle$ and $\bhat\bdot\nabla\btimes{\bf R}$ are given in App.~\ref{sec:dyadic}, and the last term in 
Eq.~\eqref{eq:Hamiltonian_constraint_2} explicitly involves the undetermined component $\vb{\Pi}_{1\bot}$.

We now note that, in contrast to first-order guiding-center Hamiltonian constraint \eqref{eq:Ham_constraint_1}, the right side of Eq.~\eqref{eq:Hamiltonian_constraint_2} contains terms that are constant, quadratic, and quartic in $p_{\|}$. Hence, since $\beta_{2\bot} \neq 0$, we cannot choose $\Psi_{2} = 0$ without making $\Pi_{2\|}$ singular in $p_{\|}$. While a purely Hamiltonian representation of guiding-center theory is possible 
($\Pi_{n\|} \equiv 0$, $n \geq 1$), a purely symplectic representation ($\Psi_{n} \equiv 0$, $n \geq 1$) is, therefore, impossible at all orders.  Here, the second-order Hamiltonian representation 
is expressed as
\begin{equation}
\left. \begin{array}{rcl}
\Pi_{2\|} & = & 0 \\
 &  & \\
 \Psi_{2} & = & \frac{1}{2}\,J\,\Omega\left( J\beta_{2\bot}/m\Omega + \varrho_{\|}^{2}\;\beta_{2\|} \right) - \left(
 \varrho_{\|}^{2}\frac{}{}|\vb{\kappa}|^{2}\right)\;p_{\|}^{2}/2m + \vb{\Pi}_{1}\bdot{\bf v}_{\rm gc}
 \end{array} \right\}.
 \label{eq:Ham_rep_second}
 \end{equation}
 In App.~\ref{sec:comp}, we present a comparison of Eq.~\eqref{eq:Ham_rep_second} with previous works by \cite{Parra_Calvo_2011}, \cite{Burby_SQ_2013}, and \cite{PCBSQ_2014}. Here, the second-order Hamiltonian representation \eqref{eq:Ham_rep_second} yields ${\sf b}^{*} = \bhat$ and $p_{\|}/m = \bhat\bdot d_{\rm gc}{\bf X}/dt - \epsilon^{2}\partial\Psi_{2}/\partial p_{\|}$, where $\partial\Psi_{2}/\partial p_{\|}$ represents a second-order Ba\~{n}os parallel drift-velocity correction.
 
 Another possible choice involves choosing $\Psi_{2}(J,{\bf X})$ as a function of the gyroaction $J$ and the guiding-center position ${\bf X}$ only \citep{Brizard_Tronko_2012}, from 
 Eq.~\eqref{eq:Hamiltonian_constraint_2}, and thus select $\Pi_{2\|}(p_{\|},J,{\bf X}) $ as a non-singular function of $p_{\|}$. This alternative choice guarantees that $p_{\|} \equiv m\,{\sf b}^{*}\bdot d_{\rm gc}{\bf X}/dt$ according to Eq.~\eqref{eq:ELgc_p} and $\Psi_{2}$ enters as either a third-order $(\epsilon_{B}^{3})$ correction to the guiding-center equations of motion for ${\bf X}$ (i.e., $-\,\nabla\Psi_{2}\btimes c{\sf b}^{*}/eB_{\|}^{**}$) and $p_{\|}$ (i.e., $-\nabla\Psi_{2}\bdot{\bf B}^{*}/B_{\|}^{**}$) or a second-order $(\epsilon_{B}^{2})$ correction (i.e., $\partial\Psi_{2}/\partial J$) to the gyrofrequency. 

\section{\label{sec:Jacobian}Guiding-center Jacobian}

So far we have derived the guiding-center transformation \eqref{eq:z_bar_z} up to second order in magnetic-field nonuniformity.  We would like to verify that the guiding-center transformation constructed so far is consistent with the guiding-center Jacobian \eqref{eq:Jac_gc} as expressed in terms of Lie-transform methods as Eq.~\eqref{eq:Jacobian_Lie}. For this purpose, we will need the gyroangle-averaged components
\begin{eqnarray}
\langle G_{1}^{p_{\|}}\rangle & = & J\;\tau, \label{eq:G1p_averaged} \\
\langle G_{1}^{J}\rangle & = & -\;J\,\varrho_{\|}\,\tau, \label{eq:G1J_averaged} \\
\langle G_{2}^{\bf x}\rangle & = & -\; \vb{\Pi}_{1}\btimes\frac{\bhat}{m\Omega} \;+\; \frac{1}{2} \left( \frac{J}{m\Omega}\;\nabla_{\bot}\ln B \;+\;
\varrho_{\|}^{2}\;\vb{\kappa} \right) \label{eq:G2x_averaged}, \\
\langle G_{2}^{p_{\|}}\rangle & = & -\;\varrho_{\|}\bhat\btimes\vb{\kappa}\bdot\vb{\Pi}_{1} + \varrho_{\|}\,\vb{\kappa}\bdot\left(J\;\nabla\ln B 
\;+\; \frac{p_{\|}^{2}\vb{\kappa}}{m\Omega}\right) \nonumber \\
 &  &-\; \Pi_{2\|} - J\varrho_{\|}\left(\frac{1}{2}\,\tau^{2} - \langle\alpha_{1}^{2}\rangle\right), 
\label{eq:G2p_averaged} \\
\langle G_{2}^{J}\rangle & = & \frac{J^{2}}{2m\Omega} \left[ \frac{1}{2}\,\tau^{2} \;+\; \bhat\bdot\nabla\btimes{\bf R} \;-\; \langle\alpha_{1}^{2}
\rangle \;-\; \frac{1}{2}\;\bhat\bdot\nabla\btimes(\bhat\btimes\nabla\ln B) \right] \nonumber \\
 &  &-\; \frac{1}{2}\;J\,\varrho_{\|}^{2} \left[ \vb{\kappa}\bdot(3\,\vb{\kappa} - \nabla\ln B) \;+\frac{}{} \nabla\bdot\vb{\kappa} \;-\; \tau^{2} 
\right],
\label{eq:G2J_averaged}
\end{eqnarray}
where Eq.~\eqref{eq:G2J_averaged} comes from from the fourth-order expression \eqref{eq:G2_J_Ham}.

The guiding-center Jacobian \eqref{eq:Jac_gc} is given by ${\cal J}_{\rm gc}/{\cal J}_{0} \equiv B_{\|}^{**}/B$:
\begin{eqnarray} 
\frac{{\cal J}_{\rm gc}}{{\cal J}_{0}} & = & 1 \;+\; \epsilon\;\varrho_{\|}\,\tau  \;+\; \epsilon^{2}\;\left[\pd{\Pi_{2\|}}{p_{\|}} \;+\; 
\varrho_{\|}\,\pd{\vb{\Pi}_{1\bot}}{p_{\|}}\bdot\nabla\btimes\bhat \right. \nonumber \\
 &  &\left. +\; \frac{c\,\bhat}{eB}\bdot\nabla\btimes(\vb{\Pi}_{1} - J\,{\bf R})\right] \;+\; \cdots \;\equiv\; 1 \;+\; \epsilon\,\frac{{\cal J}_{1}}{{\cal J}_{0}} \;+\; \epsilon^{2}\,\frac{{\cal J}_{2}}{{\cal J}_{0}} + \cdots,
\label{eq:B_star}
\end{eqnarray}
where we used the identity $\partial\Pi_{1\|}/\partial p_{\|} \equiv 0$, which follows from Eq.~\eqref{eq:Pi1||_def}, while $\partial\vb{\Pi}_{1\bot}/\partial p_{\|}$ is undetermined. At first order, using 
Eqs.~\eqref{eq:G1_p||}-\eqref{eq:G1_J} and \eqref{eq:G1_theta_def}, we find
\begin{eqnarray}
\frac{{\cal J}_{1}}{{\cal J}_{0}} & \equiv & \frac{1}{B}\;\nabla\bdot\left(B\frac{}{}\vb{\rho}_{0}\right) \;-\; \left( \pd{G_{1}^{p_{\|}}}{p_{\|}} \;+\; 
\pd{G_{1}^{J}}{J}\;+\; \pd{G_{1}^{\theta}}{\theta} \right) \nonumber \\
 & = & \varrho_{\|}\,\tau \;\equiv\; -\;\left( \pd{\langle G_{1}^{p_{\|}}\rangle}{p_{\|}} \;+\; \pd{\langle G_{1}^{J}\rangle}{J}\right).
\label{eq:Jac_1}
\end{eqnarray}
In the last equality, we have used the fact that, since the guiding-center Jacobian is gyroangle-independent, we may also gyroangle-average Eq.~\eqref{eq:Jac_1}, which greatly simplifies the calculations, since 
Eqs.~\eqref{eq:G1p_averaged}-\eqref{eq:G1J_averaged} yield $\partial\langle G_{1}^{p_{\|}}\rangle/\partial p_{\|} = 0 $ and $\partial\langle G_{1}^{J}\rangle/\partial J = -\,\varrho_{\|}\,\tau$, while the gyroangle-dependent terms cancel out exactly.

At second order, we must verify that
\begin{eqnarray}
\frac{{\cal J}_{2}}{{\cal J}_{0}} & = & \pd{\Pi_{2\|}}{p_{\|}} \;+\; \varrho_{\|}\,\pd{\vb{\Pi}_{1\bot}}{p_{\|}}\bdot\nabla\btimes\bhat \;+\; \frac{c\,\bhat}{eB}\bdot\nabla\btimes\left(\vb{\Pi}_{1} \;-\frac{}{} J\,{\bf R}\right) \label{eq:Jac_2} \\
 & \equiv & -\,\frac{1}{B}\nabla\bdot\left(B\frac{}{}\langle G_{2}^{\bf x}\rangle \right) \;-\; \pd{}{p_{\|}}\left( \langle G_{2}^{p_{\|}}\rangle \;+\;
\frac{1}{2}\;J\,\varrho_{\|}\,\tau^{2}\right) -\; \pd{}{J}\left( \langle G_{2}^{J}\rangle \;-\; \frac{1}{2}\;J\,\varrho_{\|}^{2}\,\tau^{2}\right),
\nonumber
\end{eqnarray}
where we need the gyroangle-averaged expressions for $(G_{2}^{\bf x}, G_{2}^{p_{\|}},G_{2}^{J})$, with $(\langle G_{1}^{p_{\|}}
\rangle, \langle G_{1}^{J}\rangle) = (J\,\tau, -\,J\,\varrho_{\|}\tau)$. First, using Eq.~\eqref{eq:G2x_averaged}, we find
\begin{eqnarray}
-\,\frac{1}{B}\nabla\bdot\left(B\frac{}{}\langle G_{2}^{\bf x}\rangle \right) & = & -\,\frac{J}{2m\Omega}\;\nabla\bdot\left[\left(\bhat\btimes\nabla\ln B
\right)\btimes\bhat\right] \;-\; \frac{1}{2}\,B\,\varrho_{\|}^{2}\;\nabla\bdot\left(\frac{\vb{\kappa}}{B}\right) \nonumber \\
 &  &+\; \frac{c}{eB}\;\nabla\bdot\left(\vb{\Pi}_{1}\btimes\bhat\right).
\label{eq:divG2_x}
\end{eqnarray}
Next, using Eq.~\eqref{eq:G2p_averaged}, we obtain 
\begin{eqnarray}
-\; \pd{}{p_{\|}}\left( \langle G_{2}^{p_{\|}}\rangle \;+\; \frac{1}{2}\;J\,\varrho_{\|}\,\tau^{2}\right) & = & -\;\frac{J}{m\Omega}
\left( \langle\alpha_{1}^{2}\rangle \;+\; \frac{1}{2}\,\vb{\kappa}\bdot\nabla\ln B \right) \;-\; \frac{3}{2}\,\varrho_{\|}^{2}\,|\vb{\kappa}|^{2} 
\nonumber \\
 &  &+\; \left( \frac{\vb{\Pi}_{1}}{m\Omega} \;+\; \varrho_{\|}\;\pd{\vb{\Pi}_{1}}{p_{\|}}\right) \bdot\left(\bhat\btimes\vb{\kappa}\right) \;+\; 
\pd{\Pi_{2\|}}{p_{\|}}.
\label{eq:divG2_p}
\end{eqnarray}
Lastly, using Eq.~\eqref{eq:G2J_averaged}, we obtain
\begin{eqnarray}
-\pd{}{J}\left( \langle G_{2}^{J}\rangle - \frac{J}{2}\,\varrho_{\|}^{2}\,\tau^{2}\right) & = & -\frac{J}{m\Omega} \left[ \frac{\tau^{2}}{2} + \bhat\bdot\nabla\btimes{\bf R} - \langle\alpha_{1}^{2}\rangle - \frac{\bhat}{2}\bdot\nabla\btimes\left(\bhat\btimes\nabla\ln B\right) \right] 
\nonumber \\
 &  &+\; \frac{1}{2}\;\varrho_{\|}^{2} \left[ \vb{\kappa}\bdot(3\,\vb{\kappa} - \nabla\ln B) \;+\frac{}{} \nabla\bdot\vb{\kappa} \right].
\label{eq:divG2_J}
\end{eqnarray}
By combining Eqs.~\eqref{eq:divG2_x}-\eqref{eq:divG2_J} into Eq.~\eqref{eq:Jac_2}, we obtain
\begin{eqnarray*}
\frac{{\cal J}_{2}}{{\cal J}_{0}} & = & \pd{\Pi_{2\|}}{p_{\|}} \;+\; \frac{c\,\bhat}{eB}\bdot\nabla\btimes(\vb{\Pi}_{1} - J\,{\bf R}) \;+\; \varrho_{\|}\;
\pd{\vb{\Pi}_{1\bot}}{p_{\|}}\bdot\left(\bhat\btimes\vb{\kappa}\right) \\
 &  &-\; \frac{J}{2m\Omega}\;\left\{ \nabla\bdot\left[\left(\bhat\btimes\nabla\ln B\right)\btimes\bhat\right] \;+\; \vb{\kappa}\bdot\nabla\ln B \;-\; \bhat\bdot\nabla\btimes\left(\bhat\btimes\nabla\ln B\right) \right\} \\
 &  &+\; \frac{1}{2}\;\varrho_{\|}^{2} \left[ \vb{\kappa}\bdot(3\,\vb{\kappa} - \nabla\ln B) \;+\frac{}{} \nabla\bdot\vb{\kappa} \;-\; 
\nabla\bdot\vb{\kappa} \;-\; 3\;|\vb{\kappa}|^{2} \;+\; \vb{\kappa}\bdot\nabla\ln B \right] \\
& = & \pd{\Pi_{2\|}}{p_{\|}} \;+\; \frac{c\,\bhat}{eB}\bdot\nabla\btimes(\vb{\Pi}_{1} - J\,{\bf R}) \;+\; \varrho_{\|}\;\pd{\vb{\Pi}_{1}}{p_{\|}}\bdot
\left(\bhat\btimes\vb{\kappa}\right).
\end{eqnarray*}
Hence, the guiding-center transformation derived up to second order in magnetic-field nonuniformity is consistent with the Jacobian constraint. 

We see that, while the Jacobian constraints are satisfied up to second order in magnetic-field nonuniformity, we are unable to obtain a constraint on the perpendicular component $\vb{\Pi}_{1\bot}$.  In 
Sec.~\ref{subsec:gc_pol}, however, we will show that 
$\vb{\Pi}_{1\bot} \equiv -\frac{1}{2}\,J\,\bhat\btimes\vb{\kappa}$ so that, with Eq.~\eqref{eq:Pi1||_def}, we find that, in the symplectic representation, we find
\begin{equation}
\vb{\Pi}_{1} \;\equiv\; -\;\frac{1}{2}\,J\;\nabla\btimes\bhat.
\label{eq:Pi1_constraint}
\end{equation}
We will also show that Eq.~\eqref{eq:Pi1_constraint} leads to an accurate guiding-center representation of the toroidal canonical momentum (see Sec.~\ref{sec:gc_pol_canonical}).

\section{\label{sec:push_Lag}Guiding-center Push-forward Lagrangian Constraint}

In our search for a definitive expression for $\vb{\Pi}_{1\bot}$,  we now wish to explore a new perturbation approach to guiding-center Hamiltonian theory. We begin with the following remark for the phase-space Lagrangian formulation of single-particle dynamics in a potential $U({\bf x})$, where the particle position ${\bf x}$ and its velocity ${\bf v}$ are viewed as independent phase-space coordinates. From the phase-space Lagrangian $L({\bf x},{\bf v};\dot{\bf x},\dot{\bf v}) = m{\bf v}\bdot d{\bf x}/dt - [m\,|{\bf v}|^{2}/2 + U({\bf x})]$, we first obtain the Euler-Lagrange equation for ${\bf x}$: $m\,d{\bf v}/dt = -\,\nabla U$. Since the phase-space Lagrangian is independent of $d{\bf v}/dt$, however, we immediately obtain the Lagrangian constraint
\begin{equation}
\pd{L}{\bf v} \;=\; m\;\frac{d{\bf x}}{dt} \;-\; {\bf p} \;\equiv\; 0.
\label{eq:Lag_constraint}
\end{equation}
Hence, we require that the guiding-center transformation must preserve this Lagrangian constraint.

Our new perturbation approach begins with the guiding-center version of the Lagrangian constraint \eqref{eq:Lag_constraint}. First, using the functional definition for the 
guiding-center time-evolution operator 
\begin{equation}
\frac{d_{\rm gc}}{dt} \;\equiv\; {\sf T}_{\rm gc}^{-1}\left(\frac{d}{dt}\;{\sf T}_{\rm gc}\right),
\label{eq:d_gc_def}
\end{equation}
we introduce the guiding-center Lagrangian constraint
\begin{eqnarray}
m\,{\sf T}_{\rm gc}^{-1}\left(\frac{d{\bf x}}{dt}\right) \;\equiv\; m\,\frac{d_{\rm gc}}{dt}\left({\sf T}_{\rm gc}^{-1}{\bf x} \right) \;=\; m\;\left(
\frac{d_{\rm gc}{\bf X}}{dt} + \frac{d_{\rm gc}\vb{\rho}_{\rm gc}}{dt}\right) \;\equiv\;{\sf T}_{\rm gc}^{-1}{\bf p}_{0},
\label{eq:gcLc_id} 
\end{eqnarray}
where ${\bf p}_{0}$ denotes the particle momentum expressed in terms of local coordinates $z_{0}^{\alpha}$, the guiding-center velocity 
$d_{\rm gc}{\bf X}/dt$ is defined by Eq.~\eqref{eq:gc_X_dot}, and the guiding-center displacement velocity is defined as
\begin{equation}
\frac{d_{\rm gc}\vb{\rho}_{\rm gc}}{dt} \;=\; \epsilon^{-1}\;\pd{\Psi}{J}\;\pd{\vb{\rho}_{\rm gc}}{\theta} \;+\; \frac{d_{\rm gc}{\bf X}}{dt}\bdot
\nabla^{*}\vb{\rho}_{\rm gc} \;+\; \frac{d_{\rm gc}p_{\|}}{dt}\;\pd{\vb{\rho}_{\rm gc}}{p_{\|}},
\label{eq:rho_gc_dot}
\end{equation}
which includes the guiding-center polarization velocity $d_{\rm gc}\langle\vb{\rho}_{\rm gc}\rangle/dt$. Here, the guiding-center displacement is expanded as
\begin{equation}
\vb{\rho}_{\rm gc} \;\equiv\; {\sf T}_{\rm gc}^{-1}{\bf x} \;-\; {\bf X} \;=\; \epsilon\,\vb{\rho}_{0} \;+\; \epsilon^{2}\;
\vb{\rho}_{1} \;+\; \epsilon^{3}\,\vb{\rho}_{2} + \cdots,
\label{eq:rho_gc}
\end{equation}
where the higher-order gyroradius corrections are
\begin{eqnarray}
\vb{\rho}_{1} & = & -\;G_{2}^{\bf x} \;-\; \frac{1}{2}\;{\sf G}_{1}\cdot\exd\vb{\rho}_{0}, \label{eq:rho_1} \\
\vb{\rho}_{2} & = & -\;G_{3}^{\bf x} - {\sf G}_{2}\cdot\exd\vb{\rho}_{0} + \frac{1}{6}\,{\sf G}_{1}\cdot
\exd({\sf G}_{1}\cdot\exd\vb{\rho}_{0}). \label{eq:rho_2}
\end{eqnarray}
We note that, in general, the higher-order gyroradius corrections satisfy $\langle\vb{\rho}_{n}\rangle \neq 0$ and $\vb{\rho}_{n}\bdot\bhat \neq 0$ for $n \geq 1$. In addition, we note that the particle displacement $\vb{\Delta}_{\rm gc} \equiv {\bf x} - {\sf T}_{\rm gc}{\bf X} \equiv {\sf T}_{\rm gc}^{-1}\vb{\rho}_{\rm gc}$ is identical to the guiding-center displacement $\vb{\rho}_{\rm gc}$ only at the lowest order in $\epsilon$. The guiding-center Lagrangian constraint \eqref{eq:gcLc_id} is expressed only in terms of the guiding-center displacement \eqref{eq:rho_gc}.

On the right side of Eq.~\eqref{eq:gcLc_id}, the push-forward of the particle momentum ${\sf T}_{\rm gc}^{-1}{\bf p}_{0}$ can be expanded up to second order in 
$\epsilon$ as
\begin{eqnarray}
{\sf T}_{\rm gc}^{-1}{\bf p}_{0} & = & {\bf p}_{0} \;+\; \epsilon\,\left[\vb{\rho}_{0}\bdot\nabla{\bf p}_{0} \;-\; \left( G_{1}^{p_{\|}}\;\bhat \;+\;
G_{1}^{J}\,\pd{{\bf p}_{\bot 0}}{J} \;+\; G_{1}^{\theta}\,\pd{{\bf p}_{\bot 0}}{\theta} \right) \right] \nonumber \\
 &  &-\;\epsilon^{2}\,\left[ G_{2}^{\bf x}\bdot\nabla{\bf p}_{0} \;+\; \left( G_{2}^{p_{\|}}\;\bhat \;+\; G_{2}^{J}\,\pd{{\bf p}_{\bot 0}}{J} \;+\; 
G_{2}^{\theta}\,\pd{{\bf p}_{\bot 0}}{\theta} \right) \right] \nonumber \\
 &  &+\; \frac{\epsilon^{2}}{2}\;{\sf G}_{1}\cdot\exd\left( G_{1}^{p_{\|}}\;\bhat \;+\; G_{1}^{J}\,\pd{{\bf p}_{\bot 0}}{J} \;+\; G_{1}^{\theta}\,
\pd{{\bf p}_{\bot 0}}{\theta} \;-\; \vb{\rho}_{0}\bdot\nabla{\bf p}_{0} \right) \:+\; \cdots,
\label{eq:push_p}
\end{eqnarray}
while the push-forward of the particle velocity is expanded up to second order in $\epsilon$ as
\begin{eqnarray}
\frac{d_{\rm gc}{\bf X}}{dt} \;+\; \frac{d_{\rm gc}\vb{\rho}_{\rm gc}}{dt} & = & \left(\frac{d_{0}{\bf X}}{dt} \;+\; \Omega\;\pd{\vb{\rho}_{0}}{\theta} \right) \;+\; \epsilon\;\left( \frac{d_{1}{\bf X}}{dt} \;+\; \Omega\;\pd{\vb{\rho}_{1}}{\theta} \;+\; \frac{d_{0}{\bf X}}{dt}\bdot\nabla_{0}^{*}
\vb{\rho}_{0}\right) \label{eq:gc_right} \\
 &  &+\; \epsilon^{2}\;\left[ \frac{d_{2}{\bf X}}{dt} \;+\; \Omega\;\pd{\vb{\rho}_{2}}{\theta} \;+\; \pd{}{J}\left(\Psi_{2} - \frac{p_{\|}}{m}\,\Pi_{2\|}\right)\,\pd{\vb{\rho}_{0}}{\theta} \right. \nonumber \\
 &  &\hspace*{0.5in}\left.+\; \frac{d_{1}{\bf X}}{dt}\bdot\nabla_{0}^{*}\vb{\rho}_{0} \;+\; \frac{d_{0}{\bf X}}{dt}\bdot\nabla_{0}^{*}
\vb{\rho}_{1} \;+\; \frac{d_{0}p_{\|}}{dt}\;\pd{\vb{\rho}_{1}}{p_{\|}} \right] + \cdots,
\nonumber
\end{eqnarray}
where we used $m\,d_{0}{\bf X}/dt = p_{\|}\,\bhat$, $\nabla^{*}_{0} \equiv \nabla + ({\bf R} - \partial\vb{\Pi}_{1}/\partial J)\;\partial/\partial
\theta$, and $\partial\vb{\rho}_{0}/\partial p_{\|} \equiv 0$ with the first-order symplectic representation $\Psi_{1} \equiv 0$. At the lowest order in $\epsilon$, the guiding-center Lagrangian constraint \eqref{eq:gcLc_id} yields
\begin{equation}
{\bf p}_{0} \;\equiv\; p_{\|}\,\bhat \;+\; m\,\Omega\;\pd{\vb{\rho}_{0}}{\theta},
\label{eq:p0_def}
\end{equation}
and we henceforth remove the ``local'' tag on $(p_{\|0} = p_{\|}, J_{0} = J, \theta_{0} = \theta)$. 

At the next orders, we use the expansions $d_{\rm gc}Z^{\alpha}/dt \equiv \sum_{n}^{\infty}\epsilon^{n}\;d_{n}Z^{\alpha}/dt$ (which now includes the expansion of $B_{\|}^{**}$) to obtain
\begin{eqnarray}
m\,\frac{d_{1}{\bf X}}{dt} & = & \frac{1}{\Omega} \left( \frac{p_{\|}^{2}}{m}\;\nabla\btimes\bhat + J\;\bhat\btimes\nabla\Omega\right) \;-\; p_{\|}\;
\left(\varrho_{\|}\frac{}{}\tau\right)\;\bhat \nonumber \\
 & = & \bhat\btimes\left(J\;\nabla\ln B \;+\; \frac{p_{\|}^{2}}{m\Omega}\;\vb{\kappa}\right) \;\equiv\; m\,{\bf v}_{\rm gc}, \label{eq:dx1_dt} \\
m\,\frac{d_{2}{\bf X}}{dt} & = & \left[m\,\pd{\Psi_{2}}{p_{\|}} - p_{\|}\,\left(\pd{\Pi_{2\|}}{p_{\|}} + \pd{\vb{\Pi}_{1}}{p_{\|}}\bdot\varrho_{\|}
\nabla\btimes\bhat\right)\right] \bhat \;+\; \pd{\vb{\Pi}_{1}}{p_{\|}}\btimes J\,\nabla\ln B  \\
 &  &-\; m\,(\varrho_{\|}\tau)\;{\bf v}_{\rm gc} \;+\; \varrho_{\|}\;\left[ \nabla\btimes\left(\vb{\Pi}_{1} \;-\frac{}{} J\;{\bf R}\right) - \bhat\;\bhat\bdot\nabla\btimes\left(
\vb{\Pi}_{1} \;-\frac{}{} J\;{\bf R}\right) \right], \nonumber
\end{eqnarray}
where we used the first-order symplectic representation $\Psi_{1} \equiv 0$ and $\Pi_{1\|} \equiv -\frac{1}{2}\,J\,\tau$.

\subsection{First-order constraint}

At first order, the guiding-center Lagrangian constraint \eqref{eq:gcLc_id} yields
\begin{eqnarray}
-\;{\sf G}_{1}\cdot\exd{\bf p}_{0} & = & m\,\left( {\bf v}_{\rm gc} \;+\; \Omega\;\pd{\vb{\rho}_{1}}{\theta} \;+\; \frac{d_{0}\vb{\rho}_{0}}{dt}\right),
\label{eq:gcLc_1}
\end{eqnarray}
where $G_{1}^{\bf x} = -\,\vb{\rho}_{0}$ and
\begin{eqnarray}
\frac{d_{0}\vb{\rho}_{0}}{dt} & = & \frac{p_{\|}}{m} \left[ -\;(\vb{\rho}_{0}\bdot\vb{\kappa})\;\bhat \;+\; \frac{1}{2}\,\tau\;\pd{\vb{\rho}_{0}}{\theta} \;+\; \frac{1}{2}\,\left(\nabla\bdot\bhat\right)\;\vb{\rho}_{0}\right], 
\label{eq:d0_rho0} \\
\Omega\pd{\vb{\rho}_{1}}{\theta} & \equiv & \frac{p_{\|}}{m} \left[ 2\;(\vb{\rho}_{0}\bdot\vb{\kappa}) \bhat \;-\; \frac{1}{2}\,(\tau + \alpha_{1})\;
\pd{\vb{\rho}_{0}}{\theta} \;-\; \alpha_{2}\;\vb{\rho}_{0} \right] \nonumber \\
 &  &+\; \frac{J}{m} \left( \alpha_{1}\,\bhat \;-\frac{}{} 2\,{\sf a}_{1}\bdot\nabla\ln B\right).
\label{eq:rho1_theta}
\end{eqnarray}
The first-order guiding-center Lagrangian constraint \eqref{eq:gcLc_1} yields the following component equations
\begin{eqnarray}
(\vb{\rho}_{0}\bdot\nabla{\bf p}_{0})\bdot\bhat \;-\; G_{1}^{p_{\|}} & = & m\bhat\bdot\left( {\bf v}_{\rm gc} \;+\; \Omega\;\pd{\vb{\rho}_{1}}{\theta} \;+\; \frac{d_{0}\vb{\rho}_{0}}{dt}\right), \label{eq:gcLc_1_p} \\
(\vb{\rho}_{0}\bdot\nabla{\bf p}_{0})\bdot\pd{\vb{\rho}_{0}}{\theta} \;-\; G_{1}^{J} & = & m\pd{\vb{\rho}_{0}}{\theta}\bdot\left( {\bf v}_{\rm gc} \;+\; \Omega\;\pd{\vb{\rho}_{1}}{\theta} \;+\; \frac{d_{0}\vb{\rho}_{0}}{dt}\right), \label{eq:gcLc_1_J} \\
(\vb{\rho}_{0}\bdot\nabla{\bf p}_{0})\bdot\pd{\vb{\rho}_{0}}{J} \;+\; G_{1}^{\theta} & = & m\pd{\vb{\rho}_{0}}{J}\bdot\left( {\bf v}_{\rm gc} \;+\; \Omega\;\pd{\vb{\rho}_{1}}{\theta} \;+\; \frac{d_{0}\vb{\rho}_{0}}{dt}\right), \label{eq:gcLc_1_theta}
\end{eqnarray}
where
\[ \vb{\rho}_{0}\bdot\nabla{\bf p}_{0} \;=\; J\,(\tau + 2\alpha_{1})\,\bhat \;+\; p_{\|}\,\vb{\rho}_{0}\bdot\nabla\bhat \;-\; \vb{\rho}_{0}\bdot{\bf R}\;\pd{{\bf p}_{\bot0}}{\theta} \;+\;
\frac{1}{2}\,\vb{\rho}_{0}\bdot\nabla\ln B\;{\bf p}_{\bot0}. \]
The parallel equation \eqref{eq:gcLc_1_p} becomes
\[ J\;\left(\tau \;+\frac{}{} 2\,\alpha_{1}\right) \;-\; G_{1}^{p_{\|}} \;=\; m\Omega\;\pd{\vb{\rho}_{1}}{\theta}\bdot\bhat \;-\; p_{\|}\;\vb{\rho}_{0}
\bdot\vb{\kappa}, \]
which yields the same expression \eqref{eq:G1_p||} for $G_{1}^{p_{\|}}$:
\begin{equation}
G_{1}^{p_{\|}} \;=\; -\,p_{\|}\;\vb{\rho}_{0}\bdot\vb{\kappa} \;+\; J\,\left(\tau \;+\frac{}{} \alpha_{1}\right).
\label{eq:Banos_1}
\end{equation}
The remaining equations \eqref{eq:gcLc_1_J}-\eqref{eq:gcLc_1_theta} yield the components $G_{1}^{J}$ and $G_{1}^{\theta}$. We note that Eqs.~\eqref{eq:gcLc_1_p}-\eqref{eq:gcLc_1_theta} point to a dynamical derivation of the guiding-center transformation as opposed to a Hamiltonian derivation.

\subsection{Second-order constraint}

At second order, the guiding-center Lagrangian constraint \eqref{eq:gcLc_id} yields
\begin{eqnarray}
 &  &-\;G_{2}^{\bf x}\bdot\nabla{\bf p}_{0} \;-\; \left( G_{2}^{p_{\|}}\;\bhat \;+\; G_{2}^{J}\,\pd{{\bf p}_{\bot 0}}{J} \;+\; 
G_{2}^{\theta}\,\pd{{\bf p}_{\bot 0}}{\theta} \right) + \frac{1}{2}\;{\sf G}_{1}\cdot\exd\left( {\sf G}_{1}\cdot\exd{\bf p}_{0} \right) \label{eq:gcLc_2} \\
 & &=\; m\,\frac{d_{2}{\bf X}}{dt} + m\Omega\;\pd{\vb{\rho}_{2}}{\theta} + \pd{}{J}\left(\Psi_{2} - \frac{p_{\|}}{m}\,\Pi_{2\|}\right)\,
\pd{\vb{\rho}_{0}}{\theta} + m\,\frac{d_{1}{\bf X}}{dt}\bdot\nabla_{0}^{*}\vb{\rho}_{0} + m\;\frac{d_{0}\vb{\rho}_{1}}{dt}.
\nonumber
\end{eqnarray}
The parallel component of the gyroangle-averaged second-order constraint equation \eqref{eq:gcLc_2} yields
\begin{eqnarray}
m \left( \frac{p_{\|}}{m}\,\pd{\Pi_{2\|}}{p_{\|}} \;-\; \pd{\Psi_{2}}{p_{\|}}\right) + m\,{\bf v}_{\rm gc}\bdot\pd{\vb{\Pi}_{1}}{p_{\|}} & = & \langle G_{2}^{p_{\|}}\rangle 
\;+\; m\;\frac{d_{0}\langle\vb{\rho}_{1}\rangle}{dt}\bdot\bhat \;-\; \left\langle G_{2}^{\bf x}\bdot\nabla\bhat\bdot{\bf p}_{\bot}\right\rangle \nonumber \\
 &  &-\; \frac{1}{2}\left\langle {\sf G}_{1}\cdot\frac{}{}\exd\left({\sf G}_{1}\cdot\exd{\bf p}\right)\right\rangle\bdot\bhat.
\label{eq:Pi_2_hier}
\end{eqnarray}
If we combine this equation with Eq.~\eqref{eq:G2_p_ave}:
\begin{equation}
\Pi_{2\|} \;=\; -\;\langle G_{2}^{p_{\|}}\rangle \;+\; p_{\|}\;\vb{\kappa}\bdot \langle G_{2}^{\bf x}\rangle \;+\; \bhat\bdot\left\langle
D_{1}^{2}({\bf P}_{3})\right\rangle,
\label{eq:G2_p||_ave}
\end{equation}
the contributions from $\langle G_{2}^{p_{\|}}\rangle$ cancel out when Eqs.~\eqref{eq:Pi_2_hier}-\eqref{eq:G2_p||_ave} are combined and we obtain the second-order Lagrangian constraint
\begin{eqnarray}
m\;\pd{}{p_{\|}}\left(\frac{p_{\|}}{m}\,\Pi_{2\|} \;-\; \Psi_{2} \right) + m\,{\bf v}_{\rm gc}\bdot\pd{\vb{\Pi}_{1}}{p_{\|}} & = & \left( m\;\frac{d_{0}\langle
\vb{\rho}_{1}\rangle}{dt} \;+\; \left\langle D_{1}^{2}({\bf P}_{3})\right\rangle\right) \bdot\bhat \nonumber \\
 &  &+\; p_{\|}\,\vb{\kappa}\bdot\langle G_{2}^{\bf x}\rangle \;-\; \left\langle G_{2}^{\bf x}\bdot\nabla\bhat\bdot{\bf p}_{\bot}\right\rangle \nonumber \\
 &  &-\; \frac{1}{2}\left\langle {\sf G}_{1}\cdot\frac{}{}\exd\left({\sf G}_{1}\cdot\exd{\bf p}\right)\right\rangle\bdot\bhat.
 \label{eq:second_Lag}
\end{eqnarray}
In App.~\ref{sec:Pi_2}, the right side is explicitly calculated as
\begin{eqnarray}
m\pd{}{p_{\|}}\left(\frac{p_{\|}}{m}\,\Pi_{2\|} - \Psi_{2} \right) + m\,{\bf v}_{\rm gc}\bdot\pd{\vb{\Pi}_{1}}{p_{\|}} & = & 2\,p_{\|}\varrho_{\|}^{2}\,|\vb{\kappa}|^{2}  - J\,\varrho_{\|}\beta_{2\|} - \vb{\Pi}_{1}\bdot\pd{(m{\bf v}_{\rm gc})}{p_{\|}},
\label{eq:Pi_2_final}
\end{eqnarray}
which can clearly be recovered from the second-order guiding-center Hamiltonian constraint \eqref{eq:Hamiltonian_constraint_2}.

\subsection{Guiding-center Hamiltonian constraint}

We conclude this section by applying the guiding-center Lagrangian constraint \eqref{eq:gcLc_id} on the guiding-center Hamiltonian
\begin{equation}
H_{\rm gc} \;=\; {\sf T}_{\rm gc}^{-1}\left( \frac{m}{2}\,\left|\frac{d{\bf x}}{dt}\right|^{2}\right) \;=\; \frac{m}{2}\;\left|\frac{d_{\rm gc}{\bf X}}{dt} + 
\frac{d_{\rm gc}\vb{\rho}_{\rm gc}}{dt}\right|^{2}.
\label{eq:gcLc_id_Ham}
\end{equation}
Since the guiding-center Hamiltonian must also be gyroangle-independent, the guiding-center Hamiltonian \eqref{eq:gcLc_id_Ham} must also be expressed as
\begin{equation}
H_{\rm gc} \;\equiv\; \frac{m}{2} \left\langle \left|\frac{d_{\rm gc}{\bf X}}{dt} + \frac{d_{\rm gc}\vb{\rho}_{\rm gc}}{dt}\right|^{2}\right\rangle,
\end{equation}
which means that the gyroangle-dependent terms on the right side of Eq.~\eqref{eq:gcLc_id_Ham} must vanish identically as could be readily verified explicitly. 

If we use the first-order symplectic and second-order Hamiltonian representations, defined by Eqs.~\eqref{eq:Pi1||_def} and \eqref{eq:Ham_rep_second}, we therefore find the first-order guiding-center potential energy 
\begin{equation}
\Psi_{1} \;\equiv\; \frac{d}{d\epsilon}\left( \frac{m}{2} \left\langle \left|\frac{d_{\rm gc}{\bf X}}{dt} + \frac{d_{\rm gc}\vb{\rho}_{\rm gc}}{dt}
\right|^{2}\right\rangle\right)_{\epsilon = 0} \;=\; 0,
\end{equation}
and the second-order guiding-center potential energy $\Psi_{2}$ defined as
\begin{equation}
\Psi_{2} \;\equiv\; \frac{1}{2}\,\frac{d^{2}}{d\epsilon^{2}}\left( \frac{m}{2} \left\langle \left|\frac{d_{\rm gc}{\bf X}}{dt} + \frac{d_{\rm gc}\vb{\rho}_{\rm gc}}{dt}
\right|^{2}\right\rangle\right)_{\epsilon = 0}.
\end{equation}
Hence, the physical meaning of the higher-order corrections $\Psi_{n}$ is clearly expressed in terms of the $\epsilon$-expansion of the guiding-center 
push-forward of the particle kinetic energy.

\section{\label{sec:gc_pol_canonical}Guiding-center Polarization and Toroidal Canonical Momentum}

So far we have been unable to find a way to determine the perpendicular component $\vb{\Pi}_{1\bot}$ within guiding-center Lie-transform perturbation theory. In the present section, we show how $\vb{\Pi}_{1\bot}$ 
can be determined by requiring that the guiding-center transformation yields exactly the guiding-center polarization obtained by \cite{Pfirsch_1984} and \cite{Kaufman_1986}. A recent variational derivation of the 
guiding-center magnetization by \cite{Brizard_Tronci_2016} confirms our choice for $\vb{\Pi}_{1\bot}$.

\subsection{\label{subsec:gc_pol}Guiding-center Polarization} 

The guiding-center displacement $\vb{\rho}_{\rm gc} \equiv {\sf T}_{\rm gc}^{-1}{\bf x} - {\bf X}$ is explicitly expressed as
\begin{equation}
\vb{\rho}_{\rm gc} \;=\; -\,\epsilon\,G_{1}^{\bf x} \;-\; \epsilon^{2}\,G_{2}^{\bf x} \;+\; \frac{\epsilon^{2}}{2}\,{\sf G}_{1}\cdot\exd G_{1}^{\bf x} 
\;+\; \cdots \;\equiv\; \epsilon\,\vb{\rho}_{0} \;+\; \epsilon^{2}\,\vb{\rho}_{1} \;+\; \cdots,
\label{eq:rho_gc_def}
\end{equation}
where the first-order guiding-center displacement \eqref{eq:rho_1} is expressed explicitly as
\begin{eqnarray}
\vb{\rho}_{1} & = & -\,G_{2}^{\bf x} \;+\; \frac{1}{2}\,\vb{\rho}_{0}\bdot\nabla\vb{\rho}_{0} \;-\; \frac{1}{2} \left( G_{1}^{J}\;\pd{\vb{\rho}_{0}}{J}
\;+\; G_{1}^{\theta}\;\pd{\vb{\rho}_{0}}{\theta} \right) \nonumber \\
 & = & - \left( G_{2\|}^{\bf x} \;+\; \frac{1}{2}\,\vb{\rho}_{0}\bdot\nabla\bhat\bdot\vb{\rho}_{0}\right)\,\bhat \;-\; \varrho_{\|}\tau\;\vb{\rho}_{0}
\;+\; \vb{\Pi}_{1}\btimes\frac{\bhat}{m\Omega} \nonumber \\
 &  &-\; \left[ G_{1}^{J}\;\pd{\vb{\rho}_{0}}{J} \;+\; \left( G_{1}^{\theta} \;+\frac{}{} \vb{\rho}_{0}\bdot
{\bf R}\right)\;\pd{\vb{\rho}_{0}}{\theta} \right].
\label{eq:rho1_def}
\end{eqnarray}
Using this expression, we now compute the guiding-center polarization density \citep{Brizard_2008, Brizard_2013}, which is defined up to first order in magnetic-field nonuniformity as 
\begin{equation}
\vb{\pi}_{\rm gc}^{(1)} \;\equiv\; e\,\langle\vb{\rho}_{1}\rangle \;-\; e\;\nabla\bdot\left(\left\langle\frac{\vb{\rho}_{0}\vb{\rho}_{0}}{2}\right\rangle\right),
\label{eq:gc_pol_def}
\end{equation}
which requires the gyroangle-average of Eq.~\eqref{eq:rho1_def}:
\begin{eqnarray}
\langle\vb{\rho}_{1}\rangle & = & -\;\frac{J}{m\Omega} \left[ \frac{1}{2}\,(\nabla\bdot\bhat)\,\bhat \;+\; \frac{3}{2}\,\nabla_{\bot}\ln B\right] \;-\; \varrho_{\|}^{2}\,\vb{\kappa} \;+\; \vb{\Pi}_{1}\btimes\frac{\bhat}{m\Omega} \nonumber \\
 & = & -\,\frac{1}{m\Omega}\,\left( J\;\nabla_{\bot}\ln B \;+\; \frac{p_{\|}^{2}\,\vb{\kappa}}{m\,\Omega}\right) \;+\; \nabla\bdot\left[
\frac{J}{2\,m\Omega} \left({\bf I} \,-\, \bhat\bhat\right)\right] \nonumber \\
 &  &+\; \left( \frac{J}{2} \;\bhat\btimes\vb{\kappa} \;+\; \vb{\Pi}_{1}\right)\btimes\frac{\bhat}{m\Omega}.
\label{eq:rho1_ave}
\end{eqnarray}
Hence, using $\frac{1}{2}\,\langle\vb{\rho}_{0}\vb{\rho}_{0}\rangle = ({\bf I} - \bhat\bhat)\,J/2m\Omega$, the guiding-center polarization density \eqref{eq:gc_pol_def} becomes
\begin{equation}
\vb{\pi}_{\rm gc}^{(1)} \;=\; -\,\frac{e}{m\Omega}\,\left( J\;\nabla_{\bot}\ln B \;+\; \frac{p_{\|}^{2}\,\vb{\kappa}}{m\,\Omega}\right) \;+\; \left( \frac{J}{2} \;\bhat\btimes\vb{\kappa} \;+\; \vb{\Pi}_{1}\right)\btimes\frac{\bhat}{m\Omega},
\label{eq:gc_pol}
\end{equation}
which yields the Pfirsch-Kaufman formula \citep{Pfirsch_1984, Kaufman_1986}
\begin{equation}
\vb{\pi}_{\rm gc}^{(1)} \;\equiv\; e\;\bhat\btimes\frac{{\bf v}_{\rm gc}}{\Omega},
\label{eq:gc_pol_PK}
\end{equation}
only if we use the following polarization constraint in Eq.~\eqref{eq:gc_pol_def}:
\begin{equation}
\vb{\Pi}_{1\bot} \;\equiv\; -\;\frac{J}{2}\;\bhat\btimes\vb{\kappa}.
\label{eq:Pi1_perp_choice}
\end{equation}
By combining this result with the first-order symplectic representation \eqref{eq:Pi1||_def}: $\Pi_{1\|} \equiv \bhat\bdot\vb{\Pi}_{1} = -\,\frac{1}{2}\,J\,\tau$, we, therefore, find the first-order symplectic guiding-center momentum
\begin{equation}
\vb{\Pi}_{1} \;=\; -\;\frac{J}{2} \left( \tau\;\bhat \;+\frac{}{} \bhat\btimes\vb{\kappa} \right) \;=\; -\;\frac{J}{2}\;\nabla\btimes\bhat,
\label{eq:Pi1_final}
\end{equation}
and Eq.~\eqref{eq:rho1_def} becomes
\begin{equation}
\vb{\rho}_{1} \;\equiv\; \bhat\btimes\frac{{\bf v}_{\rm gc}}{\Omega} \;+\; \nabla\bdot\left[ \frac{J}{2\,m\Omega}\;\left({\bf I} - \bhat\,\bhat\right) \right] \;+\; \wt{\vb{\rho}}_{1},
\label{eq:rho1}
\end{equation}
with the gyroangle-dependent part $\wt{\vb{\rho}}_{1} \equiv \vb{\rho}_{1} - \langle\vb{\rho}_{1}\rangle$ is
\begin{eqnarray}
\wt{\vb{\rho}}_{1} & = & -\;\varrho_{\|} \left[ 2\left(\vb{\kappa}\bdot\pd{\vb{\rho}_{0}}{\theta}\right)\,\bhat \;+\; \frac{1}{2}\,(\tau - \alpha_{1})\;
\vb{\rho}_{0} \;+\; \alpha_{2}\;\pd{\vb{\rho}_{0}}{\theta} \right] \nonumber \\
 &  &+\; \frac{J}{m\Omega}\;\left(\alpha_{2}\bhat \;-\frac{}{} 2\,{\sf a}_{2}\bdot\nabla\ln B \right).
\label{eq:rho1_tilde}
\end{eqnarray}
Since Eq.~\eqref{eq:Pi1_final} satisfies $\partial\vb{\Pi}_{1}/\partial p_{\|} \equiv 0$, then ${\sf b}^{*} = \bhat + {\cal O}(\epsilon^{2})$ according to Eq.~\eqref{eq:bstar_def}.

Lastly, the guiding-center phase-space Lagrangian is expressed as
\begin{equation}
\Gamma_{\rm gc} \;=\; \left( \frac{e}{\epsilon\,c}\;{\bf A} \;+\; p_{\|}\,\bhat \;-\; \frac{\epsilon}{2}\;J\,\nabla\btimes\bhat \right)\bdot\exd{\bf X} 
\;+\; \epsilon\,J\;\left(\exd\theta \;-\frac{}{} {\bf R}\bdot\exd{\bf X}\right),
\label{eq:Gamma_gc_primitive}
\end{equation}
when terms up to first order in magnetic-field nonuniformity are retained. In Eq.~\eqref{eq:Gamma_gc_primitive}, we have retained the guiding-center polarization contribution \eqref{eq:Pi1_perp_choice} in $\vb{\Pi}_{1} \equiv -\,\frac{1}{2}\,J\;\nabla\btimes\bhat$. We now show that this polarization correction enables us to obtain a simple expression for the guiding-center toroidal canonical momentum up to second order in $\epsilon$ (i.e., first order in magnetic-field nonuniformity).

\subsection{\label{subsec:gc_canonical}Guiding-center Toroidal Canonical Momentum}

There is now a well-established connection between polarization and the conservation of toroidal canonical momentum in an axisymmetric magnetic field \citep{Scott_Smirnov_2010, Brizard_Tronko_2011}, here represented as
\begin{equation}
{\bf B} \;=\; B_{\varphi}(\psi)\;\nabla\varphi \;+\; \nabla\varphi\btimes\nabla\psi,
\label{eq:B_axis}
\end{equation}
where $\varphi$ denotes the toroidal angle and $\psi$ denotes the magnetic flux on which magnetic-field lines lie (i.e., ${\bf B}\bdot\nabla\psi 
\equiv 0$). Note that the toroidal magnetic field $B_{\varphi}\,\nabla\varphi$ in Eq.~\eqref{eq:B_axis} is expressed with a covariant component 
$B_{\varphi}(\psi)$ that is constant on a given magnetic-flux surface.

We first calculate the guiding-center toroidal canonical momentum from the guiding-center phase-space Lagrangian \eqref{eq:Gamma_gc_primitive}:
\begin{eqnarray}
P_{{\rm gc}\varphi} & \equiv & \left[ \frac{e}{\epsilon\,c}\;{\bf A} \;+\; p_{\|}\,\bhat \;-\; \epsilon\;J\,\left({\bf R} \;+\; \frac{1}{2}\,\nabla\btimes\bhat \right) \right]\bdot\pd{\bf X}{\varphi} \label{eq:Pgc_phi_def} \\
 & = & -\;\frac{e}{\epsilon\,c}\;\psi \;+\; p_{\|}\,b_{\varphi} \;-\; \epsilon\,J \left[ b_{\sf z} \;+\; \nabla\bdot\left(\bhat\btimes \frac{1}{2}\,
{\cal R}^{2}\,\nabla\varphi\right) \;+\; \bhat\bdot\nabla\btimes\left( \frac{1}{2}\,{\cal R}^{2}\,\nabla\varphi\right) \right],
\nonumber
\end{eqnarray}
where $b_{\varphi} \equiv B_{\varphi}/B$ denotes the toroidal covariant component of the magnetic unit vector $\bhat$, we used \citep{RGL_1983}
\[ {\bf R}\bdot\pd{\bf X}{\varphi} \;=\; \pd{\wh{\sf 1}}{\varphi}\bdot\wh{\sf 2} \;=\; (\wh{\sf z}\btimes\wh{\sf 1})\bdot\wh{\sf 2} \;\equiv\; b_{\sf z} \]
(i.e., the component of $\bhat$ along the symmetry axis $\wh{\sf z}$ for toroidal rotations), we wrote $\partial{\bf X}/\partial\varphi \equiv {\cal R}^{2}\,\nabla\varphi$ in terms of the major radius ${\cal R} \equiv 
|\nabla\varphi|^{-1}$, and we used the identity ${\bf F}\bdot\nabla\btimes{\bf G} \equiv \nabla\bdot({\bf G}\btimes{\bf F}) + {\bf G}\bdot\nabla\btimes
{\bf F}$, for any two vector fields ${\bf F}$ and ${\bf G}$. Next, we use
\[ \bhat\bdot\nabla\btimes\left( \frac{1}{2}\,{\cal R}^{2}\,\nabla\varphi\right) \;=\; \bhat\bdot\left(\nabla{\cal R}\btimes{\cal R}\nabla\varphi\right)
\;=\; \bhat\bdot\left(\wh{\cal R}\btimes\wh{\varphi}\right) \;=\; b_{\sf z}, \]
and
\[ \bhat\btimes \frac{1}{2}\,{\cal R}^{2}\,\nabla\varphi \;=\; \frac{1}{2B}\;(B_{\varphi}\,\nabla\varphi + \nabla\varphi\btimes\nabla\psi)\btimes
\pd{\bf X}{\varphi} \;=\; \frac{1}{2B}\;\nabla\psi, \]
so that Eq.~\eqref{eq:Pgc_phi_def} becomes
\begin{equation}
P_{{\rm gc}\varphi} \;=\; -\;\frac{e}{\epsilon\,c}\;\psi \;+\; p_{\|}\,b_{\varphi} \;-\; \epsilon\,\left[ 2\,J\;b_{\sf z} \;+\; \nabla\bdot\left(
\frac{J}{2\,m\Omega}\;\frac{e}{c}\,\nabla\psi\right) \right].
\label{eq:Pgc_phi}
\end{equation}
Here, we suspect that the last term in Eq.~\eqref{eq:Pgc_phi} is related to the second-order finite-Larmor-radius (FLR) correction to the first term.

To prove this assertion, we introduce the guiding-center magnetic flux
\begin{eqnarray}
\psi_{\rm gc} & \equiv & \left\langle {\sf T}_{\rm gc}^{-1}\psi\right\rangle \;=\; \left\langle \psi + \epsilon\vb{\rho}_{0}\bdot\nabla\psi -
\epsilon^{2} \left[ G_{2}^{\bf x}\bdot\nabla\psi + \frac{1}{2}\,{\sf G}_{1}\cdot\exd\left(\vb{\rho}_{0}\bdot\nabla\psi\right) \right] + \cdots \right\rangle 
\nonumber \\
 & = & \psi \;+\; \epsilon^{2} \left( \langle\vb{\rho}_{1}\rangle\bdot\nabla\psi \;+\; \frac{1}{2}\,\langle\vb{\rho}_{0}\vb{\rho}_{0}\rangle:\nabla\nabla
\psi \right) + \cdots,
\end{eqnarray} 
where we used the definition \eqref{eq:rho_1} for $\vb{\rho}_{1}$. Next, using ${\bf B}\bdot\nabla\psi \equiv 0$, we obtain
\begin{eqnarray}
\psi_{\rm gc} & = & \psi + \epsilon^{2}\left\{\bhat\btimes\frac{{\bf v}_{\rm gc}}{\Omega}\bdot\nabla\psi + \nabla\bdot\left[\frac{J}{2\,m\Omega}
({\bf I} - \bhat\bhat)\right]\bdot\nabla\psi + \frac{J}{2\,m\Omega}({\bf I} - \bhat\bhat):\nabla\nabla\psi \right\} \nonumber \\
 & = & \psi \;+\; \epsilon^{2} \left[ \nabla\bdot\left( \frac{J}{2\,m\Omega}\;\nabla\psi\right) \;+\ \bhat\btimes\frac{{\bf v}_{\rm gc}}{\Omega}\bdot\nabla\psi \right].
\label{eq:psi_gc}
\end{eqnarray}
Using the identity $\nabla\psi \equiv {\bf B}\btimes\partial{\bf X}/\partial\varphi$, with $\bhat\bdot{\bf v}_{\rm gc} \equiv 0$, we now obtain
\[ \bhat\btimes\frac{{\bf v}_{\rm gc}}{\Omega}\bdot\nabla\psi \;=\; \bhat\btimes\frac{{\bf v}_{\rm gc}}{\Omega}\bdot\left({\bf B}\btimes
\pd{\bf X}{\varphi}\right) \;=\; \frac{B}{\Omega}\;\left({\bf v}_{\rm gc}\bdot\pd{\bf X}{\varphi}\right) \;\equiv\; 
\frac{B}{\Omega}\;v_{{\rm gc}\varphi}. \]
Hence, the final expression for the guiding-center toroidal canonical momentum defined by Eq.~\eqref{eq:Pgc_phi} is
\begin{equation}
P_{{\rm gc}\varphi} \;=\; -\;\frac{e}{\epsilon\,c}\;\psi_{\rm gc} \;+\; m \left( \frac{d_{0}{\bf X}}{dt} \;+\; \epsilon\;\frac{d_{1}{\bf X}}{dt}
\right)\bdot\pd{\bf X}{\varphi} \;-\; 2\,\epsilon\;J\,b_{\sf z},
\label{eq:Pgc_phi_final}
\end{equation}
where $d_{0}{\bf X}/dt \equiv (p_{\|}/m)\,\bhat$ and $d_{1}{\bf X}/dt$ given by Eq.~\eqref{eq:dx1_dt}, while
\[ m\;\left( \frac{d_{0}{\bf X}}{dt} \;+\; \epsilon\;\frac{d_{1}{\bf X}}{dt}\right)\bdot\pd{\bf X}{\varphi} \;\equiv\; m\;{\cal R}^{2}\;
\frac{d_{\rm gc}\varphi}{dt} \]
denotes the guiding-center toroidal momentum with first-order corrections due to the guiding-center magnetic-drift velocity. 

The last term in Eq.~\eqref{eq:Pgc_phi_final} might be puzzling until we consider the guiding-center transformation of the particle toroidal canonical momentum
\begin{eqnarray}
P_{{\rm gc}\varphi} & \equiv & \left\langle{\sf T}_{\rm gc}^{-1}\;P_{\varphi}\right\rangle \;=\; \left\langle {\sf T}_{\rm gc}^{-1}\left( -\,
\frac{e}{\epsilon\,c}\;\psi \;+\; m\;{\bf v}\bdot\pd{\bf x}{\varphi}\right)\right\rangle \label{eq:P_particle_phi} \\
 & = & -\,\frac{e}{\epsilon\,c}\;\langle{\sf T}_{\rm gc}^{-1}\psi\rangle \;+\; m\;\left\langle \left({\sf T}_{\rm gc}^{-1}\frac{d{\bf x}}{dt}\right)\bdot\left({\sf T}_{\rm gc}^{-1}\pd{\bf x}{\varphi}\right)\right\rangle \nonumber \\
 & = & -\,\frac{e}{\epsilon\,c}\;\psi_{\rm gc} \;+\; m\;\left\langle \left(\frac{d_{\rm gc}{\bf X}}{dt} + \frac{d_{\rm gc}\vb{\rho}_{\rm gc}}{dt}\right)\bdot\left(\frac{\partial_{\rm gc}{\bf X}}{\partial\varphi} + \frac{\partial_{\rm gc}\vb{\rho}_{\rm gc}}{\partial\varphi}\right)\right\rangle \nonumber \\
 & = & -\;\frac{e}{\epsilon\,c}\;\psi_{\rm gc} \;+\; m \left( \frac{d_{0}{\bf X}}{dt} \;+\; \epsilon\;\frac{d_{1}{\bf X}}{dt}
\right)\bdot\pd{\bf X}{\varphi} \;+\; \epsilon\;m\Omega\left\langle\pd{\vb{\rho}_{0}}{\theta}\bdot\pd{\vb{\rho}_{0}}{\varphi}\right\rangle + \cdots.
\nonumber
\end{eqnarray}
Since $\partial\vb{\rho}_{0}/\partial\varphi \equiv \wh{\sf z}\btimes\vb{\rho}_{0}$ in axisymmetric magnetic geometry, the last term becomes
\[ \epsilon\;m\Omega\left\langle\pd{\vb{\rho}_{0}}{\theta}\bdot\pd{\vb{\rho}_{0}}{\varphi}\right\rangle \;=\; \epsilon\;m\Omega\left\langle
\pd{\vb{\rho}_{0}}{\theta}\bdot\left(\wh{\sf z}\btimes\vb{\rho}_{0}\right)\right\rangle \;=\; -\; 2\,\epsilon\;J\,b_{\sf z}, \]
and we recover the guiding-center toroidal canonical momentum \eqref{eq:Pgc_phi_final} from the guiding-center transformation of the particle toroidal canonical momentum \eqref{eq:P_particle_phi}.

Lastly, we note that the guiding-center toroidal canonical momentum $P_{{\rm gc}\varphi}$ is defined as the guiding-center push-forward of the particle toroidal canonical momentum $P_{\varphi}$:
\begin{equation}
P_{{\rm gc}\varphi} \;=\; {\sf T}_{\rm gc}^{-1}P_{\varphi} \;=\; -\,\frac{e}{\epsilon c}\;{\sf T}^{-1}_{\rm gc}\psi \;+\; m \left( 
\frac{d_{\rm gc}{\bf X}}{dt} \;+\; \frac{d_{\rm gc}\vb{\rho}_{\rm gc}}{dt}\right)\bdot\left(\frac{\partial_{\rm gc}{\bf X}}{\partial\varphi} \;+\;
\frac{\partial_{\rm gc}\vb{\rho}_{\rm gc}}{\partial\varphi}\right),
\label{eq:Pgcphi_push}
\end{equation}
which guarantees the invariance of the guiding-center toroidal canonical momentum $P_{{\rm gc}\varphi}$:
\begin{equation}
\frac{d_{\rm gc}P_{{\rm gc}\varphi}}{dt} \;\equiv\; {\sf T}_{\rm gc}^{-1}\left(\frac{dP_{\varphi}}{dt}\right) \;=\; 0.
\end{equation}
We have shown in Eq.~\eqref{eq:P_particle_phi}, however, that $P_{{\rm gc}\varphi} \equiv \langle{\sf T}_{\rm gc}^{-1}P_{\varphi}\rangle$, since $P_{{\rm gc}\varphi}$ is defined as the toroidal component of the gyroangle-independent guiding-center symplectic Lagrange one-form \eqref{eq:Gamma_gc_primitive}. Hence, the gyroangle-dependent terms in ${\sf T}_{\rm gc}^{-1}P_{\varphi} - \langle{\sf T}_{\rm gc}^{-1}P_{\varphi}\rangle$ must vanish identically, which is proved up to second order in $\epsilon$ (first order in $\epsilon_{B}$) in App.~\ref{sec:Pgcphi_gyro}.

\subsection{Variational derivation of the guiding-center polarization}

We return to the guiding-center polarization \eqref{eq:gc_pol_PK} and present two alternative variational derivations based on guiding-center Lagrangian and guiding-center Hamiltonian, respectively.  First, in recent work by \cite{Brizard_Tronci_2016}, the following guiding-center magnetization is derived by variational method:
 \begin{equation}
 {\bf M}_{\rm gc} \;=\; \sum \int \pd{L_{\rm gc}}{\bf B}\;F\,dp_{\|}\,d\mu \;\equiv\; \sum \int \left( \vb{\mu}_{\rm gc} \;+\; \vb{\pi}_{\rm gc}\btimes\frac{p_{\|}\,\bhat}{mc}\right) \;F\,dp_{\|}\,d\mu
 \label{eq:Mgc_def}
 \end{equation}
 where $\sum$ denotes a summation over particle species and $F \equiv 2]pi,{\cal J}_{\rm gc}\;f$ denotes the guiding-center Vlasov phase-space density (which includes the guiding-center Jacobian ${\cal J}_{\rm gc}$ and the factor $2\pi$ replaces the integration over the gyroangle). Here, the lowest-order guiding-center (single-particle) Lagrangian is
 \begin{equation}
 L_{\rm gc}({\bf X},p_{\|},\dot{\bf X},\dot{p}_{\|}; {\bf B}) \;=\; \left( \frac{e}{c}\,{\bf A} \;+\; p_{\|}\,\bhat \right)\bdot\dot{\bf X} \;-\; \left( \frac{p_{\|}^{2}}{2m} \;+\; \mu\,B \right),
 \label{eq:Lgc_mag}
 \end{equation}
while the guiding-center magnetization \eqref{eq:Mgc_def} is expressed in terms of the intrinsic magnetic-dipole moment $\vb{\mu}_{\rm gc}$ and the moving electric-dipole moment $\vb{\pi}_{\rm gc}\btimes(p_{\|}\,
\bhat/mc)$. Here, the derivative of the 
guiding-center Lagrangian \eqref{eq:Lgc_mag} with respect to ${\bf B}$ (at constant ${\bf X}$, $p_{\|}$, and $\dot{\bf X}$) yields
\begin{equation}
\pd{L_{\rm gc}}{\bf B} \;=\; p_{\|}\;\pd{\bhat}{\bf B}\bdot\dot{\bf X} \;-\; \mu\;\pd{B}{\bf B} \;=\; \left(\frac{e\,\bhat}{\Omega}\btimes\dot{\bf X}\right)\btimes\frac{p_{\|}\,\bhat}{mc} \;-\; \mu\,\bhat,
\end{equation}
which yields the guiding-center magnetization contributions
\begin{equation}
\left. \begin{array}{rcl}
\vb{\mu}_{\rm gc} & = & -\;\mu\,\bhat \\
 &  & \\
 \vb{\pi}_{\rm gc} & = & e\,\bhat\btimes\dot{\bf X}/\Omega
 \end{array} \right\},
 \label{eq:mu_pi_gc}
 \end{equation}
where $\dot{\bf X}$ now needs to be determined from the guiding-center Euler-Lagrange equations. The guiding-center Euler-Lagrange equations derived from Eq.~\eqref{eq:Lgc_mag} are
 \begin{equation}
 \left. \begin{array}{rcl}
 0 & = & \partial L_{\rm gc}/\partial{\bf X} - d(\partial L_{\rm gc}/\partial\dot{\bf X})/dt \;=\; (e/c)\,\dot{\bf X}\btimes{\bf B}^{*} \;-\; \mu\;\nabla B \;-\; \dot{p}_{\|}\;\bhat \\
  &  & \\
 0 & = &  \partial L_{\rm gc}/\partial p_{\|} \;=\; \bhat\bdot\dot{\bf X} \;-\; p_{\|}/m
 \end{array} \right\},
 \end{equation}
 which are the lowest-order versions of Eqs.~\eqref{eq:ELgc_X}-\eqref{eq:ELgc_p}. Hence, we conclude that $\dot{\bf X} = \bhat\;p_{\|}/m + {\bf v}_{\rm gc}$, so that $\bhat\btimes\dot{\bf X} = \bhat\btimes
 {\bf v}_{\rm gc}$, and  Eq.~\eqref{eq:mu_pi_gc} yields the guiding-center polarization \eqref{eq:gc_pol_PK}.
 
A second variational derivation of the guiding-center polarization \eqref{eq:gc_pol_PK} can be more directly derived from the guiding-center Hamiltonian term $H_{\rm gc}^{\Phi} \equiv e\,\langle
{\sf T}_{\rm gc}^{-1}\Phi\rangle$ expressed in terms of the scalar potential $\Phi$. First, we expand $H_{\rm gc}^{\Phi} $ up to first order in magnetic-field nonuniformity:
\begin{eqnarray}
H_{\rm gc}^{\Phi}  & = & e\,\Phi \;+\; e\,\langle\vb{\rho}_{1}\rangle\bdot\nabla\Phi \;+\; e\;\left\langle\frac{\vb{\rho}_{0}\vb{\rho}_{0}}{2}\right\rangle:\nabla\nabla\Phi \nonumber \\
  & \equiv & e\,\Phi \;-\; e\,\langle\vb{\rho}_{1}\rangle\bdot{\bf E} \;-\; e\;\left\langle\frac{\vb{\rho}_{0}\vb{\rho}_{0}}{2}\right\rangle:\nabla{\bf E},
  \label{eq:Phi_gc}
 \end{eqnarray}
 where we substituted ${\bf E} = -\,\nabla\Phi$ into the last expression and we have neglected higher-order quadrupole contributions ($\langle\vb{\rho}_{0}\vb{\rho}_{1} + \vb{\rho}_{1}\vb{\rho}_{0}\rangle:
 \nabla\nabla\Phi$). Next, we use the traditional variational definition of the guiding-center polarization \citep{Brizard_2013}
 \begin{equation}
 \vb{\pi}_{\rm gc} \;\equiv\; -\;\pd{H_{\rm gc}^{\Phi}}{\bf E} \;+\; \nabla\bdot\left(\pd{H_{\rm gc}^{\Phi}}{(\nabla{\bf E})}\right),
 \label{eq:gcpol_var}
 \end{equation}
 which includes electric dipole and quadrupole contributions. By substituting Eq.~\eqref{eq:Phi_gc} into Eq.~\eqref{eq:gcpol_var}, we recover the Pfirsch-Kaufman formula
 \eqref{eq:gc_pol_PK}:
 \begin{equation}
 \vb{\pi}_{\rm gc}^{(1)} \;=\; e\,\langle\vb{\rho}_{1}\rangle \;-\; \nabla\bdot\left( e\;\left\langle\frac{\vb{\rho}_{0}\vb{\rho}_{0}}{2}\right\rangle\right) \;=\; e\,\bhat\btimes\frac{{\bf v}_{\rm gc}}{\Omega},
 \end{equation}
 which confirms the Lie-transform expression \eqref{eq:gc_pol_def} and is consistent with Eq.~\eqref{eq:mu_pi_gc}.

\section{\label{sec:sum}Summary}

A systematic derivation of the higher-order Hamiltonian guiding-center dynamics has been derived by Lie-transform perturbation analysis. The guiding-center Poisson bracket derived from the guiding-center phase-space Lagrangian \eqref{eq:Gamma_gc_primitive} is
\begin{eqnarray}
\left\{ F,\frac{}{} G\right\}_{\rm gc} & = & \epsilon^{-1} \left( \pd{F}{\theta}\,\pd{G}{J} \;-\; \pd{F}{J}\,\pd{G}{\theta} \right) \;+\;
\frac{{\bf B}^{*}}{B_{\|}^{*}}\bdot\left(\nabla^{*}F\;\pd{G}{p_{\|}} \;-\; \pd{F}{p_{\|}}\;\nabla^{*}G \right) \nonumber \\
 &  &-\; \frac{\epsilon\,c\bhat}{e\,B_{\|}^{*}}\bdot\nabla^{*}F\btimes\nabla^{*}G,
\label{eq:PB_gc_final}
\end{eqnarray}
where 
\begin{eqnarray}
{\bf B}^{*} & \equiv & \nabla\btimes\left( {\bf A} \;+\; \epsilon\,\frac{c}{e}\,p_{\|}\,\bhat \;-\frac{}{}
\epsilon^{2}\;\frac{c}{e}\,J\;{\bf R}^{*} \right), \label{eq:Bstar_final} \\
B_{\|}^{*} & \equiv & \bhat\bdot{\bf B}^{*} \;=\; B \left( 1 \;+\; \epsilon\,\varrho_{\|}\,\tau \;-\; \epsilon^{2}\;\frac{J}{m\Omega}\,\bhat\bdot
\nabla\btimes{\bf R}^{*} \right), \label{eq:B||star_final}
\end{eqnarray}
with $\nabla^{*} \equiv \nabla + {\bf R}^{*}\,\partial/\partial\theta$ and ${\bf R}^{*} \equiv {\bf R} + \frac{1}{2}\;\nabla\btimes\bhat$.

The guiding-center Hamiltonian, on the other hand, can be chosen as (with $\Pi_{2\|} \equiv 0$)
\begin{equation}
H_{\rm gc} \;=\; \frac{p_{\|}^{2}}{2m} \;+\; J\,\Omega \;+\; \epsilon^{2}\;\Psi_{2},
\label{eq:Ham_gc_final}
\end{equation}
where the second-order guiding-center Hamiltonian is expressed as
\begin{equation}
\Psi_{2} \;=\; J\,\Omega\left( \frac{J}{2\,m\Omega}\;\beta_{2\bot} \;+\; \frac{1}{2}\,\varrho_{\|}^{2}\;\beta_{2\|} \right) \;-\; 
\frac{p_{\|}^{2}}{2m}\;\left(\varrho_{\|}^{2}\frac{}{}|\vb{\kappa}|^{2}\right) \;+\; \vb{\Pi}_{1}\bdot{\bf v}_{\rm gc}.
\label{eq:Hamiltonian_constraint_2_final}
\end{equation}
Here, we have isolated the contribution from the perpendicular polarization component $\vb{\Pi}_{1\bot}$ and the coefficients $\beta_{2\bot}$ and 
$\beta_{2\|}$ are defined in Eqs.~\eqref{eq:beta2_perp}-\eqref{eq:beta2_par}.

We also showed that the perpendicular component $\vb{\Pi}_{1\bot}$, which could not be determined within Lie-transform perturbation theory (up to the orders considered in this work), could not be chosen to be zero in contrast to the choice made by \cite{RGL_1983}, who used a Hamiltonian representation (with $\Psi_{1} = \frac{1}{2}\,J\,\Omega\;\varrho_{\|}\,\tau$ and $\Pi_{1\|} = 0$). We showed in Sec.~\ref{sec:gc_pol_canonical} that the choice $\vb{\Pi}_{1} = -\,\frac{1}{2}\,J\;\nabla\btimes\bhat$ not only yields the standard Pfirsch-Kaufman guiding-center polarization \eqref{eq:gc_pol_PK} but also a simplified guiding-center representation of the particle toroidal canonical momentum \eqref{eq:Pgc_phi_final}. 

Lastly, we have shown that the guiding-center Hamiltonian \eqref{eq:Ham_gc_final} can be expressed as
\begin{equation}
H_{\rm gc} \;\equiv\; \left\langle {\sf T}_{\rm gc}^{-1}\left( \frac{m}{2}\,\left|\frac{d{\bf x}}{dt}\right|^{2}\right) \right\rangle \;=\; \frac{m}{2} \left\langle \left|
\frac{d_{\rm gc}{\bf X}}{dt} + \frac{d_{\rm gc}\vb{\rho}_{\rm gc}}{dt}\right|^{2}\right\rangle,
\end{equation}
which follows from the guiding-center Lagrangian constraint \eqref{eq:gcLc_id}.

\acknowledgments

Work by AJB was partially supported by a U.~S.~DoE grant under contracts No.~DE-SC0006721 and No.~DE-SC0014032. This work has been carried out within the framework of the EUROfusion Consortium and has received funding from the Euratom research and training programme 2014-2018 under grant agreement No 633053. The views and opinions expressed herein do not necessarily reflect those of the European Commission.

\appendix

\section{\label{sec:dyadic}Dyadic Calculus}

In this Appendix, we present the basic expressions associated with the gradient and curl operations on the rotating unit-vectors $\wh{{\sf u}}^{i} \equiv 
(\wh{\bot}, \wh{\rho}, \bhat \equiv \wh{\bot}\btimes\wh{\rho})$, where we shall use the identities 
\begin{eqnarray*}
{\bf p}_{\bot}\btimes\vb{\rho}_{0} & = & 2\,J\;\bhat, \\
{\bf p}_{\bot}\bdot\pd{\vb{\rho}_{0}}{\theta} & = & 2\,J \;=\; -\;\pd{{\bf p}_{\bot}}{\theta}\bdot\vb{\rho}_{0}, \\
\pd{{\bf p}_{\bot}}{J}\bdot\pd{\vb{\rho}_{0}}{\theta} & = & 1 \;=\; -\;\pd{{\bf p}_{\bot}}{\theta}\bdot\pd{\vb{\rho}_{0}}{J}.
\end{eqnarray*}

\subsection{Gyrogauge invariance}

By introducing the fixed unit-vectors $(\wh{\sf 1}, \wh{\sf 2}, \bhat \equiv \wh{\sf 1}\btimes\wh{\sf 2})$, we write the definitions for the rotating unit-vectors
\begin{equation}
\left. \begin{array}{rcl}
\wh{\rho} & \equiv & \cos\theta\;\wh{\sf 1} \;-\; \sin\theta\;\wh{\sf 2} \\
 &  & \\
\wh{\bot} & \equiv & -\;\sin\theta\;\wh{\sf 1} \;-\; \cos\theta\;\wh{\sf 2}
\end{array} \right\},
\label{eq:rho_bot_def}
\end{equation}
where the gyroangle $\theta$ is measured {\it clockwise} from the $\wh{\sf 1}$-axis, so that $\wh{\bot} \equiv \partial\wh{\rho}/\partial\theta$. We note that, while the choice of the fixed unit-vectors $(\wh{\sf 1},\wh{\sf 2})$ can be made arbitrarily in the plane locally perpendicular to $\bhat$, we must ensure that the resulting guiding-center equations of motion do not depend on a specific choice. Hence, our guiding-center Hamiltonian theory must be {\it gyrogauge}-invariant in the following sense. 

First, we allow the rotation of the unit-vectors $(\wh{\sf 1},\wh{\sf 2})$ about the magnetic unit-vector $\bhat$ by an arbitrary angle $\chi({\bf x})$ that depends on the field position ${\bf x}$, so that 
\begin{equation}
\left( \begin{array}{c}
\wh{\sf 1}^{\prime} \\
\wh{\sf 2}^{\prime}
\end{array} \right) \;=\; \left( \begin{array}{cc}
\cos\chi & \sin\chi \\
-\sin\chi & \cos\chi
\end{array} \right) \cdot \left( \begin{array}{c}
\wh{\sf 1} \\
\wh{\sf 2}
\end{array} \right).
\label{eq:gyrogauge}
\end{equation}
Next, we require that the rotating unit-vectors \eqref{eq:rho_bot_def} be invariant under this rotation, i.e., $\wh{\rho}^{\prime} = \wh{\rho}$ and $\wh{\bot}^{\prime} = \wh{\bot}$, which implies that the gyroangle $\theta$ must transform as $\theta^{\prime}(\theta,{\bf x}) = \theta + \chi({\bf x})$ under the gyrogauge rotation \eqref{eq:gyrogauge}.

Lastly, we introduce the gyrogauge vector field 
\begin{equation}
{\bf R} \;\equiv\; \nabla\wh{\sf 1}\bdot\wh{\sf 2} \;=\; \nabla\wh{\bot}\bdot\wh{\rho} \;=\; -\;\nabla\wh{\rho}\bdot\wh{\bot}, 
\label{eq:gyrogauge_R_def}
\end{equation}
which transforms as ${\bf R}^{\prime} = {\bf R} + \nabla\chi$ under the gyrogauge rotation \eqref{eq:gyrogauge}. We, therefore, readily see that a gyrogauge-invariant guiding-center theory can only include the gyrogauge vector field ${\bf R}$ either as the curl $\nabla\btimes{\bf R}$, e.g., in Eq.~\eqref{eq:Bstar_final}, the one-form 
$\exd\theta - {\bf R}\bdot\exd{\bf x}$, or the gradient operator $\nabla + {\bf R}\;\partial/\partial\theta$, e.g., in Eq.~\eqref{eq:PB_gc_final}, which are all 
gyrogauge invariant.

\subsection{Curl of rotating unit-vector basis}

In writing the expressions for $(\nabla\btimes\bhat, \nabla\btimes{\bf p}_{\bot}, \nabla\btimes\vb{\rho}_{0})$, we use the definitions
\begin{eqnarray}
\nabla\btimes\bhat & \equiv & \tau\;\bhat \;+\; \bhat\btimes\vb{\kappa}, \label{eq:curl_b} \\
\nabla\btimes\wh{\bot} & \equiv & -\;\wh{\rho}\btimes{\bf R} \;-\; C_{\rho\bot}\;\wh{\bot} \;+\; C_{\bot\bot}\;\wh{\rho}, \label{eq:curl_bot} \\
\nabla\btimes\wh{\rho} & \equiv & \wh{\bot}\btimes{\bf R} \;-\; C_{\rho\rho}\;\wh{\bot} \;+\; C_{\bot\rho}\;\wh{\rho}, \label{eq:curl_rho}
\end{eqnarray}
where the matrix elements (with the definitions $\alpha_{n} \equiv {\sf a}_{n}:\nabla\bhat$)
\begin{equation}
\left. \begin{array}{rcl}
C_{\bot\bot} & \equiv & \wh{\bot}\bdot\nabla\bhat\bdot\wh{\bot} \;=\; \frac{1}{2}\,\nabla\bdot\bhat \;+\; 2\,\alpha_{2} \\
 && \\
C_{\bot\rho} & \equiv & \wh{\bot}\bdot\nabla\bhat\bdot\wh{\rho} \;=\; \frac{1}{2}\,\tau \;-\; \alpha_{1} \\
 && \\
C_{\rho\bot} & \equiv & \wh{\rho}\bdot\nabla\bhat\bdot\wh{\bot} \;=\; -\;\frac{1}{2}\,\tau \;-\; \alpha_{1} \\
 && \\
C_{\rho\rho} & \equiv & \wh{\rho}\bdot\nabla\bhat\bdot\wh{\rho} \;=\; \frac{1}{2}\,\nabla\bdot\bhat \;-\; 2\,\alpha_{2}
\end{array} \right\},
\label{eq:Cij_def} 
\end{equation}
are  expressed in terms of the dyadic tensors
\begin{eqnarray}
{\sf a}_{1} & \equiv & -\;\frac{1}{2}\;\left(\wh{\bot}\,\wh{\rho} + \wh{\rho}\,\wh{\bot}\right) \;=\; \pd{{\sf a}_{2}}{\theta} \nonumber \\
 & = & \frac{1}{2} \left[ \sin(2\theta)\;\left(\wh{\sf 1}\,\wh{\sf 1} - \wh{\sf 2}\,\wh{\sf 2}\right) \;+\frac{}{} \cos(2\theta)\; \left(\wh{\sf 1}\,\wh{\sf 2} + 
 \wh{\sf 2}\,\wh{\sf 1}\right) \right], \label{eq:a1_def} \\
{\sf a}_{2} & \equiv & \frac{1}{4}\;\left(\wh{\bot}\,\wh{\bot} - \wh{\rho}\,\wh{\rho}\right) = -\,\frac{1}{4}\pd{{\sf a}_{1}}{\theta} \nonumber \\
 & = & \frac{1}{4} \left[ -\,\cos(2\theta)\;\left(\wh{\sf 1}\,\wh{\sf 1} - \wh{\sf 2}\,\wh{\sf 2}\right) \;+\frac{}{} \sin(2\theta)\; \left(\wh{\sf 1}\,\wh{\sf 2} + \wh{\sf 2}\,\wh{\sf 1}\right) \right],
\label{eq:a2_def}
\end{eqnarray}
so that $\partial\alpha_{2}/\partial\theta \equiv \alpha_{1}$ and $\partial\alpha_{1}/\partial\theta \equiv -\,4\;\alpha_{2}$.

Hence, we find
\begin{eqnarray}
\nabla\btimes\vb{\rho}_{0} & = & \frac{1}{2}\,\vb{\rho}_{0}\btimes\nabla\ln B \;+\; \pd{\vb{\rho}_{0}}{\theta}\btimes{\bf R} \;+\; C_{\bot\rho}\;\vb{\rho}_{0} \;-\;
C_{\rho\rho}\;\pd{\vb{\rho}_{0}}{\theta}, \label{eq:curl_rho0} \\
\nabla\btimes{\bf p}_{\bot} & = & \frac{1}{2}\,\nabla\ln B\btimes{\bf p}_{\bot} \;-\; {\bf R}\btimes\pd{{\bf p}_{\bot}}{\theta} \;-\; \left( 
C_{\rho\bot}\;{\bf p}_{\bot} \;+\; C_{\bot\bot}\;\pd{{\bf p}_{\bot}}{\theta} \right), \label{eq:curlp_perp}
\end{eqnarray}
with
\begin{eqnarray*}
\pd{\vb{\rho}_{0}}{\theta}\bdot\nabla\btimes{\bf p}_{\bot} & = & -\;J \left[ 2\,\bhat\bdot{\bf R} \;-\frac{}{} (\tau + 2\,\alpha_{1})\right], \\
\pd{\vb{\rho}_{0}}{\theta}\bdot\nabla\btimes\pd{{\bf p}_{\bot}}{J} & = & -\;\bhat\bdot{\bf R} \;+\; \left(\alpha_{1} \;+\; \frac{\tau}{2} \right), \\
\pd{\vb{\rho}_{0}}{\theta}\bdot\nabla\btimes\pd{{\bf p}_{\bot}}{\theta} & = & -\;4\,J\;\alpha_{2}.
\end{eqnarray*}

\subsection{Dyadic gradients}

We also make use of the matrix elements \eqref{eq:Cij_def} to write the components of the dyadic gradients
\begin{equation}
\left. \begin{array}{rcl}
\nabla\bhat & = & \bhat\;\vb{\kappa} \;+\; \left( C_{\rho\rho}\,\wh{\rho}\wh{\rho} \;+\; C_{\rho\bot}\,\wh{\rho}\wh{\bot} \;+\;
C_{\bot\rho}\,\wh{\bot}\wh{\rho} \;+\; C_{\bot\bot}\,\wh{\bot}\wh{\bot} \right) \\
 && \\
\nabla\wh{\bot} & = & {\bf R}\;\wh{\rho} \;-\; \left( \vb{\kappa}\bdot\wh{\bot}\right)\;\bhat\bhat \;-\; \left( C_{\bot\bot}\,\wh{\bot} \;+\; 
C_{\rho\bot}\,\wh{\rho} \right)\;\bhat \\
 && \\
\nabla\wh{\rho} & = & -\;{\bf R}\;\wh{\bot} \;-\; \left( \vb{\kappa}\bdot\wh{\rho}\right)\;\bhat\bhat \;-\; \left( C_{\bot\rho}\,\wh{\bot} \;+\; 
C_{\rho\rho}\,\wh{\rho} \right)\;\bhat 
\end{array} \right\},
\label{eq:grad_i}
\end{equation}
from which we obtain the divergence identities
\begin{equation}
\left. \begin{array}{rcl}
\nabla\bdot\bhat & = & C_{\rho\rho} \;+\; C_{\bot\bot} \\
 && \\
\nabla\bdot\wh{\bot} & = & {\bf R}\bdot\wh{\rho} \;-\; \vb{\kappa}\bdot\wh{\bot} \\
 && \\
\nabla\bdot\wh{\rho} & = & -\;{\bf R}\bdot\wh{\bot} \;-\; \vb{\kappa}\bdot\wh{\rho}
\end{array} \right\},
\label{eq:div_i}
\end{equation}
and the useful expressions
\begin{eqnarray}
\nabla\vb{\rho}_{0} & = & -\;\frac{1}{2}\,\nabla\ln B\;\vb{\rho}_{0} - {\bf R}\;\pd{\vb{\rho}_{0}}{\theta} - \left[ (\vb{\rho}_{0}\bdot
\vb{\kappa})\;\bhat + C_{\bot\rho}\;\pd{\vb{\rho}_{0}}{\theta} + C_{\rho\rho}\;\vb{\rho}_{0} \right]\;\bhat,  \label{eq:gradrho_0} \\
\nabla\bdot\vb{\rho}_{0} & = & -\;\vb{\rho}_{0}\bdot\left(\frac{1}{2}\,\nabla\ln B \;+\; \vb{\kappa} \;+\; \bhat\btimes{\bf R}\right), \label{eq:divrho_0} 
\end{eqnarray}
with
\[ \langle\vb{\rho}_{0}\bdot\nabla\vb{\rho}_{0}\rangle \;=\; \frac{J}{m\Omega} \left[ \bhat\btimes{\bf R} \;-\; (\nabla\bdot\bhat)\;\bhat \;-\; \frac{1}{2}\;
\nabla_{\bot}\ln B \right]. \]
We will also use the following expressions
\begin{eqnarray}
\frac{d_{0}\vb{\rho}_{0}}{dt} & = & \frac{p_{\|}}{m}\,\bhat\bdot\left[ \nabla\vb{\rho}_{0} \;+\; \left({\bf R} + \frac{1}{2}\,\nabla\btimes\bhat\right)
\pd{\vb{\rho}_{0}}{\theta} \right] \nonumber \\
 & = & \frac{p_{\|}}{m} \left[ \frac{1}{2}\,(\nabla\bdot\bhat)\;\vb{\rho}_{0} \;+\; \frac{1}{2}\,\tau\;
\pd{\vb{\rho}_{0}}{\theta} \;-\; (\vb{\rho}_{0}\bdot\vb{\kappa})\;\bhat \right], \label{eq:d0_rho_0} \\
\frac{d_{0}{\bf p}_{\bot}}{dt} & = & \frac{p_{\|}}{m}\,\bhat\bdot\left[ \nabla{\bf p}_{\bot} \;+\; \left({\bf R} + \frac{1}{2}\,\nabla\btimes\bhat\right)
\pd{{\bf p}_{\bot}}{\theta} \right] \nonumber \\
 & = & -\;\frac{p_{\|}}{m} \left[ \frac{1}{2}\,(\nabla\bdot\bhat)\;{\bf p}_{\bot} \;-\; \frac{1}{2}\,\tau\;
\pd{{\bf p}_{\bot}}{\theta} \;+\; ({\bf p}_{\bot}\bdot\vb{\kappa})\;\bhat \right]. \label{eq:d0_p_bot}
\end{eqnarray}

\subsection{Dyadic identities}

We conclude this Appendix by presenting the dyadic identity derived from Eq.~\eqref{eq:grad_i}:
\begin{eqnarray}
\nabla\bhat\;:\;\nabla\bhat & = & (C_{\rho\rho})^{2} \;+\; (C_{\bot\bot})^{2} \;+\; 2\;C_{\bot\rho}\;C_{\rho\bot} \nonumber \\
 & = & \frac{1}{2}\left[ \left(\nabla\bdot\bhat\right)^{2} \;-\; \tau^{2} \right] \;+\; 2 \left[ \left(\alpha_{1}\right)^{2} \;+\; 4\;\left(\alpha_{2}\right)^{2} \right],
\label{eq:nablab_nablab}
\end{eqnarray}
which implies that 
\begin{equation} 
\left(\alpha_{1}\right)^{2} \;+\; 4\;\left(\alpha_{2}\right)^{2} \;\equiv\; \left\langle\left(\alpha_{1}
\right)^{2}\right\rangle \;+\; 4\;\left\langle\left(\alpha_{2}\right)^{2}\right\rangle,
\label{eq:a12_id}
\end{equation}
as is easily demonstrated by noting that the gyroangle-derivative of the left side of Eq.~\eqref{eq:a12_id} vanishes. Next, we note that
\begin{equation}
\left\langle\left(\alpha_{1}\right)^{2}\right\rangle \;=\; \left\langle\left(\pd{\alpha_{2}}{\theta}\right)
\left(\alpha_{1}\right)\right\rangle \;=\; 4\;\left\langle\left(\alpha_{2}\right)^{2}\right\rangle,
\label{eq:alpha_12_ave}
\end{equation}
and thus the dyadic identity \eqref{eq:nablab_nablab} becomes
\begin{equation}
\nabla\bhat\;:\;\nabla\bhat \;=\; \nabla\bdot\vb{\kappa} \;-\; \bhat\bdot\nabla(\nabla\bdot\bhat) \;=\; \frac{1}{2}\left[ \left(\nabla\bdot\bhat
\right)^{2} \;-\; \tau^{2} \right] \;+\; 4\;\left\langle\left(\alpha_{1}\right)^{2}\right\rangle.
\label{eq:nablab_nablab_id}
\end{equation}
We will also need the related dyadic identity
\begin{eqnarray}
(\nabla\bhat)^{\top}\;:\;\nabla\bhat & = & |\vb{\kappa}|^{2} + (C_{\rho\rho})^{2} + (C_{\bot\bot})^{2} + (C_{\bot\rho})^{2} + (C_{\rho\bot})^{2} 
\label{eq:nablab_T_nablab_id} \\
 & = & |\vb{\kappa}|^{2} + \frac{1}{2}\left[ \left(\nabla\bdot\bhat\right)^{2} + \tau^{2} \right] + 4\;\left\langle\left(\alpha_{1}\right)^{2}\right\rangle \;=\; \nabla\bhat\;:\;\nabla\bhat + |\vb{\kappa}|^{2} + \tau^{2}.
\nonumber
\end{eqnarray}

Lastly, we give the expression for the gyrogauge-invariant vector field (Littlejohn, 1981)
\begin{eqnarray}
\nabla\btimes{\bf R} & = & \frac{1}{2} \left[ \nabla\bhat : \nabla\bhat \;-\frac{}{} (\nabla\bdot\bhat)^{2}\right]\;\bhat \;+\; (\nabla\bdot\bhat)\;
\vb{\kappa} \;-\; \vb{\kappa}\bdot\nabla\bhat,
\label{eq:b_curl_R}
\end{eqnarray}
which yields the relations
\begin{eqnarray}
\bhat\bdot\nabla\btimes{\bf R} & = & \frac{1}{2}\;\nabla\bdot\left[ \vb{\kappa} \;-\frac{}{} \bhat\;(\nabla\bdot\bhat)\right] \;=\; 2\,\langle \alpha_{1}^{2}\rangle \;-\; \frac{1}{4} \left[ \tau^{2} \;+\; \left(\nabla\bdot\bhat\right)^{2}\right],
\label {eq:alpha1_square}
\end{eqnarray}
from which we obtain an expression for $\langle\alpha_{1}^{2}\rangle$.

\section{\label{sec:D_12}Calculations of Operators $D_{1}$ and $D_{1}^{2}$}

In this Appendix, we apply the operators $D_{1}$ and $D_{1}^{2}$ on the vectors $p_{\|}\,\bhat$ and ${\bf p}_{\bot}$, whose expressions are used in Sec.~\ref{sec:second}: $D_{1}({\bf P}_{2}) \equiv D_{1}(p_{\|}\,\bhat) + \frac{1}{2}\,D_{1}({\bf p}_{\bot})$; in Sec.~\ref{sec:third}: $D_{1}({\bf P}_{3}) \equiv \frac{1}{2}\,D_{1}(p_{\|}\,\bhat) + \frac{1}{3}\,D_{1}({\bf p}_{\bot})$ and $D_{1}^{2}({\bf P}_{3})$; and in Sec.~\ref{sec:fourth}: 
$D_{1}({\bf P}_{4}) \equiv \frac{1}{3}\,D_{1}(p_{\|}\,\bhat) + \frac{1}{4}\,D_{1}({\bf p}_{\bot})$ and $D_{1}^{2}({\bf P}_{4})$. These operators are also needed in Apps.~\ref{sec:G2_J}-\ref{sec:Pi_2} and \ref{sec:Psi2_phys}.

\subsection{Operators $D_{1}$ and $D_{1}^{2}$ acting on $p_{\|}\bhat$}

We begin with the operators $D_{1}$ and $D_{1}^{2}$ acting on $p_{\|}\,\bhat$. First, we use the expression
\begin{equation}
D_{1}\left(p_{\|}\,\bhat\right) \;=\; \left( G_{1}^{p_{\|}} \;+\; p_{\|}\;\vb{\rho}_{0}\bdot\vb{\kappa}\right)\bhat \;+\; p_{\|}\;\tau\;
\pd{\vb{\rho}_{0}}{\theta} \;=\; J\,\left(\tau \;+\frac{}{} \alpha_{1}\right)\;\bhat \;+\; \varrho_{\|}\,\tau\;{\bf p}_{\bot}, 
\label{eq:D1_p_par}
\end{equation}
from which we obtain
\begin{equation}
\left. \begin{array}{rcl}
\langle D_{1}(p_{\|}\,\bhat)\rangle & = & J\,\tau\;\bhat \\
D_{1}(p_{\|}\,\bhat)\bdot\vb{\rho}_{0} & = & 0 \\
D_{1}(p_{\|}\,\bhat)\bdot\partial\vb{\rho}_{0}/\partial\theta & = & 2\,J\;\varrho_{\|}\,\tau
\end{array} \right\},
\label{eq:D1_par_rho}
\end{equation}
and, using Eq.~\eqref{eq:G2x_exp}, we obtain
\begin{eqnarray}
\left\langle D_{1}\left(p_{\|}\,\bhat\right)\bdot\pd{G_{2}^{\bf x}}{\theta}\right\rangle & = & \frac{J^{2}}{m\Omega}\;\langle\alpha_{1}^{2}\rangle \;+\; \frac{3}{2}\;J\,\varrho_{\|}^{2}\tau^{2}.
\label{eq:D1_G2_par}
\end{eqnarray}
Second, from the definition \eqref{eq:Pn_def}, we use the expression
\begin{eqnarray}
D_{1}^{2}\left(p_{\|}\,\bhat\right) & = & \left[ J\;\tau\,(\tau + \alpha_{1}) \;+\; \tau \left( G_{1}^{p_{\|}} \;+\; \frac{p_{\|}}{2J}\;G_{1}^{J}\right) \;+\; \varrho_{\|}B\;\bhat\bdot\nabla\btimes\left(\frac{\tau}{B}\;{\bf p}_{\bot}\right)\right]\;\pd{\vb{\rho}_{0}}{\theta} \nonumber \\
 &  &-\; \left( p_{\|}\,\tau\frac{}{}G_{1}^{\theta}\right)\vb{\rho}_{0} \;+\; \left[ G_{1}^{J}\;(\tau + \alpha_{1}) \;-\; 4\,J\,\alpha_{2}\,G_{1}^{\theta}\;-\; \varrho_{\|}\,\tau\;\pd{\vb{\rho}_{0}}{\theta}\bdot\nabla\btimes{\bf p}_{\bot} \right. \nonumber \\
 &  &\left.-\; J\;\pd{\vb{\rho}_{0}}{\theta}\bdot\nabla\btimes\left[(\tau + \alpha_{1})\,\bhat\right] \right]\;\bhat, 
\label{eq:D12_p_par}
\end{eqnarray}
to obtain
\begin{eqnarray}
\left\langle D_{1}^{2}\left(p_{\|}\,\bhat\right)\right\rangle\bdot\bhat  & = & -\;2\;J\,\varrho_{\|} \left( \tau^{2} \;+\; \langle\alpha_{1}^{2}\rangle
\;-\frac{}{} \tau\;\bhat\bdot{\bf R} \right),
\label{eq:D12_p_par_ave_b} \\
\left\langle D_{1}^{2}\left(p_{\|}\,\bhat\right)\bdot\pd{\vb{\rho}_{0}}{\theta}\right\rangle & = & \left(\frac{4\,J^{2}}{m\Omega} - J\,
\varrho_{\|}^{2}\right)\tau^{2},
\label{eq:D12_p_par_ave_perp}
\end{eqnarray}

\subsection{Operators $D_{1}$ and $D_{1}^{2}$ acting on ${\bf p}_{\bot}$}

Next, we consider the operators $D_{1}$ and $D_{1}^{2}$ acting on ${\bf p}_{\bot}$:
\begin{eqnarray}
D_{1}({\bf p}_{\bot}) & = & {\sf G}_{1}\cdot\exd{\bf p}_{\bot} \;+\; 2\,J\;{\bf R} \;=\; J \left[2\,{\bf R} \;-\frac{}{} (\tau + 2\,\alpha_{1})\,\bhat\right] \;+\; g_{1}^{J}\;\pd{{\bf p}_{\bot}}{J} \;+\; g_{1}^{\theta}\;\pd{{\bf p}_{\bot}}{\theta} \nonumber \\
 & = & J \left[2\,{\bf R} \;-\frac{}{} (\tau + 2\,\alpha_{1})\,\bhat\right] \;+\; \left[ \frac{p_{\|}^{2}}{m\Omega}\,(\vb{\rho}_{0}\bdot\vb{\kappa}) \;-\frac{}{} J\,\varrho_{\|}\;(\tau + \alpha_{1})\right] \pd{{\bf p}_{\bot}}{J} \nonumber \\
 &  &+\; \left[ \varrho_{\|}\alpha_{2} \;+\; \pd{\vb{\rho}_{0}}{\theta}\bdot\left(\nabla\ln B \;+\; \frac{p_{\|}^{2}\,\vb{\kappa}}{2m\Omega\;J}\right) \right] \pd{{\bf p}_{\bot}}{\theta}, 
\label{eq:D1_p_perp} \\
D_{1}^{2}({\bf p}_{\bot}) & = & 2\,G_{1}^{J}\;{\bf R} \;-\; \left[ G_{1}^{J}\;\left(\tau \;+\frac{}{} 2\,\alpha_{1}\right) \;-\frac{}{} 8\,J\alpha_{2}\;
G_{1}^{\theta} \right]\;\bhat \;+\; \vb{\rho}_{0}\btimes\nabla\btimes\left[D_{1}({\bf p}_{\bot})\right]
\nonumber \\
 &  &+\; \left[ \left(G_{1}^{a}\,\partial_{a}g_{1}^{J}\right) - 2\,J\;g_{1}^{\theta}\;G_{1}^{\theta} - \frac{1}{2J}\;g_{1}^{J}\,G_{1}^{J}\right]
\pd{{\bf p}_{\bot}}{J} \nonumber \\
 &  &+\; \left[ \left(G_{1}^{a}\,\partial_{a}g_{1}^{\theta}\right) + \frac{1}{2J}\left( g_{1}^{J}\;G_{1}^{\theta} + 
g_{1}^{\theta}\,G_{1}^{J}\right) \right] \pd{{\bf p}_{\bot}}{\theta},
\end{eqnarray}
from which we obtain
\begin{equation}
\left. \begin{array}{rcl}
\langle D_{1}({\bf p}_{\bot})\rangle & = & J\,(2{\bf R} - \tau\,\bhat) \;-\; \bhat\btimes(J\,\nabla\ln B + p_{\|}\,\varrho_{\|}\,\vb{\kappa}) \\
D_{1}({\bf p}_{\bot})\bdot\vb{\rho}_{0} & = & -\,2\,J\;G_{1}^{\theta} \\
D_{1}({\bf p}_{\bot})\bdot\partial\vb{\rho}_{0}/\partial\theta & = & 2\,J\;{\bf R}\bdot\partial\vb{\rho}_{0}/\partial\theta + (G_{1}^{J} - J\,
\vb{\rho}_{0}\bdot\nabla\ln B)
\end{array} \right\},
\label{eq:D1_bot_rho}
\end{equation}
and
\begin{eqnarray}
\left\langle D_{1}^{2}({\bf p}_{\bot})\right\rangle\bdot\bhat & = & 2\;\langle G_{1}^{J}\rangle\;\bhat\bdot{\bf R} \;-\; \left\langle G_{1}^{J}\;(\tau + 2\,\alpha_{1}) \;-\frac{}{} 8\,J\alpha_{2}\;G_{1}^{\theta}\right\rangle \nonumber \\
 &  &-\; \left\langle\pd{\vb{\rho}_{0}}{\theta}\bdot\nabla\btimes[D_{1}({\bf p}_{\bot})]\right\rangle \nonumber \\
 & = & J\,\varrho_{\|} \left( \frac{3}{2}\,\tau^{2} \;+\; 6\;\langle\alpha_{1}^{2}\rangle \;-\frac{}{} 3\,\tau\;\bhat\bdot{\bf R} \right),
\label{eq:D12_p_perp_ave_b} \\
\left\langle D_{1}^{2}({\bf p}_{\bot})\bdot\pd{\vb{\rho}_{0}}{\theta}\right\rangle & = & 2\left\langle G_{1}^{J}\frac{}{}\vb{\rho}_{0}\right\rangle
\bdot\bhat\btimes{\bf R} + 2\,J\;\frac{\bhat}{m\Omega}\bdot\nabla\btimes\left\langle D_{1}({\bf p}_{\bot})\right\rangle \nonumber \\
 &  &+\; \left\langle \left(G_{1}^{a}\,\partial_{a}g_{1}^{J}\right) - 2\,J\;g_{1}^{\theta}\;G_{1}^{\theta} - \frac{1}{2J}\;g_{1}^{J}\,G_{1}^{J}
\right\rangle,
\label{eq:D12_p_perp_ave_perp} \\
\left\langle D_{1}({\bf p}_{\bot})\bdot\pd{G_{2}^{\bf x}}{\theta}\right\rangle & = & -\;\frac{2\,J^{2}}{m\Omega}\;\langle\alpha_{1}^{2}\rangle \;-\; J\;\varrho_{\|}^{2}\,\tau^{2} \nonumber \\
 &  &+\; \left\langle J\;\left(g_{1}^{\theta}\right)^{2} \;+\; \frac{1}{4J}\;\left(g_{1}^{J}\right)^{2} \;+\; 
g_{1}^{J}\;\pd{g_{1}^{\theta}}{\theta}\right\rangle,
\label{eq:D1_G2_perp}
\end{eqnarray}
where
\begin{eqnarray*}
\bhat\bdot\nabla\btimes\langle D_{1}({\bf p}_{\bot})\rangle & = & 2\,J\;\bhat\bdot\nabla\btimes{\bf R} \;-\; J\;\tau^{2} \;-\; \bhat\bdot\nabla\btimes\left[\bhat\btimes\left( J\;\nabla\ln B \;+\; \frac{p_{\|}^{2}\,\vb{\kappa}}{m\Omega} \right) \right], \\
\left\langle\pd{\vb{\rho}_{0}}{\theta}\bdot\nabla\btimes[D_{1}({\bf p}_{\bot})]\right\rangle & = & \left\langle\pd{\vb{\rho}_{0}}{\theta}\bdot\nabla\btimes\left( g_{1}^{J}\;\pd{{\bf p}_{\bot}}{J} \;+\; g_{1}^{\theta}\;\pd{{\bf p}_{\bot}}{\theta} \right)\right\rangle \\
 & = & -\,J\,\varrho_{\|}\;\left( \frac{\tau^{2}}{2} \;+\; 2\,\langle\alpha_{1}^{2}\rangle \;-\; \tau\;\bhat\bdot{\bf R} \right).
\end{eqnarray*}

\subsection{Calculation of $F_{1\alpha} = (\partial D_{1}({\bf P}_{4})/\partial u^{\alpha})\bdot\vb{\rho}_{0}$}

Lastly, in Sec.~\ref{sec:fourth} and App.~\ref{sec:G2_J}, we need
\begin{eqnarray}
F_{1p_{\|}} \;=\; \pd{D_{1}({\bf P}_{4})}{p_{\|}}\bdot\vb{\rho}_{0} & = & -\;\frac{1}{2}\;
J\;\pd{g_{1}^{\theta}}{p_{\|}}, \label{eq:F1_p} \\
F_{1J} \;=\; \pd{D_{1}({\bf P}_{4})}{J}\bdot\vb{\rho}_{0} & = & -\;\frac{1}{2} 
\left( J\;\pd{g_{1}^{\theta}}{J} \;+\; \frac{1}{2}\,g_{1}^{\theta}\right) \;+\; \frac{1}{2}\;\vb{\rho}_{0}\bdot{\bf R}, \label{eq:F1_J} \\
F_{1\theta} \;=\; \pd{D_{1}({\bf P}_{4})}{\theta}\bdot\vb{\rho}_{0} & = & -\;\frac{1}{2}
\left( J\;\pd{g_{1}^{\theta}}{\theta} \;+\; \frac{1}{2}\,g_{1}^{J}\right) \;-\; \frac{2}{3}\,J\;\varrho_{\|}\tau, \label{eq:F1_theta}
\end{eqnarray}
and Eq.~\eqref{eq:G2_J_ave} makes use of the following expressions
\begin{eqnarray}
\pd{F_{1\theta}}{p_{\|}} \;-\; \pd{F_{1p_{\|}}}{\theta} & = & -\;\frac{1}{4}\;\pd{g_{1}^{J}}{p_{\|}} \;-\; \frac{2}{3}\;\frac{J\tau}{m\Omega}, \\
\pd{F_{1\theta}}{J} \;-\; \pd{F_{1J}}{\theta} & = & -\;\frac{1}{4} \left( \pd{g_{1}^{J}}{J} \;+\; \pd{g_{1}^{\theta}}{\theta} \right) \;-\;
\frac{1}{2}\;\pd{\vb{\rho}_{0}}{\theta}\bdot{\bf R} \;-\; \frac{2}{3}\;\varrho_{\|}\tau.
\end{eqnarray}

\section{\label{sec:G2_J}Second-order Calculations for $\langle G_{2}^{J}\rangle$}

In this Appendix, we present the detailed calculations leading to the gyroangle-averaged component $\langle G_{2}^{J}\rangle$ used in the second-order Hamiltonian constraint \eqref{eq:Ham_constraint_2_initial}.

Equation \eqref{eq:G2_J_ave} defines the gyroangle-averaged component $\langle G_{2}^{J}\rangle$:
\begin{eqnarray}
\langle G_{2}^{J}\rangle & = & \frac{1}{2} \left\langle \pd{G_{2}^{\bf x}}{\theta}\bdot D_{1}({\bf P}_{2})\right\rangle \;+\; \frac{1}{4} \left\langle 
G_{1}^{J}\;\pd{G_{1}^{\theta}}{\theta} \;-\; G_{1}^{\theta}\;\pd{G_{1}^{J}}{\theta} \;+\;
{\sf G}_{1}\cdot\exd G_{1}^{J} \right\rangle \nonumber \\
 &  &+\; \frac{1}{2} \left\langle \pd{\vb{\rho}_{0}}{\theta}\bdot D_{1}^{2}({\bf P}_{4})\right\rangle
\;-\; \frac{1}{2} \left\langle {\sf G}_{1}\cdot\exd F_{1\theta} \;-\; G_{1}^{a}\;\pd{F_{1a}}{\theta}\right\rangle,
\label{eq:G2_J_ave_app}
\end{eqnarray}
where
\begin{eqnarray}
\pd{G_{2}^{\bf x}}{\theta} & = & \left( \frac{J\,\alpha_{1}}{m\Omega} \;-\; 2\,\varrho_{\|}\;\vb{\rho}_{0}\bdot\vb{\kappa}\right)\bhat  \;+\; 
\varrho_{\|}\,\tau\;\pd{\vb{\rho}_{0}}{\theta} \;+\; \frac{1}{2} \left( \pd{g_{1}^{J}}{\theta} \;-\; 2J\;g_{1}^{\theta}\right)\;\pd{\vb{\rho}_{0}}{J} 
\nonumber \\
 &  &+\; \frac{1}{2} \left( \pd{g_{1}^{\theta}}{\theta} \;+\; \frac{1}{2J}\;g_{1}^{J}\right)\;\pd{\vb{\rho}_{0}}{\theta},
\label{eq:G2x_theta}
\end{eqnarray}
where $g_{1}^{J} \equiv G_{1}^{J} - J\;\vb{\rho}_{0}\bdot\nabla\ln B$ and $g_{1}^{\theta} \equiv G_{1}^{\theta} + \vb{\rho}_{0}\bdot{\bf R}$. We now compute each term respectively. The first and third terms are
\begin{eqnarray}
\frac{1}{2} \left\langle \pd{G_{2}^{\bf x}}{\theta}\bdot D_{1}({\bf P}_{2})\right\rangle & = & 
\frac{1}{2}\,J\;\varrho_{\|}^{2}\,\tau^{2} \;+\; \frac{1}{4} \left\langle J\;\left(g_{1}^{\theta}\right)^{2} \;+\; \frac{1}{4J}\,\left(g_{1}^{J}
\right)^{2} \;+\; g_{1}^{J}\;\pd{g_{1}^{\theta}}{\theta} \right\rangle, \\
\frac{1}{2} \left\langle \pd{\vb{\rho}_{0}}{\theta}\bdot D_{1}^{2}({\bf P}_{4})\right\rangle & = & 
\left( \frac{2\,J^{2}}{3\,m\Omega} \;-\; \frac{1}{2}\,J\,\varrho_{\|}^{2}\right)\;\tau^{2} \;+\; \frac{1}{4}\left\langle \left(G_{1}^{J} \;+\; 
\frac{1}{2}g_{1}^{J}\right)\pd{\vb{\rho}_{0}}{\theta} \;+\; J\;g_{1}^{\theta}\,\vb{\rho}_{0}\right\rangle\bdot{\bf R} \nonumber \\
 &  &+\; \frac{J\,\bhat}{4\,m\Omega}\bdot\nabla\btimes\langle D_{1}({\bf p}_{\bot})\rangle \;-\; \frac{1}{4} \left\langle J\;\left(g_{1}^{\theta}
\right)^{2} \;+\; \frac{1}{4J}\,\left(g_{1}^{J}\right)^{2} \right\rangle \nonumber \\
 &  &+\; \frac{1}{8}\left\langle G_{1}^{p_{\|}}\;\pd{g_{1}^{J}}{p_{\|}} \;+\; G_{1}^{J}\;\pd{g_{1}^{J}}{J} \;-\; g_{1}^{J}\left(\frac{1}{2}\;
\vb{\rho}_{0}\bdot\nabla\ln B \;+\; \pd{g_{1}^{\theta}}{\theta}\right) \right\rangle,
\end{eqnarray}
while the second and fourth terms are
\begin{eqnarray}
\frac{1}{4} \left\langle G_{1}^{J}\;\pd{G_{1}^{\theta}}{\theta} \;-\; G_{1}^{\theta}\;\pd{G_{1}^{J}}{\theta} \;+\;{\sf G}_{1}\cdot\exd G_{1}^{J} \right\rangle & = & 
\frac{1}{4}\left\langle G_{1}^{J} \left( \frac{3}{2}\,\vb{\rho}_{0}\bdot\nabla\ln B \;+\; \pd{g_{1}^{J}}{J} \;+\; \pd{g_{1}^{\theta}}{\theta} \right) \right\rangle 
\nonumber \\
 &  &-\;\frac{1}{4}\;\left\langle G_{1}^{J}\;\pd{\vb{\rho}_{0}}{\theta}\right\rangle \bdot{\bf R} \;+\; \frac{1}{4}\left\langle
G_{1}^{p_{\|}}\;\pd{g_{1}^{J}}{p_{\|}}\right\rangle \nonumber \\
 &  &-\; \frac{1}{4}\; \left\langle \vb{\rho}_{0}\frac{}{}\bdot\nabla G_{1}^{J} \right\rangle,
\end{eqnarray}
and
\begin{eqnarray}
-\; \frac{1}{2} \left\langle {\sf G}_{1}\cdot\exd F_{1\theta} \;-\; G_{1}^{a}\;\pd{F_{1a}}{\theta}\right\rangle & = & -\; \frac{1}{4}\left\langle 
\vb{\rho}_{0}\bdot\nabla\left( J\;\pd{g_{1}^{\theta}}{\theta} \;+\; \frac{1}{2}\;g_{1}^{J}\right)\right\rangle \nonumber \\
 &  &+\; \frac{1}{4}\;\left\langle G_{1}^{J}\;\pd{\vb{\rho}_{0}}{\theta}\right\rangle\bdot{\bf R} + \frac{1}{3} \left\langle G_{1}^{p_{\|}}\;\frac{J\tau}{m\Omega} \;+\; G_{1}^{J}\;\varrho_{\|}\,\tau \right\rangle \nonumber \\
 &  &+\; \frac{1}{8}\left\langle G_{1}^{p_{\|}}\;\pd{g_{1}^{J}}{p_{\|}} \;+\; G_{1}^{J}\left( \pd{g_{1}^{J}}{J} \;+\; \pd{g_{1}^{\theta}}{\theta}\right) \right\rangle,
\end{eqnarray}
so that Eq.~\eqref{eq:G2_J_ave_app} becomes
\begin{eqnarray}
\langle G_{2}^{J}\rangle & = & \frac{1}{3} \left[ \left( \frac{2J^{2}}{m\Omega} \;+\; J\;\varrho_{\|}^{2}\right) \tau^{2} \;+\; \left\langle
G_{1}^{p_{\|}}\;\frac{J\,\tau}{m\Omega} \;+\; G_{1}^{J}\;\varrho_{\|}\,\tau \right\rangle \right] \;+\; \frac{1}{2}\;\left\langle G_{1}^{p_{\|}}\;
\pd{g_{1}^{J}}{p_{\|}}\right\rangle \label{eq:G2_J_2} \\
 &  &+\; \frac{J\,\bhat}{4\,m\Omega}\bdot\nabla\btimes\langle D_{1}({\bf p}_{\bot})\rangle  + \frac{1}{4}\;\left\langle \left( G_{1}^{J} \;+\; \frac{1}{2}\,g_{1}^{J}\right)\pd{\vb{\rho}_{0}}{\theta} \;+\; J\,g_{1}^{\theta}\;\vb{\rho}_{0}\right\rangle \bdot{\bf R} \nonumber \\
 & &+ \frac{1}{4} \left\langle G_{1}^{J} \left[ \frac{3}{2} \left( \vb{\rho}_{0}\bdot\nabla\ln B + \pd{g_{1}^{\theta}}{\theta}\right) +
2\;\pd{g_{1}^{J}}{J} \right] + \frac{1}{2}\;g_{1}^{J} \left( \pd{g_{1}^{\theta}}{\theta} - \frac{1}{2}\,\vb{\rho}_{0}\bdot\nabla\ln B \right)
\right\rangle \nonumber \\
 &  &-\; \frac{1}{4} \left\langle\vb{\rho}_{0}\bdot\nabla\left( G_{1}^{J} \;+\; \frac{1}{2}\,g_{1}^{J} \;+\; J\;\pd{g_{1}^{\theta}}{\theta}\right)\right\rangle \nonumber
\end{eqnarray}
We now use the identity $\vb{\rho}_{0}\bdot\nabla A = \nabla\bdot(\vb{\rho}_{0}\;A) - A\;(\nabla\bdot\vb{\rho}_{0})$, where $ \nabla\bdot\vb{\rho}_{0}$ is given by Eq.~\eqref{eq:divrho_0},  so that we obtain
\begin{eqnarray}
 &  &-\; \frac{1}{4} \left\langle\vb{\rho}_{0}\bdot\nabla\left( G_{1}^{J} \;+\; \frac{1}{2}\,g_{1}^{J} \;+\; 
J\;\pd{g_{1}^{\theta}}{\theta}\right)\right\rangle \nonumber \\
 &  & \hspace*{1in}=\; -\; \frac{1}{4} \left\langle\left( G_{1}^{J} \;+\; \frac{1}{2}\,g_{1}^{J} \;+\; J\;\pd{g_{1}^{\theta}}{\theta}\right) 
\vb{\rho}_{0}\bdot\left(\frac{1}{2}\,\nabla\ln B \;+\; \vb{\kappa}\right) \right\rangle \nonumber \\
 &   &\hspace*{1.2in}-\;\frac{1}{4} \nabla\bdot\left\langle\vb{\rho}_{0}\;\left( G_{1}^{J} \;+\; 
\frac{1}{2}\,g_{1}^{J} \;+\; J\;\pd{g_{1}^{\theta}}{\theta}\right)\right\rangle \nonumber \\
 &  &\hspace*{1.2in}-\;\frac{1}{4} \left\langle\left( G_{1}^{J} \;+\; \frac{1}{2}\,g_{1}^{J}\right)
\pd{\vb{\rho}_{0}}{\theta} \;+\; J\;g_{1}^{\theta}\;\vb{\rho}_{0} \right\rangle\bdot{\bf R}.
\end{eqnarray}
By substituting this expression into Eq.~\eqref{eq:G2_J_2}, we obtain
\begin{eqnarray}
\langle G_{2}^{J}\rangle & = & \frac{1}{3} \left[ \left( \frac{2J^{2}}{m\Omega} \;+\; J\;\varrho_{\|}^{2}\right) \tau^{2} \;+\; \left\langle
G_{1}^{p_{\|}}\;\frac{J\,\tau}{m\Omega} \;+\; G_{1}^{J}\;\varrho_{\|}\,\tau \right\rangle \right] \nonumber \\
 &  &+\; \frac{J\,\bhat}{4\,m\Omega}\bdot\nabla\btimes\langle D_{1}({\bf p}_{\bot})\rangle \;+\; \frac{1}{2}\;\left\langle G_{1}^{p_{\|}}\;\pd{g_{1}^{J}}{p_{\|}}\right\rangle \nonumber \\
 &  &-\; \frac{1}{4} \nabla\bdot\left\langle\vb{\rho}_{0}\;\left( G_{1}^{J} \;+\; \frac{1}{2}\,g_{1}^{J} \;+\; J\;
\pd{g_{1}^{\theta}}{\theta}\right)\right\rangle \nonumber \\
 &  &+\; \frac{1}{4} \left\langle G_{1}^{J} \left[ \vb{\rho}_{0}\bdot\left(\nabla\ln B \;-\frac{}{}
\vb{\kappa} \right) \;+\; \frac{3}{2}\,\pd{g_{1}^{\theta}}{\theta} \;+\; 2\;\pd{g_{1}^{J}}{J} \right]\right\rangle \nonumber \\
 &  &+\; \frac{1}{8}\;\left\langle g_{1}^{J} \left[ \pd{g_{1}^{\theta}}{\theta} \;-\; \vb{\rho}_{0}\bdot\left( \nabla\ln B \;+\frac{}{} \vb{\kappa} 
\right) \right] \right\rangle \nonumber \\
 &  &-\; \frac{J}{4}\,\left\langle \pd{g_{1}^{\theta}}{\theta}\;\vb{\rho}_{0}\bdot\left(\frac{1}{2}\,\nabla\ln B \;+\;
\vb{\kappa}\right) \right\rangle.
\label{eq:G2_J_3}
\end{eqnarray}
We now substitute the definitions of the generating vector-field components and we obtain
\begin{eqnarray}
\langle G_{2}^{J}\rangle & = & \frac{J^{2}}{2m\Omega} \left[ \frac{1}{2}\,\tau^{2} \;+\; \bhat\bdot\nabla\btimes{\bf R} \;-\; \langle\alpha_{1}^{2}
\rangle \;-\; \frac{1}{2}\;\bhat\bdot\nabla\btimes\left(\bhat\btimes\nabla\ln B\right) \right] \nonumber \\
 &  &-\; \frac{1}{2}\;J\,\varrho_{\|}^{2} \left[ \vb{\kappa}\bdot(3\,\vb{\kappa} - \nabla\ln B) \;+\frac{}{} \nabla\bdot\vb{\kappa} \;-\; \tau^{2} \right],
\label{eq:G2_J_Ham_app}
\end{eqnarray} 
which is then used in Eq.~\eqref{eq:G2_J_Ham}.

\section{\label{sec:Pi_2}Calculation Details for $(p_{\|}/m)\,\Pi_{2\|} - \Psi_{2}$}

The second-order Lagrangian constraint \eqref{eq:second_Lag} is given in Sec.~\ref{sec:push_Lag} as
\begin{eqnarray}
m\;\pd{}{p_{\|}}\left(\frac{p_{\|}}{m}\,\Pi_{2\|} - \Psi_{2} \right) + m\,{\bf v}_{\rm gc}\bdot\pd{\vb{\Pi}_{1}}{p_{\|}} & = & \left( m\;
\frac{d_{0}\langle\vb{\rho}_{1}\rangle}{dt} \;+\; \left\langle D_{1}^{2}({\bf P}_{3})\right\rangle \right) \bdot\bhat \nonumber \\
 &  &+\; p_{\|}\,\vb{\kappa}\bdot\langle G_{2}^{\bf x}\rangle - \left\langle G_{2}^{\bf x}\bdot\nabla\bhat\bdot{\bf p}_{\bot}\right\rangle \nonumber \\
 &  &-\; \frac{1}{2}\left\langle {\sf G}_{1}\cdot\frac{}{}\exd\left({\sf G}_{1}\cdot\exd{\bf p}\right)\right\rangle\bdot\bhat,
\label{eq:Banos_2a_app}
\end{eqnarray}
which contains five terms that are calculated in this Appendix.  The first term is
\begin{eqnarray}
m\,\frac{d_{0}\langle\vb{\rho}_{1}\rangle}{dt}\bdot\bhat & = & \left(p_{\|}\frac{}{}\bhat\bdot\nabla\langle\vb{\rho}_{1}\rangle\right)\bdot\bhat \;=\; p_{\|}\;\bhat\bdot\nabla\left( \langle\vb{\rho}_{1}\rangle\bdot\bhat \right) \;-\; p_{\|}\;\langle\vb{\rho}_{1}\rangle\bdot\vb{\kappa} \nonumber \\
 & = & -\,\frac{1}{2}\;J\,\varrho_{\|}\;\left\{ \nabla\bdot\left[\bhat\;\left(\nabla\bdot\bhat\right)\right] \;-\; 3\;\vb{\kappa}\bdot\nabla\ln B 
\right\} \;+\; p_{\|}\;\varrho_{\|}^{2}|\vb{\kappa}|^{2} \nonumber \\
 &  &-\; \varrho_{\|}\,\bhat\btimes\vb{\kappa}\bdot\vb{\Pi}_{1},
\label{eq:first_term}
\end{eqnarray}
where $\partial(\langle\vb{\rho}_{1}\rangle\bdot\bhat)/\partial p_{\|} = 0$ follows from Eq.~\eqref{eq:rho1_ave}. For the second term in 
Eq.~\eqref{eq:Banos_2a_app}, we use Eq.~\eqref{eq:D1_3}:
\begin{equation}
\bhat\bdot\left\langle D_{1}^{2}({\bf P}_{3})\right\rangle \;=\; -\;J\,\varrho_{\|} \left( \frac{1}{2}\,\tau^{2} \;-\; \langle\alpha_{1}^{2}\rangle
\right).
\label{eq:fourth_term}
\end{equation}
The third term in Eq.~\eqref{eq:Banos_2a_app} is
\begin{equation}
p_{\|}\;\langle G_{2}^{\bf x}\rangle\bdot\vb{\kappa} \;=\; -\; \varrho_{\|}\,\bhat\btimes\vb{\kappa}\bdot\vb{\Pi}_{1} \;+\; \frac{1}{2}\,\left(
J\,\varrho_{\|}\;\vb{\kappa}\bdot\nabla\ln B \;+\frac{}{} p_{\|}\;\varrho_{\|}^{2}|\vb{\kappa}|^{2} \right),
\label{eq:second_term}
\end{equation}
where we substituted Eq.~\eqref{eq:G2_x_ave}, while the fourth term in Eq.~\eqref{eq:Banos_2a_app} is
\begin{eqnarray}
-\; \left\langle G_{2}^{\bf x}\bdot\nabla\bhat\bdot{\bf p}_{\bot}\right\rangle & = & -\;\left\langle G_{2}^{\bf x}\bdot\left[ \bhat\;(\vb{\kappa}\bdot
{\bf p}_{\bot}) \;-\; C_{\rho\bot}\;\pd{{\bf p}_{\bot}}{\theta} \;+\; C_{\bot\bot}\;{\bf p}_{\bot} \right] \right\rangle 
\label{eq:fourth_term} \\
 & = & -\;\vb{\kappa}\bdot\left\langle {\bf p}_{\bot}\;G_{2\|}^{\bf x}\right\rangle \;-\; 2\,J\;(\varrho_{\|}\,\tau)\;\langle C_{\rho\bot}\rangle \nonumber \\
 &  &-\; \frac{1}{2}\;\left\langle C_{\rho\bot}\;\left( G_{1}^{J} - J\;\vb{\rho}_{0}\bdot\nabla\ln B\right)\right\rangle \;-\; J\;
\left\langle C_{\bot\bot}\;\left(G_{1}^{\theta} \;+\frac{}{} \vb{\rho}_{0}\bdot{\bf R}\right)\right\rangle,
\nonumber 
\end{eqnarray}
Here, using Eq.~\eqref{eq:G2par_def}, we find $-\,\langle G_{2\|}^{\bf x}\,{\bf p}_{\bot}\rangle\bdot\vb{\kappa} = -\,2\,J\;\varrho_{\|}\,
|\vb{\kappa}|^{2}$, using Eq.~\eqref{eq:grad_i}, we find $-\; 2\,J\;(\varrho_{\|}\,\tau)\;\langle C_{\rho\bot}\rangle = J\,\varrho_{\|}\;\tau^{2}$, while using Eqs.~\eqref{eq:G1_J} and \eqref{eq:G1_theta_def}, we obtain
\[ -\; \frac{1}{2}\;\left\langle C_{\rho\bot}\;\left( G_{1}^{J} - J\;\vb{\rho}_{0}\bdot\nabla\ln B\right)\right\rangle \;=\; 
-\;\frac{1}{2}\;J\,\varrho_{\|} \left( \frac{1}{2}\,\tau^{2} +
\langle \alpha_{1}^{2}\rangle \right), \]
and
\[ -\;J \left\langle C_{\bot\bot}\;\left(G_{1}^{\theta} + \vb{\rho}_{0}\bdot{\bf R}\right)\right\rangle \;=\; -\;2\,J\,\varrho_{\|}\;\langle\alpha_{2}^{2}\rangle \;\equiv\; -\;\frac{1}{2}\;J\,\varrho_{\|}\;\langle\alpha_{1}^{2}\rangle, \]
so that Eq.~\eqref{eq:fourth_term} becomes
\begin{equation}
-\; \left\langle G_{2}^{\bf x}\bdot\nabla\bhat\bdot{\bf p}_{\bot}\right\rangle \;=\; J\,\varrho_{\|} \left( \frac{3}{4}\,\tau^{2}
\;-\; 2\,|\vb{\kappa}|^{2} \;-\; \langle \alpha_{1}^{2}\rangle \right).
\label{eq:third_term}
\end{equation}

Lastly, for the fifth term in Eq.~\eqref{eq:Banos_2a_app}, we begin with the identity 
\[ \left\langle {\sf G}_{1}\cdot\frac{}{}\exd\left({\sf G}_{1}\cdot\exd{\bf p}\right)\right\rangle\bdot\bhat \;\equiv\; \left\langle {\sf G}_{1}\cdot\exd
\left[({\sf G}_{1}\cdot\exd{\bf p})\bdot\frac{}{}\bhat \right] \right\rangle \;+\; \left\langle \vb{\rho}_{0}\bdot\nabla\bhat\bdot\frac{}{}
({\sf G}_{1}\cdot\exd{\bf p}) \right\rangle, \]
where
\begin{eqnarray} 
{\sf G}_{1}\cdot\exd{\bf p} & = & \left( G_{1}^{p_{\|}} \;+\frac{}{} 2\,J\,C_{\rho\bot}\right)\;\bhat \;+\; \left( G_{1}^{J} \;-\frac{}{} J\;
\vb{\rho}_{0}\bdot\nabla\ln B \right) \pd{{\bf p}_{\bot}}{J} \nonumber \\
 &  &+\; \left( G_{1}^{\theta} \;+\frac{}{} \vb{\rho}_{0}\bdot{\bf R}\right) \pd{{\bf p}_{\bot}}{\theta} \;-\; p_{\|}\;\vb{\rho}_{0}\bdot\nabla\bhat,
\label{eq:G1_dp}
\end{eqnarray}
with $\vb{\rho}_{0}\bdot\nabla\bhat \equiv C_{\rho\rho}\;\vb{\rho}_{0} + C_{\rho\bot}\;\partial\vb{\rho}_{0}/\partial\theta$. First, we find
\begin{eqnarray}
-\;\frac{1}{2}\left\langle {\sf G}_{1}\cdot\exd\left[({\sf G}_{1}\cdot\exd{\bf p})\bdot\frac{}{}\bhat \right] \right\rangle & = & \frac{1}{2} 
\left\langle G_{1}^{p_{\|}}\;\vb{\rho}_{0}\bdot\vb{\kappa} \;+\; G_{1}^{J}\;\alpha_{1} \;+\; J\;G_{1}^{\theta}\,\pd{\alpha_{1}}{\theta} \right\rangle
\nonumber \\
 &  &+\; p_{\|}\,\left( \vb{\kappa}\bdot\left\langle \frac{1}{2}\;{\sf G}_{1}\cdot\exd\vb{\rho}_{0}\right\rangle \;-\; \frac{1}{2}\left\langle
\vb{\rho}_{0}\bdot\nabla\vb{\kappa}\bdot\vb{\rho}_{0}\right\rangle \right),
\label{eq:fifth_one}
\end{eqnarray}
where we used $({\sf G}_{1}\cdot\exd{\bf p})\bdot\bhat = G_{1}^{p_{\|}} + 2\,J\,C_{\rho\bot} = -\;p_{\|}\,\vb{\rho}_{0}\bdot\vb{\kappa} - J\,\alpha_{1}$. By using 
\begin{eqnarray*} 
\frac{1}{2}\,\langle{\sf G}_{1}\cdot\exd\vb{\rho}_{0}\rangle & = & \frac{J}{m\Omega} \left[ \frac{1}{2}\,(\nabla\bdot\bhat)\,\bhat \;+\; 
\nabla_{\bot}\ln B\right] \;+\; \frac{1}{2}\;\varrho_{\|}^{2}\,\vb{\kappa},
\end{eqnarray*}
and
\[ -\; \frac{1}{2}\;p_{\|}\;\left\langle\vb{\rho}_{0}\bdot\nabla\vb{\kappa}\bdot\vb{\rho}_{0}\right\rangle \;=\; -\,\frac{1}{2}\;J\,\varrho_{\|}\;
({\bf I} - \bhat\bhat):\nabla\vb{\kappa} \;\equiv\; -\,\frac{1}{2}\;J\,\varrho_{\|}\;\left( \nabla\bdot\vb{\kappa} \;+\; |\vb{\kappa}|^{2}\right), \]
Eq.~\eqref{eq:fifth_one} becomes
\begin{eqnarray}
-\;\frac{1}{2}\;\left\langle {\sf G}_{1}\cdot\exd\left[({\sf G}_{1}\cdot\exd{\bf p})\bdot\frac{}{}\bhat \right] \right\rangle & = & J\,\varrho_{\|} 
\left(\vb{\kappa}\bdot\nabla\ln B - |\vb{\kappa}|^{2} - \frac{1}{2}\,\nabla\bdot\vb{\kappa} - \langle\alpha_{1}^{2}\rangle \right) \nonumber \\
 &  &+\; \frac{1}{2}\;p_{\|}\;\varrho_{\|}^{2}|\vb{\kappa}|^{2}.
\label{eq:fifth_a}
\end{eqnarray}
Next, we find
\begin{eqnarray}
-\;\frac{1}{2}\left\langle \vb{\rho}_{0}\bdot\nabla\bhat\bdot\frac{}{}({\sf G}_{1}\cdot\exd{\bf p}) \right\rangle & = & -\;\frac{1}{2}\left\langle
C_{\rho\rho}\;\left[\vb{\rho}_{0}\bdot({\sf G}_{1}\cdot\exd{\bf p})\right] \;+\; C_{\rho\bot}\;\left[\pd{\vb{\rho}_{0}}{\theta}\bdot
({\sf G}_{1}\cdot\exd{\bf p})\right] \right\rangle \nonumber \\
 & = & J\,\varrho_{\|} \left( \langle\alpha_{1}^{2}\rangle + \frac{1}{4}\,(\nabla\bdot\bhat)^{2} \right).
\label{eq:fifth_b}
\end{eqnarray}

If we now combine Eqs~\eqref{eq:first_term}-\eqref{eq:third_term}, and \eqref{eq:fifth_a}-\eqref{eq:fifth_b} into Eq.~\eqref{eq:Banos_2a_app}, we obtain
\begin{eqnarray}
m\,\pd{}{p_{\|}} \left( \frac{p_{\|}}{m}\;\Pi_{2\|} - \Psi_{2} \right) + m\,{\bf v}_{\rm gc}\bdot\pd{\vb{\Pi}_{1}}{p_{\|}} & = & 2\;p_{\|}\;\varrho_{\|}^{2}\,|\vb{\kappa}|^{2} \;-\; 2\,\varrho_{\|}\;
\bhat\btimes\vb{\kappa}\bdot\vb{\Pi}_{1} \nonumber \\
 &  &-\; J\,\varrho_{\|}\;\beta_{2\|},
\label{eq:Pi_2_final_app}
\end{eqnarray}
where
\[ \beta_{2\|} \;\equiv\; -\;3\,\vb{\kappa}\bdot\left(\nabla\ln B \;-\; \vb{\kappa}\right) \;-\; \frac{1}{4}\left[ \tau^{2}
\;+\; (\nabla\bdot\bhat)^{2} \right] \;+\; \frac{1}{2}\;\nabla\bdot\left[\vb{\kappa} \;+\frac{}{} \bhat\;(\nabla\bdot\bhat)\right]. \]
We note that this equation is consistent with Eq.~\eqref{eq:Hamiltonian_constraint_2}.

\section{\label{sec:comp}Comparison with Previous Higher-order Guiding-center Theories}

In this Appendix, we compare our results \eqref{eq:Ham_rep_second} with previous higher-order guiding-center theories derived by \cite{Parra_Calvo_2011}, \cite{Burby_SQ_2013}, and \cite{PCBSQ_2014}. In these works, the polarization term $\vb{\Pi}_{1\bot}$ is ignored and, consequently, these theories are incomplete as discussed in Sec.~\ref{sec:gc_pol_canonical}.

For the purpose of comparison, we summarize our results here for the second-order guiding-center Hamiltonian
\begin{equation}
\Psi_{2({\rm BT})} \;=\; \frac{1}{2}\;J\,\Omega \left( \frac{J}{m\Omega} \;\beta_{2\bot}^{({\rm BT})} \;+\; \varrho_{\|}^{2}\;\beta_{2\|}^{({\rm BT})}
\right) \;-\; \frac{p_{\|}^{2}}{2m}\;\varrho_{\|}^{2}|\vb{\kappa}|^{2} \;+\; \vb{\Pi}_{1}\bdot{\bf v}_{\rm gc},
\label{eq:BT}
\end{equation}
where
\begin{eqnarray}
\beta_{2\bot}^{({\rm BT})} & = & -\frac{1}{2}\,\tau^{2} - \bhat\bdot\nabla\btimes{\bf R} + \langle\alpha_{1}^{2}\rangle + \frac{1}{2}\bhat\bdot\nabla\btimes\left(\bhat\btimes\nabla\ln B\right) - \left|\bhat\btimes\nabla\ln B\right|^{2}, \label{eq:bot_BT} \\
\beta_{2\|}^{({\rm BT})} & = & -\;2\langle\alpha_{1}^{2}\rangle \;-\; 3\;\vb{\kappa}\bdot\left(\nabla\ln B \;-\frac{}{} \vb{\kappa}\right) \;+\; \nabla\bdot\vb{\kappa}.
\label{eq:par_BT}
\end{eqnarray}
By using the identities \eqref{eq:b_curl_R}-\eqref{eq:alpha1_square}, we obtain the following explicit expressions for the Brizard-Tronko coefficients \eqref{eq:bot_BT}-\eqref{eq:par_BT}:
\begin{eqnarray}
\beta_{2\bot}^{({\rm BT})} & = & \frac{1}{2}\bhat\bdot\nabla\btimes\left(\bhat\btimes\nabla\ln B\right) - \left|\bhat\btimes\nabla\ln B\right|^{2} 
- \frac{1}{4}\;\nabla\bdot\left[\vb{\kappa} \;-\frac{}{} \bhat\;\left(\nabla\bdot\bhat\right)\right] \nonumber \\
 &  &+\; \frac{1}{8}\,\left[ \left(\nabla\bdot\bhat\right)^{2} \;-\frac{}{} 3\;\tau^{2} \right], \label{eq:bot_BT_final} \\
\beta_{2\|}^{({\rm BT})} & = & -\; 3\;\vb{\kappa}\bdot\left(\nabla\ln B \;-\frac{}{} \vb{\kappa}\right) + \frac{1}{2}\,\nabla\bdot\left[ 
\vb{\kappa} \;+\frac{}{} \bhat\;(\nabla\bdot\bhat)\right] - \frac{1}{4}\,\left[ \left(\nabla\bdot\bhat\right)^{2} \;+\frac{}{} \tau^{2} \right].
\label{eq:par_BT_final}
\end{eqnarray}
We will now compare these coefficients with those obtained by \cite{Burby_SQ_2013} and \cite{Parra_Calvo_2011}. We note, however, that these previous results assume that $\vb{\Pi}_{1} \equiv -\,\frac{1}{2}\,J\,\tau\;\bhat$ (i.e., $\vb{\Pi}_{1\bot} \equiv 0$).

\subsection{Burby, Squire, and Qin results}

While the details of the guiding-center transformation are not explicitly presented in the work of \cite{Burby_SQ_2013}, the second-order guiding-center Hamiltonian is presented in both the first-order Hamiltonian representation [see Eqs.~(30)-(31) of \citep{Burby_SQ_2013}] and the first-order symplectic representation [see Eqs.~(33)-(35) of \citep{Burby_SQ_2013}]. In the latter case, \cite{Burby_SQ_2013} use the first-order symplectic representation $\vb{\Pi}_{1} = -\;\frac{1}{2}\,J\tau\;\bhat$, with $\vb{\Pi}_{1\bot} \equiv 0$ in agreement with Littlejohn's choice.

The second-order guiding-center Hamiltonian derived by \cite{Burby_SQ_2013} in the first-order symplectic representation is expressed as
\begin{equation}
\Psi_{2({\rm BSQ})} \;\equiv\; \frac{1}{2}\;J\,\Omega \left( \frac{J}{m\Omega} \;\beta_{2\bot}^{({\rm BSQ})} \;+\; \varrho_{\|}^{2}\;
\beta_{2\|}^{({\rm BSQ})}\right) \;-\; \frac{p_{\|}^{2}}{2m}\;\varrho_{\|}^{2}|\vb{\kappa}|^{2},
\label{eq:BSQ}
\end{equation}
with the second-order coefficients
\begin{eqnarray}
\beta_{2\bot}^{({\rm BSQ})} & = & \frac{1}{2} \left[ |\bhat\btimes\nabla\ln B|^{2} \;+\; \bhat\bdot\nabla\btimes\left(\bhat\btimes\nabla\ln B \right) 
\;-\; 3 \left( |\bhat\btimes\nabla\ln B|^{2} \;+\; \left(\nabla\bdot\bhat\right)^{2} \right) \right] \nonumber \\
 &  &+\; \frac{1}{8} \left[ \nabla\bhat:\nabla\bhat \;-\frac{}{} 3\,\nabla\bhat:\left(\nabla\bhat\right)^{\top} \;+\; 3\,|\vb{\kappa}|^{2} \;+\; 15\;
\left(\nabla\bdot\bhat\right)^{2} \right], \label{eq:bot_BSQ} \\
\beta_{2\|}^{({\rm BSQ})} & = & \frac{1}{4} \left[ 3\,\nabla\bhat:\nabla\bhat \;-\; \nabla\bhat:(\nabla\bhat)^{\top} \;+\; 
\left(\nabla\bdot\bhat\right)^{2} \;+\; |\vb{\kappa}|^{2} \right] \nonumber \\
 &  &-\; 3\,\vb{\kappa}\bdot\left(\nabla\ln B \;-\frac{}{} \vb{\kappa}\right) \;+\; \bhat\bdot\nabla\left(\nabla\bdot\bhat\right).
\label{eq:par_BSQ}
\end{eqnarray}
By using the identities \eqref{eq:nablab_nablab_id}-\eqref{eq:nablab_T_nablab_id}, we readily find
\begin{equation}
\left. \begin{array}{rcl}
\beta_{2\bot}^{({\rm BSQ})} & = & \beta_{2\bot}^{({\rm BT})} \\
 &  & \\
\beta_{2\|}^{({\rm BSQ})} & = & \beta_{2\|}^{({\rm BT})}
\end{array} \right\}.
\label{eq:BSQ-BT}
\end{equation}
Since the Burby-Squire-Qin second-order guiding-center Hamiltonian is exactly equal to ours, it can be concluded that its derivation is based on an identical set of guiding-center coordinates.

\subsection{Parra-Calvo results}

The phase-space transformation derived by \cite{Parra_Calvo_2011} proceeds by a standard iterative method that also combines elements of guiding-center and gyrocenter dynamics. This work only considers the 
first-order symplectic representation $\vb{\Pi}_{1} = -\,\frac{1}{2}\,J\,\tau\;\bhat$ [see Eq.~(104)]. The second-order guiding-center Hamiltonian derived by \cite{Parra_Calvo_2011} is expressed as
\begin{equation}
\Psi_{2({\rm PC})} \;\equiv\; \frac{1}{2}\;J\,\Omega \left( \frac{J}{m\Omega} \;\beta_{2\bot}^{({\rm PC})} \;+\; \varrho_{\|}^{2}\;
\beta_{2\|}^{({\rm PC})}\right) \;-\; \frac{p_{\|}^{2}}{2m}\;\varrho_{\|}^{2}|\vb{\kappa}|^{2},
\label{eq:PC}
\end{equation}
with the second-order coefficients
\begin{eqnarray}
\beta_{2\bot}^{({\rm PC})} & = & \frac{1}{2B}({\bf I} - \bhat\bhat):\nabla\nabla{\bf B}\bdot\bhat \;-\; \frac{3}{2B^{2}}|\nabla_{\bot}B|^{2} \;+\; 
\frac{1}{4} \nabla_{\bot}\bhat:(\nabla_{\bot}\bhat)^{\top} \nonumber \\
 &  &-\; \frac{1}{8}\,\left[(\nabla\bdot\bhat)^{2} \;+\frac{}{} \tau^{2}\right], \label{eq:bot_PC} \\
\beta_{2\|}^{({\rm PC})} & = & -\,3\vb{\kappa}\bdot(\nabla\ln B - \vb{\kappa}) \;+\; \left( \nabla\bhat:\nabla\bhat - \frac{1}{2}
\nabla_{\bot}\bhat:(\nabla_{\bot}\bhat)^{\top} \right) \nonumber \\
 &  &-\; \frac{1}{4}\,\left[ 3\;(\nabla\bdot\bhat)^{2} \;-\frac{}{} \tau^{2}\right].
\label{eq:par_PC}
\end{eqnarray}
In order to compare the Parra-Calvo second-order Hamiltonian \eqref{eq:PC} with our second-order Hamiltonian, we will need the identities
\eqref{eq:nablab_nablab_id}-\eqref{eq:nablab_T_nablab_id} and the following identities
\begin{eqnarray*}
B^{-1}({\bf I} - \bhat\bhat):\nabla\nabla{\bf B}\bdot\bhat & = & |\bhat\btimes\nabla\ln B|^{2} \;+\; \bhat\bdot\nabla\btimes\left(\bhat\btimes\nabla\ln B\right) \;-\; \left(\nabla\bdot\bhat\right)^{2} \nonumber \\
 &  &-\; \nabla\bhat:(\nabla\bhat)^{\top} \;+\; |\vb{\kappa}|^{2},
\end{eqnarray*}
and
\[ \nabla_{\bot}\bhat:(\nabla_{\bot}\bhat)^{\top} \;=\; \nabla\bhat:(\nabla\bhat)^{\top} \;-\; |\vb{\kappa}|^{2} \;=\; \nabla\bhat:(\nabla\bhat) 
\;+\; \tau^{2}. \]
By using these identities, we obtain the following explicit expressions for the Parra-Calvo coefficients \eqref{eq:bot_PC}-\eqref{eq:par_PC}:
\begin{eqnarray}
\beta_{2\bot}^{({\rm PC})} & = & \frac{1}{2}\bhat\bdot\nabla\btimes\left(\bhat\btimes\nabla\ln B\right) - \left|\bhat\btimes\nabla\ln B\right|^{2} 
- \frac{1}{4}\;\nabla\bdot\left[\vb{\kappa} \;-\frac{}{} \bhat\;\left(\nabla\bdot\bhat\right)\right] \nonumber \\
 &  &-\; \frac{1}{8}\,\left[ 7\;\left(\nabla\bdot\bhat\right)^{2} \;+\frac{}{} 3\;\tau^{2} \right], \label{eq:bot_PC_final} \\
\beta_{2\|}^{({\rm PC})} & = & -\; 3\;\vb{\kappa}\bdot\left(\nabla\ln B \;-\frac{}{} \vb{\kappa}\right) + \frac{1}{2}\,\nabla\bdot\left[ 
\vb{\kappa} \;-\frac{}{} \bhat\;(\nabla\bdot\bhat)\right] - \frac{1}{4}\,\left[ \left(\nabla\bdot\bhat\right)^{2} \;+\frac{}{} \tau^{2} \right].
\label{eq:par_PC_final}
\end{eqnarray}
By comparing Eqs.~\eqref{eq:bot_PC_final}-\eqref{eq:par_PC_final} with Eqs.~\eqref{eq:bot_BT_final}-\eqref{eq:par_BT_final}, we obtain the differences
\begin{eqnarray}
\beta_{2\bot}^{({\rm BT})} - \beta_{2\bot}^{({\rm PC})} & = & \left(\nabla\bdot\bhat\right)^{2}, \label{eq:BT-PC_bot} \\
\beta_{2\|}^{({\rm BT})} - \beta_{2\|}^{({\rm PC})} & = & \nabla\bdot\left[\bhat\frac{}{}\left(\nabla\bdot\bhat\right)\right] \;=\;
\left(\nabla\bdot\bhat\right)^{2} \;+\; \bhat\bdot\nabla\left(\nabla\bdot\bhat\right). \label{eq:BT-PC_par}
\end{eqnarray}

In more recent work, \cite{PCBSQ_2014} showed that the second-order Hamiltonian difference between works of \cite{Parra_Calvo_2011} and \cite{Burby_SQ_2013}:
\begin{eqnarray}
\Psi_{2({\rm PC})} - \Psi_{2({\rm BSQ})} & = & -\;\frac{J^{2}}{2m}\left(\nabla\bdot\bhat\right)^{2} - \frac{1}{2}\,J\Omega\varrho_{\|}^{2}\;
\nabla\bdot\left[\bhat\;\left(\nabla\bdot\bhat\right)\right] \nonumber \\
 & = & -\;\frac{d_{0}}{dt}\left[ \frac{J}{2}\,\varrho_{\|}\;\left(\nabla\bdot\bhat\right)\right] \equiv -\;\frac{d_{0}\langle\sigma_{3}\rangle}{dt},
\label{eq:PC_gauge}
\end{eqnarray}
could be explained, using our notation, by adding the gyroangle-independent gauge function 
\begin{equation}
\langle\sigma_{3}\rangle \;\equiv\; \frac{1}{2}\,J\,\varrho_{\|}\;(\nabla\bdot\bhat) \;=\; \frac{d_{0}}{dt}\left(\frac{J}{2\,\Omega}\right)
\label{eq:sigma3_ave}
\end{equation}
in Eq.~\eqref{eq:sigma3_sol}. Hence, according to Eq.~\eqref{eq:Gamma_3_p}, this new gauge term introduces the following change in $G_{2\|}^{\bf x}$, according to Eq.~\eqref{eq:G2_xpar_eq}: 
\begin{equation}
G_{2\|}^{\bf x} \;\rightarrow\; \ov{G}_{2\|}^{\bf x} \;\equiv\; G_{2\|}^{\bf x} \;-\; \pd{\langle\sigma_{3}\rangle}{p_{\|}} \;=\; G_{2\|}^{\bf x} \;-\; \frac{J}{2\,m\Omega}\;\left(\nabla\bdot\bhat\right),
\label{eq:G2par_x_new}
\end{equation}
so that Eq.~\eqref{eq:G2x_exp} yields the change
\begin{equation}
\ov{G}_{2}^{\bf x} \;\equiv\; G_{2}^{\bf x} \;-\; \frac{J\,(\nabla\bdot\bhat)}{2m\,\Omega}\;\bhat, 
\end{equation}
and, thus, the new first-order gyroradius is now given as
\begin{equation}
\ov{\vb{\rho}}_{1} \;\equiv\; \vb{\rho}_{1} \;+\; \frac{J\,(\nabla\bdot\bhat)}{2m\,\Omega}\;\bhat.
\label{eq:ov_rho_1}
\end{equation}
We note that, according to Eq.~\eqref{eq:rho1_ave}, we now find $\bhat\bdot\langle\ov{\vb{\rho}}_{1}\rangle \equiv 0$. 

Lastly, the gyroangle-independent gauge function \eqref{eq:sigma3_ave} also yields the following change in $G_{2}^{p_{\|}}$, according to 
Eq.~\eqref{eq:G2_p||}:
\begin{equation}
G_{2}^{p_{\|}} \;\rightarrow\; \ov{G}_{2}^{p_{\|}} \;\equiv\; G_{2}^{p_{\|}} \;+\; \bhat\bdot\nabla\langle\sigma_{3}\rangle \;=\;  G_{2}^{p_{\|}} \;+\; \frac{1}{2}\;J\,\varrho_{\|}\;\nabla\bdot\left[\bhat\frac{}{} \left(\nabla\bdot\bhat\right)\right],
\label{eq:G2p_new}
\end{equation}
while $G_{1}^{J}$ is unchanged, according to Eq.~\eqref{eq:Gamma_3_p}, because $\partial\langle\sigma_{3}\rangle/\partial\theta \equiv 0$. With 
$\ov{\Psi}_{2} \equiv \Psi_{2({\rm PC})}$, we immediately note that $G_{2}^{J}$ remains unchanged according to Eq.~\eqref{eq:G_2_mu_tilde}, and that the Jacobian constraint \eqref{eq:Jac_2} is still satisfied since
\[ \frac{1}{B}\;\nabla\bdot\left(\langle \ov{G}_{2}^{\bf x}\rangle\;B\right) \;+\; \pd{\ov{G}_{2}^{p_{\|}}}{p_{\|}} \;=\;
\frac{1}{B}\;\nabla\bdot\left(\langle G_{2}^{\bf x}\rangle\;B\right) \;+\; \pd{G_{2}^{p_{\|}}}{p_{\|}}. \]
Hence, by extending the class of Lie-transform perturbation theories with the inclusion of gyroangle-independent gauge functions (i.e.,
$\langle\sigma_{n}\rangle \neq 0$) in Sec.~\ref{sec:Lie}, we introduce an additional degree of freedom in the equivalence between guiding-center Hamiltonian theories.

\section{\label{sec:Psi2_phys}Physical Interpretation of the Second-order Guiding-center Hamiltonian $\Psi_{2}$}

In this Appendix, we provide a physical interpretation of the second-order guiding-center Hamiltonian $\Psi_{2}$. Using the guiding-center push-forward Lagrangian constraint (see Sec.~\ref{sec:push_Lag}), we begin with the definition of the guiding-center Hamiltonian through the guiding-center push-forward 
\begin{equation}
H_{\rm gc} \;\equiv\; {\sf T}_{\rm gc}^{-1}\left(\frac{|{\bf p}|^{2}}{2m}\right) \;=\; \frac{|{\sf T}_{\rm gc}^{-1}{\bf p}|^{2}}{2\;m} \;=\;
\frac{p_{\|}^{2}}{2m} \;+\; J\,\Omega \;+\; \epsilon^{2}\,\Psi_{2},
\label{eq:Hgc_push}
\end{equation}
where we are using the first-order symplectic representation $\Psi_{1} \equiv 0$. While the definition does not require a gyroangle average, we shall use one here in order to remove terms that will cancel out anyway. Using the identity \eqref{eq:gcLc_id}, we therefore obtain
\begin{eqnarray}
H_{\rm gc} & \equiv & \frac{m}{2} \left\langle \left|\frac{d_{\rm gc}{\bf X}}{dt} + \frac{d_{\rm gc}\vb{\rho}_{\rm gc}}{dt}\right|^{2}\right\rangle 
\nonumber \\
 & = & \frac{m}{2} \left(\left|\frac{d_{\rm gc}{\bf X}}{dt}\right|^{2} + \left\langle \left|\frac{d_{\rm gc}\vb{\rho}_{\rm gc}}{dt}\right|^{2}\right\rangle \right) + m\;\frac{d_{\rm gc}{\bf X}}{dt}\bdot\left\langle\frac{d_{\rm gc}\vb{\rho}_{\rm gc}}{dt}\right\rangle.
 \label{eq:gc_H_push}
\end{eqnarray}
where the guiding-center kinetic energy
\begin{equation}
\frac{m}{2}\,\left|\frac{d_{\rm gc}{\bf X}}{dt}\right|^{2} \;=\; \frac{p_{\|}^{2}}{2m} \;+\; \epsilon^{2} \left( \frac{m}{2}\,|{\bf v}_{\rm gc}|^{2}
\;+\; p_{\|}\;\pd{\Psi_{2}}{p_{\|}} \right) 
\label{eq:gc_kinetic}
\end{equation}
includes the second-order guiding-center kinetic energy associated with the guiding-center drift velocity and the second-order Ba\~{n}os parallel drift $p_{\|}\,\partial\Psi_{2}/\partial
p_{\|}$, while 
\begin{equation} 
m\;\frac{d_{\rm gc}{\bf X}}{dt}\bdot\left\langle\frac{d_{\rm gc}\vb{\rho}_{\rm gc}}{dt}\right\rangle  \;=\; \epsilon^{2}\;
p_{\|}\,\bhat\bdot\frac{d_{0}\langle\vb{\rho}_{1}\rangle}{dt}.
\label{eq:X_rho_dt}
\end{equation}
Lastly, the ``gyration'' kinetic energy is
\begin{eqnarray}
\frac{m}{2} \left\langle \left|\frac{d_{\rm gc}\vb{\rho}_{\rm gc}}{dt}\right|^{2}\right\rangle & = & J\,\Omega \;+\; \epsilon\;\left\langle
{\bf p}_{\bot}\bdot\left( \Omega\,\pd{\vb{\rho}_{1}}{\theta} \;+\; \frac{d_{0}\vb{\rho}_{0}}{dt}\right) \right\rangle \\
 &  &+\; \epsilon^{2} \left\langle {\bf p}_{\bot}\bdot\left( \Omega\,\pd{\vb{\rho}_{2}}{\theta} \;+\; \pd{\Psi_{2}}{J}\,\pd{\vb{\rho}_{0}}{\theta}
\;+\; \frac{d_{0}\vb{\rho}_{1}}{dt} \;+\; \frac{d_{1}\vb{\rho}_{0}}{dt} \right)\right\rangle \nonumber \\
 &  &+\; \epsilon^{2}\,\frac{m}{2}\;\left\langle\left|\Omega\,\pd{\vb{\rho}_{1}}{\theta} \;+\; \frac{d_{0}\vb{\rho}_{0}}{dt}\right|^{2}\right\rangle, \nonumber 
\end{eqnarray}
where
\begin{equation} 
\left\langle {\bf p}_{\bot}\bdot\left( \pd{\Psi_{2}}{J}\,\pd{\vb{\rho}_{0}}{\theta}\right)\right\rangle \;=\; 2\,J\;\pd{\Psi_{2}}{J},
\label{eq:Omega_2}
\end{equation}
and
\begin{eqnarray} 
\Omega\;\pd{\vb{\rho}_{1}}{\theta} \;+\; \frac{d_{0}\vb{\rho}_{0}}{dt} & = & \frac{p_{\|}}{m} \left[ (\vb{\rho}_{0}\bdot\vb{\kappa})\,\bhat \;+\;
\left( \frac{1}{2}\,\nabla\bdot\bhat \;-\; \alpha_{2} \right) \vb{\rho}_{0} \;-\;  \frac{1}{2}\,\alpha_{1}\;\pd{\vb{\rho}_{0}}{\theta} \right] 
\nonumber \\
 &  &+\; \frac{J}{m} \left( \alpha_{1}\;\bhat \;-\frac{}{} 2\;{\sf a}_{1}\bdot\nabla\ln B \right),
\label{eq:v_perp_1}
\end{eqnarray}
Here, it is a simple task to show that the first-order terms vanish
\[ \left\langle{\bf p}_{\bot}\bdot\left( \Omega\,\pd{\vb{\rho}_{1}}{\theta} \;+\; \frac{d_{0}\vb{\rho}_{0}}{dt}\right) \right\rangle \;=\; 0. \]
Hence, we see that the higher-order terms associated with magnetic nonuniformity enter at the second order.

By combining the remaining components in Eqs.~\eqref{eq:gc_kinetic}-\eqref{eq:v_perp_1}, we now obtain the second-order equation
\begin{eqnarray}
\Psi_{2} & \equiv & \frac{m}{2}\,|{\bf v}_{\rm gc}|^{2} \;+\; \left( p_{\|}\;\pd{\Psi_{2}}{p_{\|}} \;+\; 2\,J\;\pd{\Psi_{2}}{J}\right) \nonumber \\
 &  &+\; \left[ \Psi_{2(A)} \;+\frac{}{} \Psi_{2(B)} \;+\; \Psi_{2(C)} \;+\; \Psi_{2(D)} \;+\; \Psi_{2(E)} \right], 
\label{eq:Psi2_func} 
\end{eqnarray}
where we defined
\begin{eqnarray}
\Psi_{2(A)} & \equiv & \frac{m}{2}\;\left\langle \left|\Omega\,\pd{\vb{\rho}_{1}}{\theta} \;+\; \frac{d_{0}\vb{\rho}_{0}}{dt}\right|^{2}\right\rangle,
\label{eq:Psi2_A} \\
\Psi_{2(B)} & \equiv & p_{\|}\;\bhat\bdot\frac{d_{0}\langle\vb{\rho}_{1}\rangle}{dt}, \label{eq:Psi2_B} \\
\Psi_{2(C)} & \equiv & \left\langle {\bf p}_{\bot}\bdot\frac{d_{1}\vb{\rho}_{0}}{dt} \right\rangle, \label{eq:Psi2_C} \\
\Psi_{2(D)} & \equiv & \left\langle {\bf p}_{\bot}\bdot\frac{d_{0}\vb{\rho}_{1}}{dt} \right\rangle, \label{eq:Psi2_D} \\
\Psi_{2(E)} & \equiv & \left\langle {\bf p}_{\bot}\bdot\left(\Omega\,\pd{\vb{\rho}_{2}}{\theta}\right)\right\rangle. \label{eq:Psi2_E}
\end{eqnarray}

First, using Eq.~\eqref{eq:v_perp_1}, we find
\begin{eqnarray}
\Psi_{2(A)} & = & \frac{J^{2}}{2m} \left( \langle\alpha_{1}^{2}\rangle \;+\frac{}{} |\nabla_{\bot}\ln B|^{2} \right) \;+\; \frac{1}{2}\,J\Omega\;
\varrho_{\|}^{2} \left[ |\vb{\kappa}|^{2} \;+\; \langle\alpha_{1}^{2}\rangle \;+\; \frac{1}{2}\,(\nabla\bdot\bhat)^{2} \right].
\label{eq:one_Psi2} 
\end{eqnarray}
Second, using Eq.~\eqref{eq:rho1_ave}, we find
\begin{eqnarray}
\Psi_{2(B)}  & = & \frac{p_{\|}^{2}}{m} \left[ \bhat\bdot\nabla\left(\langle\vb{\rho}_{1}\rangle
\bdot\bhat\right) \;-\frac{}{} \vb{\kappa}\bdot\langle\vb{\rho}_{1}\rangle \right] \nonumber \\
 & = & -\,\frac{1}{2}\;J\Omega\,\varrho_{\|}^{2}\;\left\{ \nabla\bdot\left[\bhat\;\left(\nabla\bdot\bhat\right)\right] \;-\; 
3\;\vb{\kappa}\bdot\nabla\ln B \right\} \;+\; \frac{p_{\|}^{2}}{m}\;\varrho_{\|}^{2}|\vb{\kappa}|^{2} \nonumber \\
 &  &-\; \vb{\Pi}_{1}\bdot\frac{\bhat}{m}\btimes\left( \frac{p_{\|}^{2}}{m\Omega}\;\vb{\kappa} \right).
\label{eq:two_Psi2}
\end{eqnarray}
Third, using
\begin{eqnarray*}
\frac{d_{1}\vb{\rho}_{0}}{dt} \;\equiv\; {\bf v}_{\rm gc}\bdot\nabla_{0}^{*}\vb{\rho}_{0} & = & -\,J\,{\bf v}_{\rm gc}\bdot\nabla\ln B\;
\pd{\vb{\rho}_{0}}{J} \;+\; \frac{1}{2}\;{\bf v}_{\rm gc}\bdot\nabla\btimes\bhat\;\pd{{\rho}_{0}}{\theta} \nonumber \\
 &  &-\; {\bf v}_{\rm gc}\bdot\left( C_{\bot\rho}\,\pd{{\rho}_{0}}{\theta} + C_{\rho\rho}\,\vb{\rho}_{0}\right)\;\bhat, 
\end{eqnarray*}
we find
\begin{eqnarray} 
\Psi_{2(C)} & = & \left( \frac{1}{2}\,{\bf v}_{\rm gc}\bdot\nabla\btimes\bhat\right)\left\langle\pd{\vb{\rho}_{0}}{\theta}\bdot{\bf p}_{\bot}
\right\rangle \;=\; \left( J\frac{}{}\nabla\btimes\bhat\right)\bdot{\bf v}_{\rm gc} \;\equiv\; -2\,\vb{\Pi}_{1}\bdot{\bf v}_{\rm gc}.
\label{eq:three_Psi2}
\end{eqnarray}
Hence, both Eqs.~\eqref{eq:two_Psi2}-\eqref{eq:three_Psi2} contain direct contributions of the polarization term $\vb{\Pi}_{1\bot}$ to the second-order guiding-center Hamiltonian. 

Fourth, we find
\begin{eqnarray*}
\Psi_{2(D)} & = & \frac{d_{0}}{dt}\;\langle{\bf p}_{\bot}\bdot\vb{\rho}_{1}\rangle \;-\; \left\langle \frac{d_{0}{\bf p}_{\bot}}{dt}\bdot\vb{\rho}_{1} \right\rangle \;=\; -\;\left\langle \frac{d_{0}{\bf p}_{\bot}}{dt}\bdot\vb{\rho}_{1} \right\rangle \\
 & = & -\,p_{\|}\Omega\;\left[ \left\langle \left(\vb{\kappa}\bdot\vb{\rho}_{0}\right)\;\bhat\bdot\pd{\vb{\rho}_{1}}{\theta} \right\rangle \;-\; 
\frac{1}{2}\,\tau\;\left\langle \pd{\vb{\rho}_{0}}{\theta}\bdot\pd{\vb{\rho}_{1}}{\theta}\right\rangle \right],
\end{eqnarray*}
where we used $\langle{\bf p}_{\bot}\bdot\vb{\rho}_{1}\rangle \;=\; -\;m\Omega\,\langle\vb{\rho}_{0}\bdot\partial\vb{\rho}_{1}/\partial\theta\rangle \equiv 0$, and 
\[ \frac{d_{0}{\bf p}_{\bot}}{dt} \;\equiv\; \frac{p_{\|}}{m}\bhat\bdot\nabla_{0}^{*}{\bf p}_{\bot} \;=\; -\;\frac{p_{\|}}{m} \left[ \frac{1}{2}\,
(\nabla\bdot\bhat)\;{\bf p}_{\bot} \;-\; \frac{1}{2}\,\tau\;\pd{{\bf p}_{\bot}}{\theta} \;+\; ({\bf p}_{\bot}\bdot\vb{\kappa})\;\bhat \right]. \]
Here,
\[ -\;p_{\|}\,\Omega\;\left\langle \left(\vb{\kappa}\bdot\vb{\rho}_{0}\right)\;\bhat\bdot\pd{\vb{\rho}_{1}}{\theta} \right\rangle \;=\; 
-\;2\;J\,\Omega\;\varrho_{\|}^{2}\,|\vb{\kappa}|^{2}\]
and
\[ \frac{1}{2}\,p_{\|}\,\Omega\;\tau\;\left\langle \pd{\vb{\rho}_{0}}{\theta}\bdot\pd{\vb{\rho}_{1}}{\theta}\right\rangle \;=\; -\;\frac{1}{2}\,
J\,\Omega\;\varrho_{\|}^{2}\,\tau^{2}. \]
Hence, we find
\begin{equation}
\Psi_{2(D)} \;=\; -\;J\Omega\;\varrho_{\|}^{2} \left( 2\,|\vb{\kappa}|^{2} \;+\; \frac{1}{2}\,\tau^{2} \right).
\label{eq:four_Psi2}
\end{equation}

Lastly, we write
\begin{eqnarray}
\Psi_{2(E)} & = &  m\Omega^{2}\, \langle\vb{\rho}_{0}\bdot\vb{\rho}_{2}\rangle \;\equiv\; \Psi_{2(E)}^{(1)} + \Psi_{2(E)}^{(2)} + \Psi_{2(E)}^{(3)},
\label{eq:pperp_rho2}
\end{eqnarray}
where
\begin{eqnarray}
\Psi_{2(E)}^{(1)} \;\equiv\; -\,m\Omega^{2}\langle\vb{\rho}_{0}\bdot G_{3}^{\bf x}\rangle  & = & -\,p_{\|}\Omega \left\langle \left(\vb{\kappa}\bdot
\pd{\vb{\rho}_{0}}{\theta}\right)\;G_{2\|}^{\bf x} - \tau\;\left(\vb{\rho}_{0}\bdot G_{2}^{\bf x}\right) \right\rangle \nonumber \\
 &  &+\; \left\langle \Omega\,\pd{\vb{\rho}_{0}}{\theta}\bdot\left[ D_{1}^{2}({\bf P}_{3}) + \nabla\sigma_{3}\right] \right\rangle, 
\label{eq:E1}
\end{eqnarray}
which makes use of $G_{3}^{\bf X}$,
\begin{eqnarray}
 \Psi_{2(E)}^{(2)} & \equiv & -\,m\Omega^{2} \left\langle \vb{\rho}_{0}\bdot\left({\sf G}_{2}\cdot\frac{}{}\exd\vb{\rho}_{0}\right) \right\rangle \;=\;
-\;\frac{m\Omega^{2}}{2}\;\left\langle {\sf G}_{2}\cdot\exd\left(\frac{2\,J}{m\Omega}\right) \right\rangle \label{eq:E2} \\
 & = & -\,\Omega\;\langle G_{2}^{J}\rangle \;+\; J\,\Omega\;\langle G_{2}^{\bf x}\rangle\bdot\nabla\ln B, \nonumber
\end{eqnarray}
which makes use of the components of ${\sf G}_{2}$, and
\begin{eqnarray}
\Psi_{2(E)}^{(3)} & \equiv & \frac{1}{6}\,m\Omega^{2} \left\langle \vb{\rho}_{0}\bdot\left[{\sf G}_{1}\cdot\frac{}{}\exd\left({\sf G}_{1}\cdot\exd
\vb{\rho}_{0}\right) \right] \right\rangle \label{eq:E3} \\
 & = & \frac{1}{6}\;B\Omega \left\langle {\sf G}_{1}\cdot\exd\left[ B^{-1}\;\left( G_{1}^{J} \;+\frac{}{} J\;\vb{\rho}_{0}\bdot\nabla\ln B \right) 
\right] \right\rangle \nonumber \;-\; \frac{1}{6}\,m\, \Omega^{2} \left\langle \left|{\sf G}_{1}\cdot\exd\vb{\rho}_{0}\right|^{2} \right\rangle.
\nonumber
\end{eqnarray}

In Eq.~\eqref{eq:E1}, we find
\begin{equation}
-\,p_{\|}\Omega \left\langle \left(\vb{\kappa}\bdot\pd{\vb{\rho}_{0}}{\theta}\right)\;G_{2\|}^{\bf x} - \tau\;\left(\vb{\rho}_{0}\bdot G_{2}^{\bf x}
\right) \right\rangle \;=\; -\;J\,\Omega\;\varrho_{\|}^{2} \left( 2\,|\vb{\kappa}|^{2} \;-\; \frac{3}{2}\,\tau^{2} \right).
\label{eq:E1a}
\end{equation}
Using Eq.~\eqref{eq:sigma3_sol}, we also find
\begin{eqnarray}
\left\langle\pd{\vb{\rho}_{0}}{\theta}\bdot\nabla\sigma_{3}\right\rangle & = & \nabla\bdot\left\langle\pd{\vb{\rho}_{0}}{\theta}\;\sigma_{3}\right\rangle
\;-\; \left\langle \sigma_{3}\;\left(\nabla\bdot\pd{\vb{\rho}_{0}}{\theta}\right)\right\rangle 
\label{eq:E1c} \\
 & = & -\;\frac{1}{3}\,p_{\|} \left[ \nabla\bdot\left\langle\pd{\vb{\rho}_{0}}{\theta}\;G_{2\|}^{\bf x}\right\rangle
\;-\; \left\langle G_{2\|}^{\bf x}\;\left(\nabla\bdot\pd{\vb{\rho}_{0}}{\theta}\right)\right\rangle \right] \nonumber \\
 & = & -\;\frac{2}{3}\,J\,\varrho_{\|}^{2} \left[ B^{2}\;\nabla\bdot\left(\frac{\vb{\kappa}}{B^{2}}\right) \;+\; \vb{\kappa}\bdot\left( \frac{1}{2}\,
\nabla\ln B \;+\; \vb{\kappa} \;+\; \bhat\btimes{\bf R} \right) \right]. \nonumber 
\end{eqnarray}
Next, we need
\begin{eqnarray*}
\left\langle \pd{\vb{\rho}_{0}}{\theta}\bdot D_{1}^{2}({\bf P}_{3})\right\rangle & = & \frac{2J}{m\Omega}\;\bhat\bdot\nabla\btimes 
\langle D_{1}({\bf P}_{3})\rangle \;+\;  \left\langle G_{1}^{p_{\|}}\;\pd{}{p_{\|}}\left(D_{1}({\bf P}_{3})\bdot\pd{\vb{\rho}_{0}}{\theta}\right)\right\rangle \nonumber \\
 &  &+\;  \left\langle G_{1}^{J}\;\left[\pd{}{J}\left(D_{1}({\bf P}_{3})\bdot\pd{\vb{\rho}_{0}}{\theta}\right) \;-\; D_{1}({\bf P}_{3})\bdot
\frac{\partial^{2}\vb{\rho}_{0}}{\partial J\partial\theta}\right] \right\rangle \\
 &  &+\; \left\langle G_{1}^{\theta}\;\left[\pd{}{\theta}\left(D_{1}({\bf P}_{3})\bdot\pd{\vb{\rho}_{0}}{\theta}\right) \;+\; D_{1}({\bf P}_{3})\bdot
\vb{\rho}_{0}\right]\right\rangle, 
\end{eqnarray*}
where
\begin{eqnarray*}
\langle D_{1}({\bf P}_{3})\rangle & = & \frac{1}{3}\,J \left( 2\,{\bf R} \;+\; \frac{1}{2}\,\tau\;\bhat\right) \;-\; \frac{\bhat}{3}\btimes\left( J\;
\nabla\ln B \;+\; \frac{p_{\|}^{2}}{m\Omega}\;\vb{\kappa}\right), \\
D_{1}({\bf P}_{3})\bdot\pd{\vb{\rho}_{0}}{\theta} & = & \frac{1}{3}\,J\,\varrho_{\|} (2\,\tau - \alpha_{1}) \;+\; \frac{1}{3}\,\vb{\rho}_{0}\bdot
\left( 2\,J\;\bhat\btimes{\bf R} \;+\; \frac{p_{\|}^{2}}{m\Omega}\;\vb{\kappa} \right), \\
D_{1}({\bf P}_{3})\bdot\vb{\rho}_{0} & = & \frac{1}{3}\;D_{1}({\bf p}_{\bot})\bdot\vb{\rho}_{0} \;=\; -\;\frac{2}{3}\,J\; G_{1}^{\theta}.
\end{eqnarray*}
First, we find
\begin{eqnarray} 
\frac{2J}{m\Omega}\;\bhat\bdot\nabla\btimes\langle D_{1}({\bf P}_{3})\rangle & = & \frac{J^{2}}{3\;m\Omega}\;\left( 4\;\bhat\bdot\nabla\btimes{\bf R} 
\;+\; \tau^{2} \right) \nonumber \\
 &  &-\; \frac{2J}{3\,m\Omega}\;\bhat\bdot\nabla\btimes\left[\bhat\btimes \left( J\;\nabla\ln B \;+\; \frac{p_{\|}^{2}}{m\Omega}\;
\vb{\kappa}\right) \right].
\label{eq:E1b1} 
\end{eqnarray}
Next, we find
\begin{eqnarray}
\left\langle G_{1}^{p_{\|}}\;\pd{}{p_{\|}}\left(D_{1}({\bf P}_{3})\bdot\pd{\vb{\rho}_{0}}{\theta}\right)\right\rangle & = & \frac{J^{2}}{3\,m\Omega}\;\left(2\,\tau^{2} \;-\frac{}{} \langle\alpha_{1}^{2}\rangle \right) \;-\; \frac{2}{3}\,J\;\varrho_{\|}^{2}\,
|\vb{\kappa}|^{2},
\label{eq:E1b2} 
\end{eqnarray}
and
\begin{eqnarray}
 &  &\left\langle G_{1}^{J}\left[\pd{}{J}\left(D_{1}({\bf P}_{3})\bdot\pd{\vb{\rho}_{0}}{\theta}\right) - D_{1}({\bf P}_{3})\bdot
\frac{\partial^{2}\vb{\rho}_{0}}{\partial J\partial\theta}\right] \right\rangle \nonumber \\
 &  &\hspace*{1in}=\; -\,\frac{1}{6}\,J\,\varrho_{\|}^{2}\left(2\,\tau^{2} \;-\frac{}{} \langle\alpha_{1}^{2}\rangle \right) 
\label{eq:E1b3} \\
 &  &\hspace*{1.2in}+\; \frac{2\,J}{3\,m\Omega}\;\bhat\btimes{\bf R}\bdot\left( J\,\nabla\ln B \;+\; \frac{p_{\|}^{2}}{m\Omega}\;\vb{\kappa}\right) \nonumber
\end{eqnarray}
and
\begin{eqnarray}
 &  &\left\langle G_{1}^{\theta}\;\left[\pd{}{\theta}\left(D_{1}({\bf P}_{3})\bdot\pd{\vb{\rho}_{0}}{\theta}\right) \;+\; D_{1}({\bf P}_{3})\bdot
\vb{\rho}_{0}\right]\right\rangle \nonumber \\
 &  &\hspace*{1in}=\; \frac{1}{6}\,J\;\varrho_{\|}^{2}\;\left( \langle \alpha_{1}^{2}\rangle \;-\frac{}{} 2\;\vb{\kappa}\bdot\nabla\ln B \right)
\label{eq:E1b4} \\
 &  &\hspace*{1.2in}-\; \frac{2\,J^{2}}{3\,m\Omega}\;\nabla\ln B \bdot\left(\nabla\ln B + \bhat\btimes{\bf R} \right) \nonumber
\end{eqnarray}
Hence, by combining Eqs.~\eqref{eq:E1a}-\eqref{eq:E1b4} into Eq.~\eqref{eq:E1}, we obtain
\begin{eqnarray}
\Psi_{2(E)}^{(1)} & = & \frac{J^{2}}{3\,m} \left[ 4\,\bhat\bdot\nabla\btimes{\bf R} \;+\; 3\;\tau^{2} \;-\; \langle\alpha_{1}^{2}\rangle \;-\; 
2\;|\bhat\btimes\nabla\ln B|^{2} \;-\; 2\;\bhat\bdot\nabla\btimes\left(\bhat\btimes\nabla\ln B\right) \right] \nonumber \\
 &  &-\; \frac{1}{3}\,J\Omega\;\varrho_{\|}^{2} \left[ 4\,\nabla\bdot\vb{\kappa} \;-\; 4\,\vb{\kappa}\bdot\nabla\ln B \;+\; 12\,|\vb{\kappa}|^{2} \;-\;
\frac{7}{2}\,\tau^{2} \;-\; \langle\alpha_{1}^{2}\rangle \right],
\label{eq:Psi2_3}
\end{eqnarray}
where the gyrogauge-dependent terms cancel out. Next, using the definition \eqref{eq:Ham_constraint_2_initial}, Eq.~\eqref{eq:E2} can be expressed as
\begin{eqnarray}
\Psi_{2(E)}^{(2)} & = & \Psi_{2} \;+\; \frac{m}{2}\,|{\bf v}_{\rm gc}|^{2} \;-\; \vb{\Pi}_{1}\bdot{\bf v}_{\rm gc} - J\Omega\;\varrho_{\|}^{2} \left( \frac{1}{2}\,\tau^{2} \;-\; \langle\alpha_{1}^{2}\rangle\right) \label{eq:Psi2_2} \\
 &  &+\; J\,\Omega\;\nabla\ln B\bdot\left[-\; \vb{\Pi}_{1}\btimes\frac{\bhat}{m\Omega} \;+\; \frac{1}{2} \left( \frac{J}{m\Omega}\;\nabla_{\bot}\ln B \;+\;
\varrho_{\|}^{2}\;\vb{\kappa} \right) \right]
\nonumber
\end{eqnarray}

Lastly, the two terms in Eq.~\eqref{eq:E3} are
\begin{eqnarray}
 & &\frac{1}{6}\;B\Omega \left\langle {\sf G}_{1}\cdot\exd\left[ B^{-1}\;\left( G_{1}^{J} \;+\frac{}{} J\;\vb{\rho}_{0}\bdot\nabla\ln B \right) \right] \right\rangle 
\label{eq:E3a} \\
 & = & \frac{J}{6\,m} \left( \frac{1}{2}\,\nabla\ln B - \vb{\kappa} - \bhat\btimes{\bf R}\right) \bdot\left( 2J\;\nabla\ln B \;+\; 
\frac{p_{\|}^{2}}{m\Omega}\;\vb{\kappa} \right) \nonumber \\
 &  &-\; \frac{\Omega}{6}\;\left\{ \nabla\bdot\left[ \frac{J}{m\Omega}\;\left(2\,J\;\nabla_{\bot}\ln B \;+\; \frac{p_{\|}^{2}}{m\Omega}\;\vb{\kappa} \right)\right] \;+\; \frac{J^{2}}{m\Omega}\;\left(\tau^{2} + \langle\alpha_{1}^{2}\rangle \right) \;+\; 2\,J\;\varrho_{\|}^{2}\,|\vb{\kappa}|^{2}
\right\} \nonumber \\
 &  &+\; \frac{1}{6\,m} \left( J\;\nabla\ln B \;+\; \frac{p_{\|}^{2}}{m\Omega}\;\vb{\kappa} \right) \bdot\left( 3J\,\nabla\ln B \;+\; 
\frac{p_{\|}^{2}}{2\;m\Omega}\;\vb{\kappa} \right) \;+\; \frac{1}{6}\,J\Omega\;\varrho_{\|}^{2}\,\left(\tau^{2} \;+\frac{}{} \langle\alpha_{1}^{2}\rangle
\right) \nonumber \\
 &  &+\; \frac{J}{6\,m} \left( \nabla\ln B \;+\; \frac{p_{\|}^{2}\,\vb{\kappa}}{2J\,m\Omega} \;+\; \bhat\btimes{\bf R}\right)\bdot\left( 2J\,\nabla\ln B \;+\; \frac{p_{\|}^{2}}{m\Omega}\;\vb{\kappa} \right) \;+\; \frac{1}{6}\;J\,\Omega\;\varrho_{\|}^{2}\,\langle\alpha_{1}^{2}\rangle, \nonumber
\end{eqnarray}
and
\begin{eqnarray}
-\; \frac{1}{6}\,m\, \Omega^{2} \left\langle \left|{\sf G}_{1}\cdot\exd\vb{\rho}_{0}\right|^{2} \right\rangle   & = & -\frac{J^{2}}{6\,m} \left[ \left(\nabla\bdot\bhat\right)^{2} + 4\;\langle\alpha_{1}^{2}\rangle \right] - \frac{1}{12\,m} \left|
2\,J\;\nabla\ln B + \frac{p_{\|}^{2}}{m\Omega}\;\vb{\kappa}\right|^{2} \nonumber \\
 &  &- \frac{1}{12}J\Omega\;\varrho_{\|}^{2} \left( \tau^{2} +\frac{}{} 2\;\langle\alpha_{1}^{2}\rangle \right) - \frac{J^{2}}{3\,m} 
\left|\nabla_{\bot}\ln B + \frac{p_{\|}^{2}\,\vb{\kappa}}{2J\;m\Omega} \right|^{2}. \label{eq:E3b}
\end{eqnarray}
Hence, combining Eqs.~\eqref{eq:E3a}-\eqref{eq:E3b} into Eq.~\eqref{eq:E3}, we obtain
\begin{eqnarray}
\Psi_{2(E)}^{(3)} & = & \frac{J^{2}}{3\,m} \left[ |\bhat\btimes\nabla\ln B|^{2} \;-\; \vb{\kappa}\bdot\nabla\ln B \;-\; B\;\nabla\bdot\left(B^{-1}\;
\nabla_{\bot}\ln B\right) \right. \nonumber \\
 &  &\left.-\; \frac{1}{2} \left( 5\,\langle\alpha_{1}^{2}\rangle \;+\frac{}{} \tau^{2} \;+\; (\nabla\bdot\bhat)^{2} \right) \right] \;+\; \left.
\frac{1}{3}\,J\Omega\;\varrho_{\|}^{2} \right[ \vb{\kappa}\bdot\nabla\ln B  \nonumber \\
 &  &\left. -\; \frac{1}{2}\,B^{2}\;\nabla\bdot\left(\frac{\vb{\kappa}}{B^{2}}\right) \;-\; \frac{1}{2} \left( 3\,|\vb{\kappa}|^{2} \;-\; \frac{1}{2}\,\tau^{2} \;-\; \langle\alpha_{1}^{2}\rangle\right) \right],
\label{eq:Psi2_1}
\end{eqnarray}
and the gyrogauge-dependent terms have once again cancelled each other. 

We now combine Eqs.~\eqref{eq:one_Psi2}-\eqref{eq:four_Psi2}, \eqref{eq:Psi2_3}-\eqref{eq:Psi2_2}, and \eqref{eq:Psi2_1}, so that we obtain 
\begin{eqnarray*}
\Psi_{2} & \equiv & \frac{m}{2}\,|{\bf v}_{\rm gc}|^{2} \;+\; p_{\|}\;\pd{\Psi_{2}}{p_{\|}} \;+\; 2\,J\;\pd{\Psi_{2}}{J} \;+\; \frac{J^{2}}{2m} \left( \langle\alpha_{1}^{2}\rangle \;+\frac{}{} |\nabla_{\bot}\ln B|^{2} \right) \nonumber \\
 &  &+\; \frac{1}{2}\,J\Omega\;\varrho_{\|}^{2} \left[ |\vb{\kappa}|^{2} \;+\; \langle\alpha_{1}^{2}\rangle \;+\; \frac{1}{2}\,(\nabla\bdot\bhat)^{2} \right] \nonumber \\
 &  &-\,\frac{1}{2}\;J\Omega\,\varrho_{\|}^{2}\;\left\{ \nabla\bdot\left[\bhat\;\left(\nabla\bdot\bhat\right)\right] \;-\; 
3\;\vb{\kappa}\bdot\nabla\ln B \right\} \;+\; \frac{p_{\|}^{2}}{m}\;\varrho_{\|}^{2}|\vb{\kappa}|^{2} \nonumber \\
 &  &-\; \vb{\Pi}_{1}\bdot\frac{\bhat}{m}\btimes\left( \frac{p_{\|}^{2}}{m\Omega}\;\vb{\kappa} \right) \;-\; 2\,\vb{\Pi}_{1}\bdot{\bf v}_{\rm gc} \;-\; J\Omega\;\varrho_{\|}^{2} \left( 2\,|\vb{\kappa}|^{2} + \frac{1}{2}\,\tau^{2} \right) \nonumber \\
 &  &+\; \frac{J^{2}}{3\,m} \left[ 4\,\bhat\bdot\nabla\btimes
{\bf R} + 3\;\tau^{2} - \langle\alpha_{1}^{2}\rangle - 2\;|\bhat\btimes\nabla\ln B|^{2} - 2\;\bhat\bdot\nabla\btimes\left(\bhat\btimes\nabla\ln B\right) \right] \nonumber \\
 &  &-\; \frac{1}{3}\,J\Omega\;\varrho_{\|}^{2} \left[ 4\,\nabla\bdot\vb{\kappa} \;-\; 4\,\vb{\kappa}\bdot\nabla\ln B \;+\; 12\,|\vb{\kappa}|^{2} \;-\;
\frac{7}{2}\,\tau^{2} \;-\; \langle\alpha_{1}^{2}\rangle \right] \nonumber \\
 &  &+\; \Psi_{2} \;+\; \frac{m}{2}\,|{\bf v}_{\rm gc}|^{2} \;-\; \vb{\Pi}_{1}\bdot{\bf v}_{\rm gc} - J\Omega\;\varrho_{\|}^{2} \left( \frac{1}{2}\,
\tau^{2} \;-\; \langle\alpha_{1}^{2}\rangle\right) \nonumber \\
 &  &+\; J\,\Omega\;\nabla\ln B\bdot\left[-\; \vb{\Pi}_{1}\btimes\frac{\bhat}{m\Omega} \;+\; \frac{1}{2} \left( \frac{J}{m\Omega}\;\nabla_{\bot}\ln B \;+\;
\varrho_{\|}^{2}\;\vb{\kappa} \right) \right] \nonumber \\
 &  &+\; \frac{J^{2}}{3\,m} \left[ |\bhat\btimes\nabla\ln B|^{2} \;-\; \vb{\kappa}\bdot\nabla\ln B \;-\; B\;\nabla\bdot\left(B^{-1}\;
\nabla_{\bot}\ln B\right) \right. \nonumber \\
 &  &\left.-\; \frac{1}{2} \left( 5\,\langle\alpha_{1}^{2}\rangle \;+\frac{}{} \tau^{2} \;+\; (\nabla\bdot\bhat)^{2} \right) \right]
\nonumber \\
 &  &+\; \frac{1}{3}\,J\Omega\;\varrho_{\|}^{2} \left[ \vb{\kappa}\bdot\nabla\ln B \;-\; \frac{1}{2}\,B^{2}\;\nabla\bdot\left(\frac{\vb{\kappa}}{B^{2}}
\right) \;-\; \frac{1}{2} \left( 3\,|\vb{\kappa}|^{2} \;-\; \frac{1}{2}\,\tau^{2} \;-\; \langle\alpha_{1}^{2}\rangle\right) \right],
\end{eqnarray*}
which can be simplified to the final expression
\begin{eqnarray}
-\,p_{\|}\,\pd{\Psi_{2}}{p_{\|}} - 2\,J\,\pd{\Psi_{2}}{J} & = & 2\;\frac{p_{\|}^{2}}{m}\;\varrho_{\|}^{2}\,|\vb{\kappa}|^{2} \;-\; 
4\;\vb{\Pi}_{1}\bdot{\bf v}_{\rm gc} \\
 &  &+\; \frac{J^{2}}{3\,m} \left[ 5\,|\nabla_{\bot}\ln B|^{2} \;-\; 2\,\langle\alpha_{1}^{2}\rangle \;+\; 4\;\bhat\bdot\nabla\btimes{\bf R} \;-\; 2\,\bhat\bdot\nabla\btimes\left(\bhat\btimes\nabla\ln B\right) \right. \nonumber \\
 &  &\hspace*{0.5in}\left.-\; B\;\nabla\bdot\left(\frac{\nabla_{\bot}\ln B}{B}\right) \;-\; \vb{\kappa}\bdot
\nabla\ln B \;+\; \frac{1}{2} \left( 5\;\tau^{2} \;-\; (\nabla\bdot\bhat)^{2} \right) \right] \nonumber \\
 &  &+\; \frac{1}{3}\,J\,\Omega\;\varrho_{\|}^{2} \left[ 17\;\vb{\kappa}\bdot\nabla\ln B \;-\; \frac{B^{2}}{2}\;\nabla\bdot\left( 
\frac{\vb{\kappa}}{B^{2}} \right) \;-\; 18\;|\vb{\kappa}|^{2} \;-\; 4\;\nabla\bdot\vb{\kappa} \right. \nonumber \\
 &  &\hspace*{0.5in}\left.+\; \frac{3}{4} \left( \tau^{2} \;+\; (\nabla\bdot\bhat)^{2} \right) \;+\; 6\;\langle\alpha_{1}^{2}\rangle 
\;-\; \frac{3}{2}\;\nabla\bdot\left[\bhat\;(\nabla\bdot\bhat)\right] \right] \nonumber
\end{eqnarray}
Here, using Eq.~\eqref{eq:Hamiltonian_constraint_2_final}, we find
\begin{eqnarray*}
-\,p_{\|}\;\pd{\Psi_{2}}{p_{\|}} & = & -\;2\,J\Omega\;\varrho_{\|}^{2}\,\beta_{2\|} \;+\; 2\;\frac{p_{\|}^{2}}{m}\;\varrho_{\|}^{2}\,|\vb{\kappa}|^{2} 
\;-\; 2\;\vb{\Pi}_{1}\bdot\left( \frac{\bhat}{m}\btimes \frac{p_{\|}^{2}}{m\Omega}\;\vb{\kappa} \right), \\
-\,2\,J\,\pd{\Psi_{2}}{J} & = & -\,2\;\frac{J^{2}}{m}\;\beta_{2\bot} \;-\; 2\,\vb{\Pi}_{1}\bdot{\bf v}_{\rm gc} \;-\; 2\;\vb{\Pi}_{1}\bdot\left( \frac{\bhat}{m}\btimes J\;\nabla\ln B \right),
\end{eqnarray*}
so that
\[ -\,p_{\|}\,\pd{\Psi_{2}}{p_{\|}} - 2\,J\,\pd{\Psi_{2}}{J} \;-\; 2\;\frac{p_{\|}^{2}}{m}\;\varrho_{\|}^{2}\,|\vb{\kappa}|^{2} \;+\; 4\;\vb{\Pi}_{1}\bdot{\bf v}_{\rm gc} \;=\; -\;2\,J\Omega\;\varrho_{\|}^{2}\,\beta_{2\|} \;-\; 2\;\frac{J^{2}}{m}\;\beta_{2\bot}. \]
Hence, we finally obtain
\begin{eqnarray}
-\,6\beta_{2\bot} & = & 5\,|\nabla_{\bot}\ln B|^{2} \;-\; 2\,\langle\alpha_{1}^{2}\rangle \;+\; 4\;\bhat\bdot\nabla\btimes{\bf R} \;-\; 2\,\bhat\bdot\nabla\btimes\left(\bhat\btimes\nabla\ln B\right) \nonumber \\
 &  &-\; B\;\nabla\bdot\left(\frac{\nabla_{\bot}\ln B}{B}\right) \;-\; \vb{\kappa}\bdot\nabla\ln B \;+\; \frac{1}{2} \left( 5\;\tau^{2} \;-\; (\nabla\bdot\bhat)^{2} \right) 
 \label{eq:beta2_bot_constraint} \\
 & = & -6\left\{ -\,\frac{1}{2}\,\tau^{2} - \bhat\bdot\nabla\btimes{\bf R} + \langle\alpha_{1}^{2}\rangle + \frac{1}{2}\bhat\bdot\nabla\btimes\left(\bhat\btimes\nabla\ln B\right) - \left|\bhat\btimes\nabla\ln B\right|^{2} 
 \right\},
 \nonumber
\end{eqnarray}
where we used
\begin{eqnarray*} 
-\,B\;\nabla\bdot\left(\frac{\nabla_{\bot}\ln B}{B}\right) & = & -\;\bhat\bdot\nabla\btimes\left(\bhat\btimes\ln B\right) \;+\; \nabla\ln B\bdot\left(\vb{\kappa} \;+\frac{}{} \nabla_{\bot}\ln B\right) \\
-\,4\;\langle \alpha_{1}^{2}\rangle & = & -\,2\;\bhat\bdot\nabla\btimes{\bf R} \;-\; \frac{1}{2} \left[ \tau^{2} \;+\; 
\left(\nabla\bdot\bhat\right)^{2} \right],
\end{eqnarray*}
and
\begin{eqnarray}
-\,6\;\beta_{2\|} & = & 17\;\vb{\kappa}\bdot\nabla\ln B \;-\; \frac{B^{2}}{2}\;\nabla\bdot\left( \frac{\vb{\kappa}}{B^{2}} \right) \;-\; 
18\;|\vb{\kappa}|^{2} \;-\; 4\;\nabla\bdot\vb{\kappa} \nonumber \\
 &  &+\; \frac{3}{4} \left( \tau^{2} \;+\; (\nabla\bdot\bhat)^{2} \right) \;+\; 6\;\langle\alpha_{1}^{2}\rangle 
\;-\; \frac{3}{2}\;\nabla\bdot\left[\bhat\;(\nabla\bdot\bhat)\right] \nonumber \\
 & = & -\,6\;\left\{ -\,2\;\langle\alpha_{1}^{2}\rangle \;-\; 3\;\vb{\kappa}\bdot\left(\nabla\ln B \;-\frac{}{} \vb{\kappa}\right) \;+\; 
\nabla\bdot\vb{\kappa} \right\},
\label{eq:beta2_par_constraint}
\end{eqnarray}
where we used
\begin{eqnarray*}
-\;\frac{B^{2}}{2}\;\nabla\bdot\left( \frac{\vb{\kappa}}{B^{2}} \right) & = & -\;\frac{1}{2}\;\nabla\bdot\vb{\kappa} \;+\; \vb{\kappa}\bdot\nabla\ln B \\
6\,\langle\alpha_{1}^{2}\rangle & = & \frac{3}{2}\;\nabla\bdot\left[\vb{\kappa} \;-\frac{}{} \bhat\,(\nabla\bdot\bhat)\right] \;+\; \frac{3}{4}
\left[\tau^{2} \;+\; (\nabla\bdot\bhat)^{2} \right].
\end{eqnarray*}
The equalities \eqref{eq:beta2_bot_constraint}-\eqref{eq:beta2_par_constraint}, therefore, validate the push-forward representation \eqref{eq:gc_H_push} of the guiding-center Hamiltonian \eqref{eq:Hgc_push}.

\section{\label{sec:Pgcphi_gyro}Constraint due to the Guiding-center Toroidal Canonical Momentum}

In this last Appendix, we show that the guiding-center toroidal canonical momentum $P_{{\rm gc}\varphi} \equiv {\sf T}_{\rm gc}^{-1}P_{\varphi}$, which is defined as the guiding-center push-forward of the particle toroidal canonical momentum, is explicitly gyroangle-independent, i.e., $P_{{\rm gc}\varphi} \equiv \langle{\sf T}_{\rm gc}^{-1}P_{\varphi}\rangle$. Hence, the gyroangle-dependent terms in ${\sf T}_{\rm gc}^{-1}P_{\varphi}$ must vanish exactly at all orders in $\epsilon$. The importance of this constraint lies in the fact that, in order for the guiding-center toroidal canonical momentum $P_{{\rm gc}\varphi}$ to be an exact (and faithful) invariant, it must satisfy the conservation law 
\begin{equation}
\frac{d_{\rm gc}P_{{\rm gc}\varphi}}{dt} \;\equiv\; {\sf T}_{\rm gc}^{-1}\left(\frac{dP_{\varphi}}{dt}\right) \;=\; 0.
\label{eq:P_phi_dot_exact}
\end{equation}

First, we begin with the expression for the particle toroidal canonical momentum $P_{\varphi} \equiv -\,e\psi/\epsilon\,c + (m\,d{\bf x}/dt)\bdot\partial{\bf x}/\partial\varphi$, whose guiding-center push-forward yields
\begin{equation}
P_{{\rm gc}\varphi} \;=\; -\,\frac{e}{\epsilon c}\;{\sf T}^{-1}_{\rm gc}\psi \;+\; m \left( \frac{d_{\rm gc}{\bf X}}{dt} \;+\; 
\frac{d_{\rm gc}\vb{\rho}_{\rm gc}}{dt}\right)\bdot\left(\frac{\partial_{\rm gc}{\bf X}}{\partial\varphi} \;+\;
\frac{\partial_{\rm gc}\vb{\rho}_{\rm gc}}{\partial\varphi}\right),
\label{eq:Pgcphi_push_App}
\end{equation}
where the guiding-center push-forward of $\psi$ is expressed as
\begin{equation}
{\sf T}^{-1}_{\rm gc}\psi \;=\; \psi \;+\; \epsilon\;\vb{\rho}_{0}\bdot\nabla\psi \;+\; \epsilon^{2}\left[ \vb{\rho}_{1}\bdot\nabla\psi \;+\;
\frac{1}{2}\;\left(\vb{\rho}_{0}\vb{\rho}_{0}\right):\nabla\nabla\psi \right] \;+\; \cdots,
\label{eq:psi_push}
\end{equation}
the guiding-center push-forward of the particle velocity is
\begin{equation}
\frac{d_{\rm gc}{\bf X}}{dt} \;+\; \frac{d_{\rm gc}\vb{\rho}_{\rm gc}}{dt} \;=\; \left( \frac{d_{0}{\bf X}}{dt} \;+\; \Omega\;\pd{\vb{\rho}_{0}}{\theta}
\right) \;+\; \epsilon \left( \frac{d_{1}{\bf X}}{dt} \;+\; \frac{d_{0}\vb{\rho}_{0}}{dt} \;+\; \Omega\;\pd{\vb{\rho}_{1}}{\theta}\right) + \cdots,
\label{eq:velocity_push}
\end{equation}
and the guiding-center push-forward of $\partial{\bf x}/\partial\varphi$ is
\begin{equation}
\frac{\partial_{\rm gc}{\bf X}}{\partial\varphi} \;+\; \frac{\partial_{\rm gc}\vb{\rho}_{\rm gc}}{\partial\varphi} \;=\; {\sf T}_{\rm gc}^{-1}\left(
\pd{\bf x}{\varphi}\right) \;=\; \pd{\bf X}{\varphi} \;+\; \epsilon\;\pd{\vb{\rho}_{0}}{\varphi} \;+\; \cdots.
\label{eq:toroidal_push}
\end{equation}
When Eqs.~\eqref{eq:psi_push}-\eqref{eq:toroidal_push} are inserted in Eq.~\eqref{eq:Pgcphi_push_App}, we obtain the expression (up to second order in $\epsilon$)
\begin{eqnarray}
P_{{\rm gc}\varphi} & = & \left\langle{\sf T}_{\rm gc}^{-1}P_{\varphi}\right\rangle \;+\; \left( m\Omega\,\pd{\vb{\rho}_{0}}{\theta}\bdot
\pd{\bf X}{\varphi} \;-\; \frac{e}{c}\;\vb{\rho}_{0}\bdot\nabla\psi \right) \label{eq:Pgcphi_id} \\
 &  &+\; \epsilon\,p_{\|} \left[ \bhat\bdot\pd{\vb{\rho}_{0}}{\varphi} \;+\; (\vb{\rho}_{0}\bdot\vb{\kappa})\,\bhat\bdot\pd{\bf X}{\varphi} \;+\;
\left( C_{\bot\rho}\;\pd{\vb{\rho}_{0}}{\theta} + C_{\rho\rho}\;\vb{\rho}_{0}\right)\bdot\pd{\bf X}{\varphi} \right] \nonumber \\
 &  &+\; \epsilon\,J \left[ 2({\sf a}_{2}\bdot\nabla\ln B)\bdot\frac{\nabla\psi}{B} + 2\,\frac{{\sf a}_{2}}{B}:\nabla\nabla\psi -
2({\sf a}_{1}\bdot\nabla\ln B)\bdot\pd{\bf X}{\varphi} + \alpha_{1}\;\bhat\bdot\pd{\bf X}{\varphi} \right],
\nonumber
\end{eqnarray}
where gyroangle-dependent terms are shown explicitly up to first order in $\epsilon_{B}$. Since we have the identity
\begin{equation}
P_{{\rm gc}\varphi} \;\equiv\; \left\langle{\sf T}_{\rm gc}^{-1}P_{\varphi}\right\rangle,
\end{equation}
we must, therefore, show that all remaining gyroangle-dependent terms in Eq.~\eqref{eq:Pgcphi_id} must vanish identically.

At zeroth order in $\epsilon_{B}$, we use the magnetic identity
\begin{equation}
{\bf B}\btimes\pd{\bf X}{\varphi} \;\equiv\; \nabla\psi,
\label{eq:mag_id}
\end{equation}
and obtain
\[ \frac{e}{c}\;\vb{\rho}_{0}\bdot\nabla\psi \;=\; m\Omega\;\vb{\rho}_{0}\bdot\bhat\btimes\pd{\bf X}{\varphi} \;=\; m\Omega\,\pd{\vb{\rho}_{0}}{\theta}\bdot\pd{\bf X}{\varphi}, \]
and thus the gyroangle-dependent zeroth-order terms in Eq.~\eqref{eq:Pgcphi_id} cancel each other out.

At first order in $\epsilon_{B}$, we discuss the terms proportional to $p_{\|}$ and $J$ in Eq.~\eqref{eq:Pgcphi_id} separately. First, for the 
$p_{\|}$-terms, using the definitions \eqref{eq:Cij_def}, we find
\[ C_{\bot\rho}\;\pd{\vb{\rho}_{0}}{\theta} + C_{\rho\rho}\;\vb{\rho}_{0} \;\equiv\; \left({\bf I} \;-\frac{}{}\bhat\,\bhat\right)\bdot\nabla\bhat\bdot
\vb{\rho}_{0} \;=\; \nabla\bhat\bdot\vb{\rho}_{0} \;-\; (\vb{\kappa}\bdot\vb{\rho}_{0})\;\bhat, \]
so that
\[ \left[ C_{\bot\rho}\;\pd{\vb{\rho}_{0}}{\theta} + C_{\rho\rho}\;\vb{\rho}_{0} \;+\; (\vb{\kappa}\bdot\vb{\rho}_{0})\;\bhat\right]\bdot
\pd{\bf X}{\varphi} \;=\; \pd{\bf X}{\varphi}\bdot\nabla\bhat\bdot\vb{\rho}_{0} \;\equiv\; \pd{\bhat}{\varphi}\bdot\vb{\rho}_{0}, \]
which combines with the remaining $p_{\|}$-term in Eq.~\eqref{eq:Pgcphi_id} to yield
\[ \epsilon\,p_{\|} \left( \pd{\bhat}{\varphi}\bdot\vb{\rho}_{0} \;+\; \bhat\bdot\pd{\vb{\rho}_{0}}{\varphi} \right) \;=\; \epsilon\,p_{\|}\;\pd{}{\varphi}
\left(\bhat\bdot\vb{\rho}_{0}\right) \;\equiv\; 0, \]
since $\bhat\bdot\vb{\rho}_{0} \equiv 0$.

Next, for the $J$-terms in Eq.~\eqref{eq:Pgcphi_id}, we use the identity
\[ 2\,({\sf a}_{2}\bdot\nabla\ln B)\bdot\frac{\nabla\psi}{B} \;=\; \frac{1}{2} \left[\left(\wh{\bot}\wh{\bot} - \wh{\rho}\wh{\rho}\right)\bdot\nabla\ln B 
\right]\bdot\bhat\btimes\pd{\bf X}{\varphi} \;=\; ({\sf a}_{1}\bdot\nabla\ln B)\bdot\pd{\bf X}{\varphi}, \]
to obtain
\[ 2\,\frac{{\sf a}_{2}}{B}:\nabla\nabla\psi \;-\; 2\,({\sf a}_{2}\bdot\nabla\ln B)\bdot\frac{\nabla\psi}{B} \;+\; \alpha_{1}\;\bhat\bdot
\pd{\bf X}{\varphi} \;=\; 2\,{\sf a}_{2}:\nabla\left(\frac{\nabla\psi}{B}\right) \;+\; \alpha_{1}\;\bhat\bdot\pd{\bf X}{\varphi}. \]
Here, we find
\begin{eqnarray*} 
2\,{\sf a}_{2}:\nabla\left(\frac{\nabla\psi}{B}\right) & = & 2\,{\sf a}_{2}:\nabla\left(\bhat\btimes\pd{\bf X}{\varphi}\right) \;=\;
{\sf a}_{1}:\nabla\left(\pd{\bf X}{\varphi}\right) \;+\; \frac{1}{2}\,\left(C_{\bot\rho} \;+\frac{}{} C_{\rho\bot}\right)\bhat\bdot\pd{\bf X}{\varphi}
\\
 & = & {\sf a}_{1}:\nabla\left(\pd{\bf X}{\varphi}\right) \;-\; \alpha_{1}\;\bhat\bdot\pd{\bf X}{\varphi}.
\end{eqnarray*}
which combines with the remaining $J$-term in Eq.~\eqref{eq:Pgcphi_id} to yield
\[ \epsilon\,J\;{\sf a}_{1}:\nabla\left(\pd{\bf X}{\varphi}\right) \;\equiv\; 0, \]
since ${\sf a}_{1}$ is a symmetric matrix and $\nabla(\partial{\bf X}/\partial\varphi)$ is an antisymmetric matrix so that their trace vanishes.

\bibliographystyle{jpp}

\bibliography{gc_second}

\end{document}